\documentclass[journal,twocolumn,10pt,twoside]{IEEEtran}
\normalsize

\ifCLASSINFOpdf
\else
\fi

\usepackage{amsmath}
\usepackage{amssymb}
\usepackage{cite}
\usepackage{graphicx}
\usepackage{algorithm}
\usepackage{algpseudocode}
\usepackage{subfigure}
\usepackage{multirow}
\usepackage{longtable}
\usepackage{threeparttable}
\usepackage{caption3}
\usepackage{array}
\usepackage{cases}
\usepackage{xcolor}
\usepackage[T1]{fontenc}
\usepackage[utf8]{inputenc}
\usepackage{authblk}
\usepackage{subfigmat}
\usepackage{fixltx2e}
\include{header}

\newtheorem{remark}{Remark}
\newtheorem{example}{Example}

\newcommand{\R}{\textnormal{R}}

\begin{document}

\title{A Primer on Rate-Splitting Multiple Access: \\ Tutorial, Myths, and Frequently Asked Questions}

\author{Bruno Clerckx, \textit{Fellow, IEEE}, Yijie Mao, \textit{Member, IEEE}, Eduard A. Jorswieck, \textit{Fellow, IEEE}, Jinhong Yuan, \textit{Fellow, IEEE}, David J. Love, \textit{Fellow, IEEE}, Elza Erkip, \textit{Fellow, IEEE}, and Dusit Niyato, \textit{Fellow, IEEE} 
\thanks{
Manuscript received August 25, 2022; accepted Dec 15, 2022. This work has been supported in part by the National Nature Science Foundation of China under Grant 62201347.
\textit{(Corresponding author: Yijie Mao.)}

\par B. Clerckx is with the Department of Electrical and Electronic Engineering at Imperial College London, London SW7 2AZ, UK and with Silicon Austria Labs (SAL), Graz A-8010, Austria (email: b.clerckx@imperial.ac.uk).
\par Y. Mao is with the School of Information Science and Technology, ShanghaiTech University, Shanghai 201210, China (e-mail:
maoyj@shanghaitech.edu.cn).
\par E. A. Jorswieck is with the Institute for Communications Technology at TU Braunschweig, Brunswick, Germany (email: e.jorswieck@tu-bs.de).
\par J. Yuan is with the School of Electrical Engineering and
Telecommunications, University of New South Wales, Sydney, NSW 2052, Australia (e-mail: j.yuan@unsw.edu.au).
\par D. J. Love is with the School of Electrical and Computer Engineering, Purdue University, West Lafayette, IN, USA (email: djlove@purdue.edu). 
\par E. Erkip is with the Electrical and Computer Engineering Department, New York University Tandon School of Engineering, Brooklyn, NY 11201, USA (email: elza@nyu.edu).
\par D. Niyato is with the School of Computer Engineering, Nanyang
Technological University, Singapore (email: dniyato@ntu.edu.sg).
}
\\
{\sublargesize\textit{(Invited Paper)}}	
}


\maketitle

\begin{abstract}
Rate-Splitting Multiple Access (RSMA) has emerged as a powerful multiple access, interference management, and multi-user strategy for next generation communication systems. In this tutorial, we depart from the orthogonal multiple access (OMA) versus non-orthogonal multiple access (NOMA) discussion held in 5G, and the conventional multi-user linear precoding approach used in space-division multiple access (SDMA), multi-user and massive MIMO in 4G and 5G, and show how multi-user communications and multiple access design for 6G and beyond should be intimately related to the fundamental problem of interference management. We start from foundational principles of interference management and rate-splitting, and progressively delineate RSMA frameworks for downlink, uplink, and multi-cell networks. We show that, in contrast to past generations of multiple access techniques (OMA, NOMA, SDMA), RSMA offers numerous benefits: 1) enhanced spectral, energy and computation \textit{efficiency}; 2) \textit{universality} by unifying and generalizing OMA, SDMA, NOMA, physical-layer multicasting, multi-user MIMO under a single framework that holds for any number of antennas at each node (SISO, SIMO, MISO, and MIMO settings); 3) \textit{flexibility} by coping with any interference levels (from very weak to very strong), network loads (underloaded, overloaded), services (unicast, multicast), traffic, user deployments (channel directions and strengths); 4) \textit{robustness} to inaccurate channel state information (CSI) and \textit{resilience} to mixed-critical quality of service;  5) \textit{reliability} under short channel codes and \textit{low latency}. We then discuss how those benefits translate into numerous opportunities for RSMA in over forty different applications and scenarios of 6G, e.g., multi-user MIMO with statistical/quantized CSI, FDD/TDD/cell-free massive MIMO, millimeter wave and terahertz, cooperative relaying, physical layer security, reconfigurable intelligent surfaces, cloud-radio access network, internet-of-things, massive access, joint communication and jamming, non-orthogonal unicast and multicast, multigroup multicast, multibeam satellite, space-air-ground integrated networks, unmanned aerial vehicles, integrated sensing and communications, grant-free access, network slicing, cognitive radio, optical/visible light communications, mobile edge computing, machine/federated learning, etc. We finally address common myths and answer frequently asked questions, opening the discussions to interesting future research avenues. Supported by the numerous benefits and applications, the tutorial concludes on the underpinning role played by RSMA in next generation networks, which should inspire future research, development, and standardization of RSMA-aided communication for 6G.
\end{abstract}


\IEEEpeerreviewmaketitle

\section{Introduction}\label{Intro_section}

\IEEEPARstart{C}ommunication systems are inherently multi-user systems. Multiple access (MA) techniques play the crucial role of deciding how to make use of the resources (e.g., time, frequency, power, antenna, and code) to serve those multiple users. Next generation communications systems, e.g., 6G and beyond, will have to cope with increasing demands for high throughput, reliability, heterogeneity of quality of service (QoS), and massive connectivity to satisfy the requirements of further-enhanced mobile broadband (FeMBB), extremely ultra reliable and low-latency communication (eURLLC), ultra massive machine type
communication (umMTC), mixture thereof, and of new services such as integrated sensing and communications (ISAC), integrated satellite-terrestrial, and extended reality. To that end, it is critical to understand how next generation MA can fulfill those demands and requirements by going beyond the conventional orthogonal versus non-orthogonal discussion held in 5G. 

\par Table \ref{tab:Acronyms} details the main abbreviations used throughout this work.

\begin{table*}
\caption{List of abbreviations.}
\label{tab:Acronyms}
\centering
\begin{tabular}{|l|l||l|l|}
\hline

BC    & Broadcast Channel                            & MU--LP   & Multi-User Linear Precoding                               \\
CDMA  & Code Division Multiple Access                & MU-MIMO & Multi-User Multiple-Input Multiple-Output                 \\
CoMP  & Coordinated Multi-Point                & NOMA     & Non-Orthogonal Multiple Access                            \\
CSI   & Channel State  Information                      & NOUM     & Non-Orthogonal Unicast and Multicast                      \\
CSIT/R  & Channel State Information at the Transmitter/Receiver & OFDMA    & Orthogonal Frequency Division Multiple Access             \\
C-RAN & Cloud-Radio  Access  Networks                & OMA      & Orthogonal Multiple Access                                \\
DoF   & Degree-of-Freedom                            & QoS      & Quality of Service                                                 \\
DPC   & Dirty Paper Coding                           &  RF       & Radio Frequency                                       \\
DPCRS & Dirty Paper Coded Rate-Splitting             &   RIS     &      Reconfigurable Intelligent Surfaces                               \\
D2D   & Device-to-Device                             &  RS       & Rate-Splitting                                      \\
EE    & Energy Efficiency                            &  RSMA     & Rate-Splitting Multiple Access                           \\
(F)eMBB  & (further-)enhanced Mobile Broadband Service       &   SC       & Superposition Coding        \\
ER    & Ergodic Rate                                 &  SDMA     & Space Division Multiple Access                     \\
ESR   & Ergodic Sum Rate                             &  SE       & Spectral Efficiency                     \\
FDD   & Frequency  Division  Duplex                  &  SIC      & Successive Interference Cancellation                                    \\
FDMA  & Frequency Division Multiple Access           & SISO     & Single-Input Single-Output                       \\
F-RAN & Fog-Radio Access Networks                    & SNR      & Signal-to-Noise Ratio                                       \\
HK    & Han  and  Kobayashi               &  SWIPT    & Simultaneous  Wireless  Information  and  Power  Transfer                     \\
IC    & Interference Channel         &     SIMO & Single-Input Multiple-Output                            \\
IRS   & Intelligent  Reconfigurable   Surface        &  TDD      & Time  Division  Duplex                                    \\
LLS    &  Link-Level Simulation                   & TDMA     & Time-Division Multiple Access \\
MA    & Multiple Access       & THz   &   TeraHertz \\
MAC     &  Multiple Access Channel &  UAV      & Unmanned Aerial Vehicles                                   \\
MIMO   &    Multiple-Input Multiple-Output       &   (e)URLLC    & (extremely) Ultra-Reliable  Low-Latency  Communication                           \\
MISO  &   Multiple-Input Single-Output                    &  VLC      & Visible Light Communication                            \\
MMF   &    Max-Min  Fairness                           &   V2X      & Vehicle-to-Everything                                \\
(u)mMTC  & (ultra) massive Machine-Type Communication   &   WSR      & Weighted Sum Rate              \\ 
mmWave   & millimeter-Wave                          &    ZFBF     & Zero-Forcing Beamforming                             \\
\hline
\end{tabular}
\end{table*}

\subsection{Beyond Orthogonal versus Non-Orthogonal}

\par In the past decade, MA schemes have often been classified into two categories, namely orthogonal (serving a single user per resource) versus non-orthogonal (serving multiple users per resource). This classification has triggered the question \textit{Is a non-orthogonal approach to MA better than an orthogonal approach?} and has led to the emergence of the rich literature on non-orthogonal multiple access (NOMA) versus orthogonal multiple access (OMA) \cite{noma2013saito,NOMAsurvey2015}. This question was motivated by the claim that prior generations of cellular communication networks are based on OMA serving multiple users on orthogonal resources using time division multiple access (TDMA), frequency division multiple access (FDMA), code division multiple access (CDMA), or orthogonal frequency division multiple access (OFDMA) \cite{multipleAccessReview2017}. 
\par Classifying MA schemes into non-orthogonal vs orthogonal is however over-simplistic as it does not fully reflect modern communication designs. Indeed, those systems are equipped with multiple antennas, and spatial domain processing in the form of multi-user linear precoding (MU--LP), space division multiple access (SDMA), multi-user  multiple-input multiple-output (MU-MIMO), massive MIMO, is an integral part of 4G and 5G for twenty years. Importantly, MU-LP/SDMA/MU-MIMO serves users in a non-orthogonal manner since multiple users are allocated different precoders/beams in the same time-frequency grid and interfere with each other in the same cell. Hence both 4G and 5G already use a combination of orthogonal (in the time-frequency domains) and non-orthogonal (in the spatial domain) approaches in the form of OFDMA combined with SDMA/MU-MIMO/massive MIMO. It is well documented that non-orthogonality can be beneficial in both single and multi-antenna settings \cite{tsefundamentalWC2005, Goldsmithbook2005,clerckx2013mimo}.

\par A major drawback of such classification is that it tends to amalgamate many different MA schemes under the non-orthogonal umbrella without contrasting them or truly understanding the essence of those schemes. This has caused unnecessary confusions and misunderstandings in the past few years \cite{bruno2021MISONOMA}. For instance, SDMA and power-domain NOMA\footnote{In the sequel, we simply use NOMA to refer to power-domain NOMA.} are two different non-orthogonal approaches to MA but are fundamentally different. Indeed, SDMA and NOMA can be seen as two extreme \textit{interference} management strategies where the former \textit{treats interference as noise} and the latter \textit{fully decodes interference} \cite{mao2017rate}. An alternative interpretation of this difference is to note that SDMA (and other forms of linearly precoded and non-linearly precoded MU-MIMO schemes) relies on a \textit{transmit-side interference
cancellation} strategy while NOMA can be seen as a \textit{receive-side interference cancellation} strategy \cite{mao2019beyondDPC}. Such major differences have however not been captured and not been addressed when answering the aforementioned question, though they lead to drastic performance and complexity gaps between the two schemes \cite{bruno2021MISONOMA}. Hence, instead of contrasting orthogonal vs non-orthogonal, a different classification should be considered in next generation wireless networks. In this paper, we will show that the fundamental question behind MA design should instead be \textit{how to manage multi-user interference?} Answering this question will shed the light on the differences between non-orthogonal approaches to MA designs and on a new classification of MA schemes based on how the interference is managed. Even more importantly, this exercise will bring to light the powerful and newly emerging Rate-Splitting Multiple Access (RSMA) for downlink and uplink communications. 

\subsection{Toward Rate-Splitting Multiple Access}

\par RSMA refers to a broad class of multi-user schemes whose commonality is to rely on the \textit{rate-splitting} (RS) principle \cite{carleial1978RS,TeHan1981,EL2011networkIT}. RS consists in splitting a user message (e.g., information bits) into two or multiple parts such that each of those parts can be decoded flexibly at one or multiple receivers. A receiver would have to retrieve each part to reconstruct the original message. A key benefit of RS and its message splitting capability is to flexibly manage inter-user interference, as it will appear clear throughout this paper.

\par Though the RS principle first appeared in the information theoretic literature in the late 70's and early 80's, RS (and consequently the emerging RSMA) has received a renewed interest in the past decade in the broader communication community. What triggered this renewed interest is a different line of research, seemingly unrelated to the MA literature at first, on understanding the fundamental limits of robust interference management, i.e., how to manage interference in a multi-user multi-antenna communications system in the presence of imperfect channel state information at the transmitter (CSIT) \cite{AG2015}. 

\par Conventional multi-user multi-antenna approaches such as SDMA/MU-MIMO heavily rely on timely and highly-accurate CSIT. Unfortunately, in practice, CSIT is always imperfect due to pilot reuse, channel estimation errors, pilot contamination, limited and quantized feedback accuracy, delay/latency, mobility (due to ever-increasing speeds of vehicle/trains/satellite/flying objects and emerging applications as Vehicle-to-Everything), radio frequency (RF) impairments (e.g., phase noise), inaccurate calibrations of RF chains, subband level estimation \cite{RSintro16bruno}. 
Consequently SDMA/MU-MIMO are inherently non-robust. The classical approach to dealing with this practical limitation takes a “robustification” stance – techniques that have been developed under the assumption of perfect CSIT are tweaked to account for imperfect CSIT \cite{Davidson_2008,Vucic_2008}.

\par The challenge is that any CSIT inaccuracy results in a residual multi-user interference that needs proper management instead of “robustification”. Despite its significance in realistic wireless deployments, the fundamental limits of multi-user multi-antenna communications with imperfect CSIT are still an open problem, i.e., the capacity and capacity achieving schemes remain to be found. Consequently, this lack of understanding of how to design robust communication schemes has led to modern systems like 4G and 5G designs to be fundamentally designed for perfect CSIT (using SDMA and treat interference as noise) instead of being designed from scratch to be truly robust to imperfect CSIT \cite{RSintro16bruno,mao2019beyondDPC}. 

\par Interestingly, we are now in a much better position to design truly robust MIMO wireless networks accounting for imperfect CSIT and its resulting multi-user interference \cite{AG2015,RSintro16bruno,mao2019beyondDPC}. It is indeed known that to benefit from imperfect CSIT and tackle the multi-user interference, the transmitter should take an RS approach that splits the messages into common and  private parts, encodes the common parts into a common stream, and private parts into private streams and superposes in a non-orthogonal manner the common stream on top of all private streams \cite{RSintro16bruno}. The common stream is decodable by all receivers, while the private streams are to be decoded by their corresponding receivers only. Such an approach is optimal from an information theoretic perspective (Degrees-of-Freedom, DoF\footnote{The Degrees-of-Freedom (DoF), or  multiplexing gain, is a first-order approximation
of the rate at high signal-to-noise ratio (SNR). It can be viewed as the pre-log factor of the rate at high SNR and
be interpreted as the number or fraction of interference-free
stream(s) that can be simultaneously communicated to a user (or multiple users). The DoF achieved depends on the communication strategy used. The larger the DoF,
the faster the rate increases with the SNR. Hence, ideally a communication strategy should achieve the highest DoF possible. Readers are referred to \cite{bruno2021MISONOMA} for more details on various definitions used to assess the DoF performance of multiple access schemes.}) for downlink multi-user mutliple-input single-output (MISO) and MIMO transmissions with imperfect CSIT \cite{Davoodi_gdof_2018,RS2016hamdi,enrico2017bruno,hamdi2019spawc,Davoodi2021DoF,chenxi2017bruno}.

\par This literature opens various doors that have a major impact on MA designs. First, since imperfect CSIT is more general than perfect CSIT\footnote{Perfect CSIT is obtained by setting the channel estimation error in the imperfect CSIT model to zero.}, finding efficient schemes for imperfect CSIT leads to discovering a broader and general class of communication strategies that would subsume perfect CSIT strategies as particular instances. Second, it gives communication engineers clear and fundamentally grounded guidelines on how to design robust schemes. Third, it provides refreshing and new thoughts about low complexity non-orthogonal scheme that are applicable and beneficial even in perfect CSIT settings. Fourth, it brings RS, originally developed for the two-user single-antenna interference channel \cite{carleial1978RS,TeHan1981}, into MU-MIMO, which was never investigated despite the rich literature on MU-MIMO schemes in the past two decades \cite{clerckx2013mimo}. Fifth, RS serves users in a non-orthogonal manner by \textit{partially treating interference as noise} and \textit{partially decoding interference}. This highlights the usefulness to depart from the extremes of (fully) treat interference as noise (as in SDMA) and (fully) decode interference (as in NOMA) in multi-antenna networks, but also the usefulness to bridge and unify those two extremes \cite{mao2017rate}. Sixth, RS can be seen as a \textit{smart combination of transmit-side and receive-side interference cancellation} strategy where the contribution of the common stream is adjusted to benefit from the best balance between transmit and receive cancellation. This departs from the transmit-side only and receive-side only interference cancellation strategies of SDMA (and the rich literature on MU-MIMO) and NOMA, respectively \cite{mao2019beyondDPC}.

\par The above design makes RS a fundamental building block
of a powerful MA framework for downlink and uplink communications, namely, RSMA \cite{mao2017rate,Rimo1996}. 
\par In the \textit{downlink}, RSMA uses linearly or non-linearly precoded RS at the transmitter to split each user message into one or multiple common messages and a private message. The common messages are combined and encoded into common streams for the intended users. Successive interference cancellation (SIC) - or any other forms of joint decoding - is used at each user to sequentially decode the intended common streams (and therefore decode part of the interference). Such linearly or non-linearly precoded generalized RSMA has been demonstrated to be a powerful framework to bridge, reconcile, and generalize SDMA, NOMA, OMA, and physical-layer multicasting and further boost system spectral and energy efficiencies for downlink transmissions with both perfect and partial CSIT \cite{mao2017rate,bruno2019wcl,Maosurvey}. 
\par In the \textit{uplink}, users split their message into multiple streams and allocate proper transmit power to each stream. The receiver at the base station performs SIC to retrieve each stream and reconstruct the original messages. By splitting messages, inter-user interference can be dynamically managed without utilizing time sharing among users to achieve the capacity. Here again, NOMA is a subset of RSMA and is obtained when user messages are not split. This message splitting property can enable core services where users have intermittent access behaviour such as URLLC and mMTC, as well as enable different combinations of core services to serve users with heterogeneous profiles \cite{Rimo1996,CL1_2022}.

     



\subsection{Challenges and Opportunities for RSMA}

In our view, MA schemes will have to address new challenges and requirements in next generation networks:
\begin{itemize}
    \item \textit{Efficient}: Time, frequency, power, spatial, and code-domain resources should be used to serve users in the most efficient way, from both spectral efficiency and energy efficiency perspectives, while enhancing QoS and fairness across devices and services (e.g., unicast, multicast) and making the best use of the available computational resources (transmitter and receiver complexity). This calls for performance evaluations of MA schemes beyond OMA vs NOMA \cite{bruno2021MISONOMA}.
    \item \textit{Universal}: Simpler is better, i.e., a single unified and general MA scheme would be easier to implement and optimize than a combination of multiple MA schemes, each optimized for specific conditions. This calls for a deeper understanding of non-orthogonality in MA designs and of how non-orthogonal approaches to MA schemes relate to each other \cite{bruno2021MISONOMA,bruno2019wcl}.
    \item \textit{Flexible}: Wireless networks are dynamic and MA schemes should be flexible or versatile enough to cope with various interference levels, network density and load (underloaded, overloaded), topology, services, user distribution, channel directions and strengths in general deployments. This calls for MA schemes that depart from those two extreme interference management strategies, namely fully treat interference as noise and fully decode interference \cite{mao2017rate}.
    \item \textit{Robust and Resilient}:  
    MA designs need to depart from the conventional aforementioned “robustification” stance and adopt a true robustness to practical scenarios subject to imperfect CSIT \cite{RSintro16bruno}. Aside being robust, MA schemes need to also become increasingly resilient to emerging services involving mixed-criticality, i.e., services with different priorities. This is particularly relevant for applications in safety-critical contexts such as driverless traffic \cite{Reifert_2022_MC}. 
    \item \textit{Low Latency and Reliable}: Providing reliable and low-latency communications for intelligent transportation and industrial automation is key in next generation communications. Critical sources of latency relate to link establishment, packet re-transmissions, and data blocklength. Hence, an MA scheme needs to be able to perform reliably under short channel codes and decrease the number of re-transmissions \cite{onur2020sixG,onur2021sixG}.
\end{itemize}

RSMA uniquely appeared in recent years to fulfill all these requirements \cite{Maosurvey}, thanks to its inherent message splitting capability, which is not featured in any other MA schemes. This split capability provides several benefits at once as it allows to: 1) partially decode interference and partially treat interference as noise (hence its efficiency, flexibility, reliability, and resilience), 2) reconcile the two extreme strategies of interference management and multiple MA schemes into a single framework (hence its generality/universality), 3) achieve the optimal DoF in practical scenarios subject to imperfect CSIT (hence its robustness). This contrasts with conventional approaches like OMA, SDMA, and NOMA, that achieve only some of the aforementioned features \cite{bruno2021MISONOMA}. Indeed, SDMA has low complexity and works well in perfect CSIT conditions especially in underloaded regimes, while downlink multi-antenna NOMA is inefficient in general as it incurs a severe DoF loss despite an increased receiver complexity due to an inefficient use of SIC receivers \cite{bruno2021MISONOMA}. OMA, SDMA, and NOMA are not general as they are suited for particular propagation conditions \cite{bruno2019wcl}. Similarly they are not as flexible as RSMA since they rely on eliminating interference by orthogonalization (OMA), treat interference as noise (SDMA), and decode interference (NOMA), which are suboptimal strategies in general settings \cite{mao2017rate,Tse2008}. Though NOMA can be made more robust than SDMA in the presence of imperfect CSIT, the DoF of NOMA and SDMA are both suboptimal, therefore incurring a rate loss compared to RSMA \cite{bruno2021MISONOMA}. Similarly SDMA has also been shown to be less resilient than RSMA \cite{Reifert_2022_EE_MC}. Finally, SDMA and NOMA are less reliable with finite block code lengths \cite{yunnuo2021FBL,Onur2020LLS,yunnuo2022FBL}.

\subsection{Objectives and Contributions}

\par It is worth to point out (again) that SDMA, NOMA, and RSMA are all non-orthogonal approaches to MA design. However, classifying them all as non-orthogonal does not give us a clue as to where the difference in efficiency, universality, flexibility, robustness and resilience, reliability and low latency comes from. Answering the fundamental question \textit{how to manage multi-user interference?} is crucial because it helps us to design efficient MA schemes, to reveal how MA schemes manage interference, and identify and predict under what conditions a given MA scheme may be efficient, universal, flexible, robust, resilient, reliable and have low latency.

\par In this paper, we provide a tutorial on RSMA and address common myths and frequently asked questions. To that end, we make the following contributions:
\begin{itemize}
    \item We start this tutorial by departing from the orthogonal vs non-orthogonal classification and rather classify schemes as a function of how they manage interference. We go back to the basics of RS proposed for the two-user interference channel, and show step by step, how to build downlink, uplink, and multi-cell RSMA and its various schemes. We also show how interference management and MA designs are closely related and how OMA, SDMA, NOMA, and RSMA schemes differ in terms of how to manage interference. We then explain that by understanding the interference management capability of a scheme, we can understand the drawbacks and benefits of the MA schemes. 
    \item We generalize the downlink, uplink, and multi-cell RSMA frameworks to MIMO settings with multiple antennas at all nodes. This builds upon recent efforts to design MIMO RSMA \cite{anup2021MIMO}, contrast with recent survey and works \cite{Maosurvey,bruno2021MISONOMA} that were primarily limited to MISO settings, and confirm that the generality and universality of RSMA are not limited to single-input single-output (SISO), MISO, or single-input multiple-output (SIMO) settings, but hold for general MIMO settings. Consequently, recent MIMO NOMA schemes \cite{schober2021MUMIMONOMA1,schober2021MUMIMONOMA2} are shown particular instances of MIMO RSMA.   
    \item We demonstrate that, in contrast to previous generation MA schemes (as OMA, SDMA, NOMA), RSMA is spectrally and energy efficient, universal, flexible, robust and resilient, reliable and low latency. Those unique features of RSMA are discussed in over 40 applications of RSMA in different systems, scenarios, and services. We briefly describe how RSMA is beneficial for each of those applications to demonstrate how powerful RSMA is for next generation networks.
    \item We discuss numerous myths/misunderstandings and frequently asked questions about RSMA that we have personally witnessed. Myths and questions cover a wide range of topics, ranging from RSMA relationship with NOMA, gain over MU-MIMO and massive MIMO, role and content of common streams, standardization status of RSMA, implementation complexity of RSMA, interplay with other techniques, role of machine learning in RSMA, etc. Those myths and frequently asked questions about RSMA nicely complement the tutorial part.
\end{itemize}
\par This paper also differs from the recent survey paper on RSMA \cite{Maosurvey} by providing a tutorial flavour instead of survey, by diving further into the key role of interference management, by addressing myths and frequently asked questions, and by expanding the framework to general MIMO settings. It also differs from \cite{CL1_2022,CL2_2022,CL3_2022} that give an overview of RSMA and focus on the specific interplay between RSMA and ISAC and reconfigurable intelligent surfaces (RIS). 
\par 
\subsection{Organization and Notations}
The remainder of this paper is organized as follows.
Section \ref{section_interference} introduces fundamentals about how to manage interference, delineates the RS strategy for the 2-user interference channel, relates the problem of MA design with interference management, and draws important observations.   
Section \ref{section_twouser_RSMA} builds upon important lessons drawn from previous section and introduces various architectures of 2-user RSMA for both downlink and uplink for general MIMO settings at all nodes. Important lessons about the relationships between existing MA schemes and RSMA are summarized.
Section \ref{section_Kuser} extends the RSMA design to general $K$-user settings for downlink, uplink and multi-cell deployments.
Section \ref{section_applications} shows how the RSMA architectures find applications in 40 different scenarios relevant in 6G. The breadth and depth of those applications and scenarios showcase how powerful and underpinning RSMA is for 6G.
Section \ref{section_myths} debunks some myths that have appeared in the RSMA literature. 
Section \ref{section_asked_questions} presents answers to several common questions that are often asked about RSMA. 
Section \ref{conclusions} concludes this tutorial overview. 
\par \textit{Notations:} Bold upper and lower case letters denote matrices and column vectors, respectively.
{$(\cdot)^{T}$,   $(\cdot)^{H}$}, $|\cdot|$, $\|\cdot\|$, $\mathbb{E}\{\cdot\}$, and $\mathrm{tr}(\cdot)$ represent the transpose, Hermitian, absolute value, Euclidean norm, expectation, and trace  operators, respectively.  $\mathcal{CN}(0,\sigma^2 )$ denotes the circularly symmetric complex Gaussian (CSCG) distribution with zero mean and variance $\sigma^2$.
 
\section{Key Question behind Multiple Access Design: How to Manage Interference?}\label{section_interference}

\par In this section, we present fundamentals about how to manage interference. To that end, we focus on an example of a symmetric two-user interference channel, which is the simplest scenario that enables to capture the essence of interference management. We then use that simple example to draw major lessons for MA design.

\subsection{Interference Channel}

\begin{figure}
	\centering
	\includegraphics[width=0.9\columnwidth]{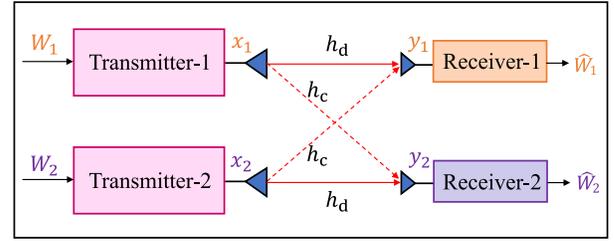}
	\caption{Two-user symmetric Gaussian interference channel.}
	\label{fig_2user_IC}
\end{figure}

\par Let us consider a two-user symmetric Gaussian interference channel (IC) of Fig. \ref{fig_2user_IC}. We have four nodes, two transmitters (Tx) and two receivers (Rx), each equipped with a single antenna. Tx-1 (resp. Tx-2) wants to transmit a message $W_1$ (resp. $W_2$) to Rx-1 (resp. Rx-2). To simplify the number of parameters and obtain some insight into the role of interference, we assume some symmetry in that the direct channel from a Tx to its intended Rx (Tx-1 $\rightarrow$ Rx-1, Tx-2 $\rightarrow$ Rx-2) is $h_{\mathrm{d}}$ and the cross channel (Tx-1 $\rightarrow$ Rx-2, Tx-2 $\rightarrow$ Rx-1) is $h_{\mathrm{c}}$. Hence, the stronger $h_{\mathrm{c}}$ compared to $h_{\mathrm{d}}$, the stronger the interference from Tx-$k$ to Rx-$j$, $j \neq k$.
\par A few strategies come to mind to manage this interference:
\begin{itemize}
    \item \textit{Orthogonalize}: Tx-1 and Tx-2 transmit on orthogonal resources (e.g., time or frequency) so that that they do not interfere with each other at all. This is suboptimal since this strategy does not care about how weak the interference is.
    \item \textit{Treat Interference as} (additive white Gaussian) \textit{Noise}: This is a natural approach whenever $h_{\mathrm{c}}$ is very weak compared to $h_{\mathrm{d}}$ since the interference created by Tx-$k$ would be buried in the noise of Rx-$j$, $j \neq k$. The drawback of this strategy is that the interference actually carries information and has a structure that could potentially be exploited in mitigating its effect \cite{Tse2008}. 
    \item \textit{Decode Interference}: Fully decode the interference at the receivers is especially relevant and actually optimal when the interference is strong enough that $h_{\mathrm{c}}$ is large compared to $h_{\mathrm{d}}$ \cite{carleial1975}. In that case, Rx-$k$ has a better reception of Tx-$j$'s signal than the intended receiver Rx-$j$, $k \neq j$.
\end{itemize}
However, we can do more than those seemingly three different strategies by adopting a \textit{rate-splitting} (RS) approach \cite{carleial1978RS,TeHan1981,Tse2008}. RS splits the transmitted information into two parts, namely a common part to be decoded by both receivers and a private part to be decoded only by the intended receiver. The key benefit of RS is that by enabling a receiver to decode the common part, part of the interference is cancelled off, while the remaining private part from the other transmitter is treated as noise. This enables RS (by properly adjusting the power to the common and private parts) to \textit{partially decode interference} and \textit{partially treat the remaining interference as noise}, therefore bridging, unifying, and actually outperforming the two extremes of treat interference as noise and decode interference, as it will appear clearer in the sequel.

\subsection{Rate-Splitting for Two-User Interference Channel}

\begin{figure}
	\centering
\includegraphics[width=0.45\textwidth]{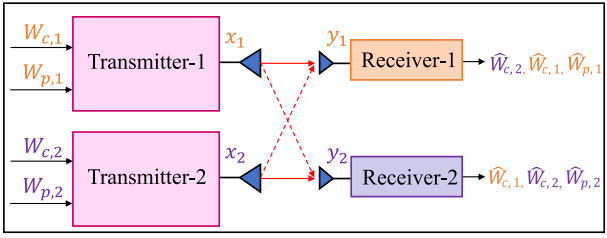}%
	\caption{RS for two-user SISO IC (HK scheme) \cite{TeHan1981}.}
	\label{fig:2UserRSMA_IC}
\end{figure}

\par We adopt an RS strategy\footnote{In the two-user IC, RS is also known as Han-Kobayashi strategy \cite{TeHan1981}.} to split each transmitter message $W_k$ into two parts denoted as a \textit{common} part $W_{\mathrm{c},k}$ and a \textit{private} part $W_{\mathrm{p},k}$ as per Fig. \ref{fig:2UserRSMA_IC} \cite{carleial1978RS,TeHan1981,Tse2008}. On the one hand, a codebook shared between both transmitters\footnote{This is not an issue in modern systems since, for example, in an 4G Long Term Evolution (LTE)}/5G New Radio (NR) system, all codebooks are shared since all users use the same family of modulation and coding schemes (MCS) specified in the standard. is used to encode the common parts $W_{\mathrm{c},k}$, $k=1,2$, into the common stream $s_{\mathrm{c},k}$, $k=1,2$, such that they are decodable by both receivers. The private parts $W_{\mathrm{p},k}$, $k=1,2$ are on the other hand encoded into private streams $s_k$, $k=1,2$, constructed from independent codebooks. 
Tx-$k$ then transmits the superposition of its common stream $s_{\mathrm{c},k}$ and private stream $s_{k}$ with proper power allocation to both streams as
\begin{equation}\label{RS_IC_eq}
x_k=\sqrt{P_{\mathrm{c}}} s_{\mathrm{c},k} + \sqrt{P_k} s_{k}.
\end{equation}
By defining $\mathbf{s}_k=[s_{\mathrm{c},k},s_k]^T$ and assuming that $\mathbb{E}[\mathbf{s}_k\mathbf{s}_k^H]=\mathbf{I}$, the average transmit (sum) power constraint at each transmitter is $P$ with $P_{\mathrm{c}}$ and $P_k=P-P_{\mathrm{c}}$ being the power allocated to the common and private stream respectively. Let us also denote by $t=\frac{P_{\mathrm{c}}}{P}$ the fraction of the transmit power allocated to the common stream.

\par The received signals at both receivers can be written as 
\begin{align}
    y_1=h_{\mathrm{d}} x_1 + h_{\mathrm{c}} x_2 +n_1,\\
    y_2=h_{\mathrm{c}} x_1 + h_{\mathrm{d}} x_2 +n_2,
\end{align}
where $n_k\sim\mathcal{CN}(0,\sigma_{n,k}^2)$ is the additive white Gaussian noise (AWGN). Without loss of generality, we assume the variance $\sigma_{n,k}^2=1$. We also assume perfect CSIT and perfect CSI at the receivers (CSIR).



\par Each Rx decodes the common streams and its intended private stream as illustrated in Fig. \ref{fig:2UserRSMA_IC}. To that end, each Rx jointly decodes the common streams into messages $\widehat{W}_{\mathrm{c},1}$ and $\widehat{W}_{\mathrm{c},2}$ 
by treating the private streams as noise. Then Rx-$k$ cancels the common streams from the received signal and decodes the intended private stream $s_k$ into message $\widehat{W}_{\mathrm{p},k}$ by treating the other private stream $s_j$, $j\neq k$, as noise. Rx-$k$ finally reconstructs the original message by recombining $\widehat{W}_{\mathrm{p},k}$ and $\widehat{W}_{\mathrm{c},k}$ into $\widehat{W}_{k}$, which is the same as $W_{k}$ in the absence of any decoding error. By doing so, Rx-$k$ has partially decoded the message of Rx-$j$, namely $\widehat{W}_{\mathrm{c},j}$, and therefore cancelled off part of the interference.
\par Because of symmetry, the rates of the common streams $s_{\mathrm{c},1}$ and $s_{\mathrm{c},2}$ are the same and are simply denoted as $R_{\mathrm{c}}$. Similarly, the rates of the private streams $s_{1}$ and $s_{2}$ are also the same and are denoted as $R_{\mathrm{p}}$. Assuming Gaussian signaling and infinite blocklength, $R_{\mathrm{c}}$ must satisfy the following inequalities at Rx-1 (and equivalently at Rx-2 due to symmetry)
\begin{align}
    R_{\mathrm{c}}&\leq \log_2\left(1+\frac{t P\left|h_{\mathrm{d}}\right|^2}{1+I}\right),\label{rate_Rc_1}\\
    R_{\mathrm{c}}&\leq \log_2\left(1+\frac{t P\left|h_{\mathrm{c}}\right|^2}{1+I}\right),\label{rate_Rc_2}\\
    2 R_{\mathrm{c}}&\leq\log_2\left(1+\frac{t P\left[\left|h_{\mathrm{d}}\right|^2+\left|h_{\mathrm{c}}\right|^2\right]}{1+I} \right),\label{rate_Rc_3}
\end{align}
where $I=\left(1-t\right)P\left[\left|h_{\mathrm{d}}\right|^2+\left|h_{\mathrm{c}}\right|^2\right]$. Inequalities \eqref{rate_Rc_1}, \eqref{rate_Rc_2}, and \eqref{rate_Rc_3} originate from the fact that the common streams are jointly decoded first at each receiver by treating the private streams as noise. Those three inequalities are obtained by writing the rate constraints of a two-user multiple access channel (MAC) formed by two virtual transmitters sending respectively $s_{\mathrm{c},1}$ and $s_{\mathrm{c},2}$ to Rx-1 (or 2 due to symmetry) subject to the additional noise power $I$ created by the private streams $s_1$ and $s_2$. Hence, \eqref{rate_Rc_1} refers to the individual rate constraint of decoding $s_{\mathrm{c},1}$ at Rx-1 by treating $s_1$ and $s_2$ as noise. Similarly, \eqref{rate_Rc_2} refers to the individual rate constraint of decoding $s_{\mathrm{c},2}$ at Rx-2 by treating $s_1$ and $s_2$ as noise. The sum-rate constraint writes as \eqref{rate_Rc_3}. 
\par The rate $R_{\mathrm{p}}$ of the private streams in \eqref{rate_Rp} is obtained by noticing that a private stream is decoded after cancelling the common streams while treating the other interfering private stream as noise. We can therefore write
\begin{equation}
 R_{\mathrm{p}}\leq \log_2\left(1+\frac{\left(1-t\right)P\left|h_{\mathrm{d}}\right|^2}{1+\left(1-t\right)P\left|h_{\mathrm{c}}\right|^2}\right)\label{rate_Rp},
\end{equation}
where the presence of $\left(1-t\right)P\left|h_{\mathrm{c}}\right|^2$ at the denominator expresses the interference from the private stream of Tx-$k$ owed to Rx-$j$ private stream.
\par Combining \eqref{rate_Rc_1}-\eqref{rate_Rc_3} and \eqref{rate_Rp}, the achievable (symmetric) rate in this two-user IC is written as the sum of the private and common rates $R_{\mathrm{sym}}=R_{\mathrm{p}}+R_{\mathrm{c}}$ in \eqref{R_sym}. The terminology rate-splitting appears clearly here where we note that $R_{\mathrm{sym}}$ is split into two parts, namely, the private rate $R_{\mathrm{p}}$ and the common rate $R_{\mathrm{c}}$. The benefit of this RS architecture is the ability to adjust the power allocation to the common and private streams through parameter $t$ (and therefore the amount of information to be carried by the common and private streams) to maximize $R_{\mathrm{sym}}$ depending on the channel strengths of $h_{\mathrm{d}}$ and $h_{\mathrm{c}}$.
\begin{table*}[!t]
\begin{multline}
    R_{\mathrm{sym}}=\log_2\left(1+\frac{\left(1-t\right)P\left|h_{\mathrm{d}}\right|^2}{1+\left(1-t\right)P\left|h_{\mathrm{c}}\right|^2}\right)+\min\left\{\log_2\left(1+\frac{t P\left|h_{\mathrm{d}}\right|^2}{1+I}\right),\log_2\left(1+\frac{t P\left|h_{\mathrm{c}}\right|^2}{1+I}\right),\frac{1}{2}\log_2\left(1+\frac{t P\left[\left|h_{\mathrm{d}}\right|^2+\left|h_{\mathrm{c}}\right|^2\right]}{1+I} \right)\right\}\label{R_sym}
\end{multline}
\hrulefill
\end{table*}

\begin{remark} The terminology \textit{common} and \textit{private} is taken from the information theory literature. In simple words, common (also sometimes called \textit{public}) simply refers to a message or stream that is to be decoded by multiple users though the content of the message is not necessarily intended to all those users. Indeed, $s_{\mathrm{c},k}$, $k=1,2$, are decoded by both receivers though the content of $s_{\mathrm{c},k}$ is only intended to Rx-$k$. Common also contrasts with a multicast message that is a message decoded by multiple users but that is genuinely intended to all those users. Private on the other hand refers to a message or stream that is only decoded by its intended receiver and is treated as noise at other receivers. Indeed, $s_k$ is decoded only by Rx-$k$ and is treated as noise by Rx-$j$ with $j \neq k$. Messages $W_k,k=1,2$, are unicast messages (each intended to a single user), but they are split into a common and private parts for interference management benefits.
\end{remark}

\subsection{Interference Regimes}\label{subsection_Interf_regimes}

\par We can identify four interference regimes depending on the relative strengths\footnote{In the asymmetric IC, the interference regimes take a more complicated form \cite{Tse2008}.} of $h_{\mathrm{d}}$ and $h_{\mathrm{c}}$:

    \par \textit{Very weak} ($\left|h_{\mathrm{c}}\right|\ll\left|h_{\mathrm{d}}\right|$): Whenever $\left|h_{\mathrm{c}}\right|$ is much smaller than $\left|h_{\mathrm{d}}\right|$, the maximization of $R_{\mathrm{sym}}$ ends up allocating all the transmit power to the private stream such that $t=0$ and $R_{\mathrm{sym}}=\log_2\left(1+\frac{P\left|h_d\right|^2}{1+P\left|h_c\right|^2}\right)$. In such a regime,  message $W_k$ is entirely encoded in the private stream $s_k$ (i.e., no splitting of the messages occurs) such that any interference from Tx-$k$ to Rx-$j$ is treated as noise. This corresponds to the regime where the \textit{Treat Interference as Noise} strategy is optimal.
    \par \textit{Weak} ($\left|h_{\mathrm{c}}\right|\leq\left|h_{\mathrm{d}}\right|$): As the strength of $h_{\mathrm{c}}$ increases relatively to $h_{\mathrm{d}}$ but as long as $\left|h_{\mathrm{c}}\right|$ remains smaller than $\left|h_{\mathrm{d}}\right|$, one enters the weak interference regime where splitting the messages becomes necessary to maximize $R_{\mathrm{sym}}$. In that regime, inequality \eqref{rate_Rc_1} is inactive (since r.h.s. of \eqref{rate_Rc_2} < r.h.s. of \eqref{rate_Rc_1}) and $t>0$ is to be chosen, i.e.,\ a non-zero power is allocated to the common stream. There exists a tradeoff between achieving a large private rate and minimizing the interference caused to the other receiver. A qualitative and insightful way of allocating power to the common stream is to choose $t$ so that the interference level caused by the private stream has roughly the same level as the other receiver's noise level, i.e., from \eqref{rate_Rp} choose $t$ such that $\left(1-t\right)P\left|h_{\mathrm{c}}\right|^2 \approx 1$ or equivalently $t\approx \frac{P\left|h_{\mathrm{c}}\right|^2-1}{P\left|h_{\mathrm{c}}\right|^2}$ \cite{Tse2008}. By doing so, the interference caused by a private stream has little impact on the other receiver's performance (compared to the impairments already caused by the noise). At the same time, it does not prevent each transmitter from experiencing a relatively large private rate as long as $\left|h_{\mathrm{c}}\right|\leq\left|h_{\mathrm{d}}\right|$.
    \par \textit{Strong} ($\left|h_{\mathrm{d}}\right|^2\leq \left|h_{\mathrm{c}}\right|^2 \leq \left|h_{\mathrm{d}}\right|^2\big(1+P\left|h_{\mathrm{d}}\right|^2\big)$): As the strength of $h_c$ further increases relatively to $h_{\mathrm{d}}$ and enters the regime where $\left|h_{\mathrm{c}}\right|$ is larger than $\left|h_{\mathrm{d}}\right|$, each receiver is able to decode both the interfering signal and the desired signal by performing for instance SIC or joint decoding. In other words, $R_{\mathrm{sym}}$ is maximized by choosing $t=1$ and messages $W_k$, $k=1,2$, are entirely encoded in the common streams $s_{\mathrm{c},k}$, respectively, so that they are both decoded by both receivers (i.e.,\ there are no private streams in this regime and $I=0$). With $\left|h_{\mathrm{c}}\right|\geq\left|h_{\mathrm{d}}\right|$, inequality \eqref{rate_Rc_2} is inactive and $R_{\mathrm{sym}}=\frac{1}{2}\log_2\big(1+ P\big[\left|h_d\right|^2+\left|h_c\right|^2\big] \big)$ as long as $\frac{1}{2}\log_2\big(1+ P\big[\left|h_{\mathrm{d}}\right|^2+\left|h_{\mathrm{c}}\right|^2\big] \big)\leq\log_2\big(1+ P\left|h_d\right|^2\big)$, i.e.,\ $P\left|h_{\mathrm{d}}\right|^2\leq P\left|h_{\mathrm{c}}\right|^2 \leq P\left|h_{\mathrm{d}}\right|^2\big(1+P\left|h_{\mathrm{d}}\right|^2\big)$. In other words, the sum-rate constraint \eqref{rate_Rc_3} is active and forces each transmitter to transmit at a rate smaller than $\log_2\big(1+ P\left|h_d\right|^2\big)$, i.e., smaller than the rate achievable without any interference. It is worth noting that $\frac{1}{2}\log_2\big(1+ P\big[\left|h_{\mathrm{d}}\right|^2+\left|h_{\mathrm{c}}\right|^2\big] \big)\leq\log_2\big(1+ P\left|h_{\mathrm{d}}\right|^2\big)$ can equivalently be written as $\log_2\big(1+ \frac{P\left|h_{\mathrm{c}}\right|^2}{1+P\left|h_{\mathrm{d}}\right|^2} \big)\leq\log_2\big(1+ P\left|h_{\mathrm{d}}\right|^2\big)$, which expresses that when Rx-$k$ performs SIC to decode the interfering signal $s_{\mathrm{c},j}$ ($j \neq k$) before decoding $s_{\mathrm{c},k}$, the decodability of $s_{\mathrm{c},j}$ at Rx-$k$ puts a constraint on the transmission rate of $s_{\mathrm{c},j}$ at Tx-$j$. 
    \par \textit{Very strong} ($\left|h_{\mathrm{c}}\right|^2 \geq \left|h_{\mathrm{d}}\right|^2\big(1+P\left|h_{\mathrm{d}}\right|^2\big)$): In this regime, the interference link is even stronger such that $\frac{1}{2}\log_2\big(1+ P\big[\left|h_d\right|^2+\left|h_c\right|^2\big] \big)\geq\log_2\big(1+ P\left|h_d\right|^2\big)$ and the sum-rate constraint \eqref{rate_Rc_3} becomes inactive. In other words, when performing SIC, the decodability of $s_{\mathrm{c},j}$ at Rx-$k$ does not restrict the transmission rate of $s_{\mathrm{c},j}$ at transmitter $j$. Instead the rates are only limited by the direct links and each transmitter can transmit at a rate $R_{\mathrm{sym}}=\log_2\big(1+ P\left|h_d\right|^2\big)$ equal to the one achievable without any interference. The strong and very strong interference regimes correspond to the regimes where the \textit{Decode Interference} strategy is optimal.

\par What distinguishes the weak interference regime from the strong (and very strong) regime is that the interference in the former is not strong enough so that only a part of the interference (through the common message) can be decoded by a receiver while the interference in the latter is so strong that the receiver can fully decode the interference. Consequently, moving from the very weak interference regime to the strong interference regime, the messages are respectively encoded into private streams only, a mixture of common and private streams, and into common streams only as further summarized in Table \ref{fig_mapping}. 

\begin{table}
	\caption{Messages-to-streams mapping in the two-user IC.}
	\centering
\begin{tabular}{|p{.2\textwidth}|p{.6\textwidth}|p{2.3cm}|}
\hline
\multicolumn{1}{|c|}{}             & \multicolumn{1}{c|}{$s_k$}    & \multicolumn{1}{c|}{$s_{\mathrm{c},k}$}          \\ \hline
\multicolumn{1}{|c|}{ Very weak}         & \multicolumn{1}{c|}{\color{blue} $\,\,\quad \quad W_1\quad\quad\,\,$}     & \multicolumn{1}{c|}{\color{red} --}                 \\ \hline
\multicolumn{1}{|c|}{Weak}           & \multicolumn{1}{c|}{\color{blue} $W_{\mathrm{p},k}$} & \multicolumn{1}{c|}{\color{red} $W_{\mathrm{c},k}$} \\ \hline
\multicolumn{1}{|c|}{Strong/Very strong}         &  \multicolumn{1}{c|}{\color{blue} --}        & \multicolumn{1}{c|}{\color{red} $W_2$}             \\ \hline

 \multicolumn{1}{c}{}  & \multicolumn{1}{c}{\color{blue} decoded by its intended Rx-$k$ and}            &  \multicolumn{1}{c}{\color{red} decoded by}                         \\
 \multicolumn{1}{c}{}  &\multicolumn{1}{c}{\color{blue} treated as noise by Rx-$j$}          & \multicolumn{1}{c}{ \color{red} both Rx   }                         
\end{tabular}
	\label{fig_mapping}
\end{table}

\par Though the capacity of the two-user IC remains unknown, RS can achieve a rate that is within a single bit per second per hertz (bit/s/Hz) of the capacity for all values of the channel parameters (even in the non-symmetric case) and is information-theoretically optimal at asymptotic high SNR regimes \cite{Tse2008}.

\begin{figure}
	\centering
	\includegraphics[width=0.5\textwidth]{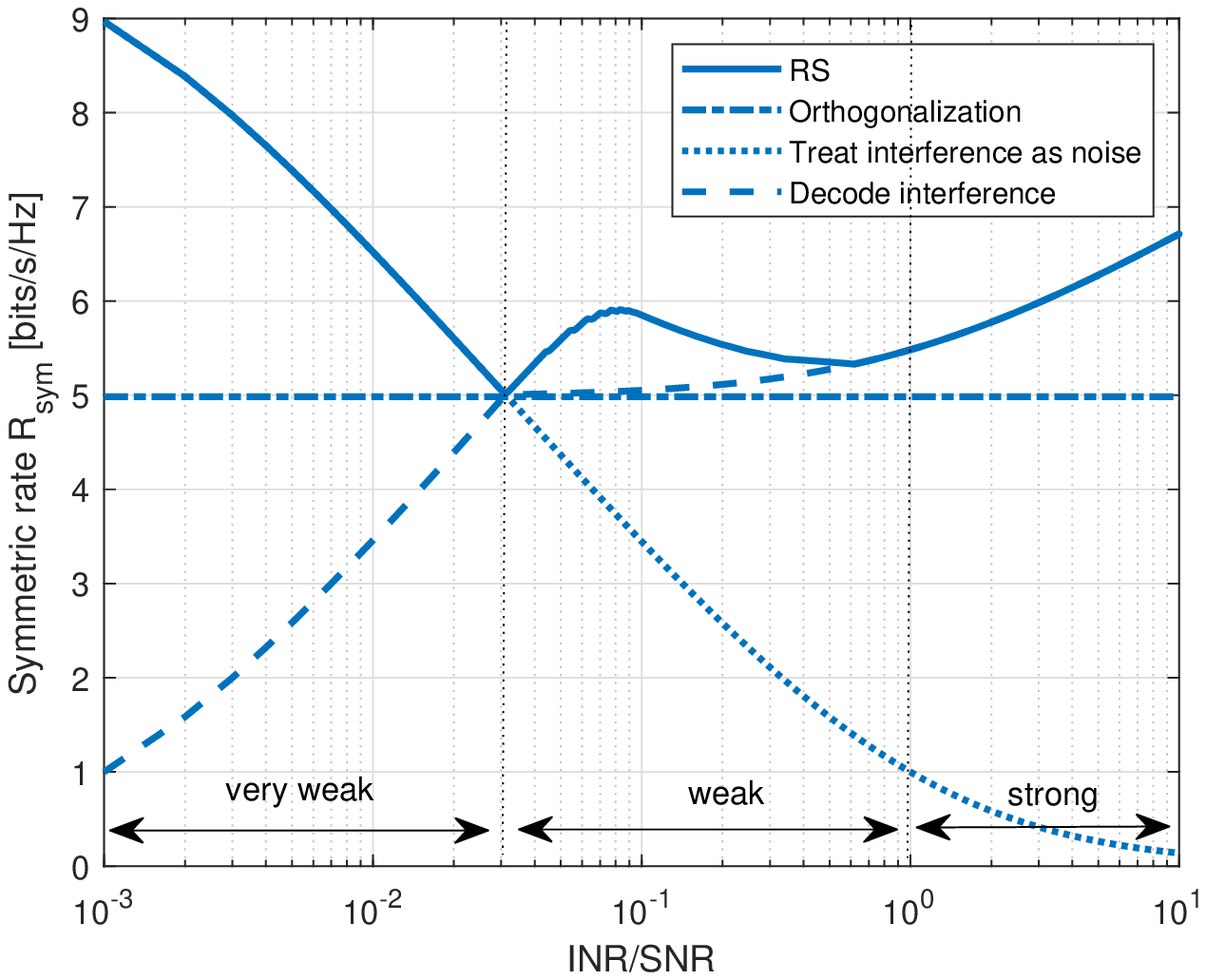}
	\caption{Symmetric rate versus INR/SNR. INR=$P\left|h_{\mathrm{c}}\right|^2$, SNR=$P\left|h_{\mathrm{d}}\right|^2$=1000.}
	\label{fig_2user_rate_simu}
\end{figure}

\par Fig. \ref{fig_2user_rate_simu} illustrates the symmetric rate of the various interference management strategies using a concrete numerical example. The x-axis is the ratio between the interference-to-noise ratio (INR) and SNR; hence the higher INR/SNR, the stronger the interference. The different interference regimes are clearly visible. We note how RS can softly bridge all strategies and outperform them in the weak interference regime\footnote{Note that if the two-user IC is used to represent a two-cell network, the inter-cell interference level is unlikely to be in the strong regime since a user is associated with its closest base station.}. In contrast, the treat interference as noise strategy (resp. decode interference) quickly becomes suboptimal as INR/SNR increases (resp. decreases).

\begin{remark} If each transmitter is equipped with a directive antenna, each transmitter could increase the direct channel $h_{\mathrm{d}}$ and decrease the cross channel $h_\mathrm{c}$ by suitably pointing the directive antenna. If this operation can be done accurately, the interference level is more likely to shift toward weaker regimes. This suggests that it would make more sense to combine multi-antenna processing (enabling directive beams) with an interference management strategy designed for weaker interference regimes (treat interference as noise and RS) rather than stronger interference regimes (decode interference).
\end{remark}

\subsection{Lessons Learned}

\noindent\fbox{%
    \parbox{0.97\columnwidth}{%
\begin{itemize}
\item Conventional interference management strategies rely on orthogonalization, treat interference as noise, or decode interference.
    \item Orthogonalize the resources so as to completely eliminate multi-user interference is clearly suboptimal.
    \item Treat interference as noise is an efficient strategy whenever the interference level is very weak but is inefficient whenever the interference level grows to weak, strong or very strong.
    \item Decode interference is an efficient strategy whenever the interference level is strong and very strong but is inefficient whenever the interference level is very weak or weak.
    \item RS splits messages into common and private parts so as to partially decode interference and partially treat interference as noise. This allows RS to bridge, unify, and generalize the two extremes of treat interference as noise and decode interference and being efficient in all four interference regimes. 
    \item RS is a superset of treat interference as noise and decode interference strategies, and can specialize to each of them depending on how messages are mapped to streams.
\end{itemize}
    }%

}
    \vspace{0.1cm}
\par
\begin{figure}[!h]
	\centering
\includegraphics[width=0.45\textwidth]{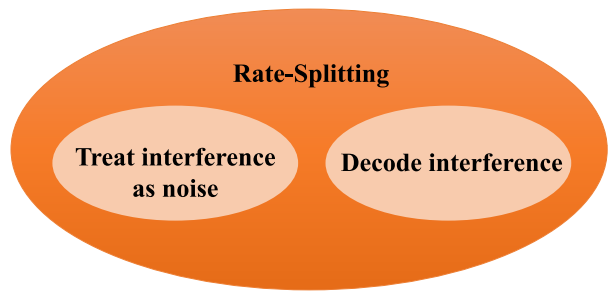}%
	\caption{The relationship between RS, treat interference as noise and decode interference strategies. Each set illustrates the optimization space of the corresponding interference management strategy. The
optimization space of RS is larger such that other strategies are just subsets.}
	\label{fig:bubble_interf_manag}
\end{figure}

 Those lessons are summarized in Table \ref{tab:lessonFromIT} and Fig. \ref{fig:bubble_interf_manag}.

\begin{table}
\centering
\caption{Comparison of preferable interference levels of different interference management strategies.}
\label{tab:lessonFromIT}
\begin{tabular}{|c|c|c|c|c|}
\hline 
\textbf{Interference levels}
 & \textbf{Very weak} & \textbf{Weak} & \textbf{Strong} & \textbf{Very strong} \\ \hline \hline
\textbf{Orthogonalize} & $\times$       & $\times$          & $\times$ & $\times$      \\ \hline
\textbf{Treat interf. as noise} & $\surd$       & $\times$         & $\times$   & $\times$     \\ \hline
\textbf{Decode interference} & $\times$       & $\times$         & $\surd$    & $\surd$    \\ \hline
\textbf{Rate-Splitting} & $\surd$       & $\surd$         & $\surd$   & $\surd$     \\ \hline
\end{tabular}
\vspace{0.1cm}

{Notations: $\surd$: Suited. $\times$: Not well suited. }
\end{table}

\section{Two-User Rate-Splitting Multiple Access}\label{section_twouser_RSMA}
Building upon the RS scheme for two-user IC, we can obtain some insight into how to design RSMA for two-user downlink (broadcast channel) and uplink (multiple access channel) in the next two sub-sections. We can also relate the above discussion to MA design and show a direct relationship/analogy between interference management and conventional MA designs such as OMA, SDMA, and NOMA.

\subsection{Two-User Downlink Rate-Splitting Multiple Access}
\label{section_2user_dl}

\begin{figure}
	\centering
\includegraphics[width=0.45\textwidth]{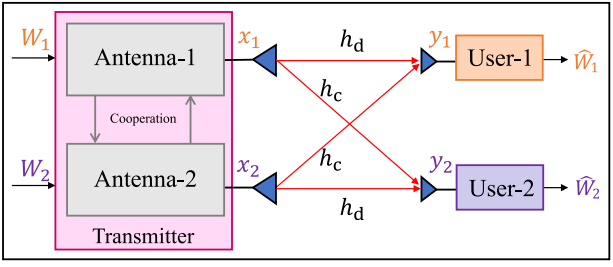}%
	\caption{Two-user MISO downlink (Gaussian MISO broadcast channel). }
	\label{fig_2user_BC}
\end{figure}

Let us consider a single transmitter potentially equipped with $M$ antennas serving two users. To that end, suppose that the two transmit antennas in Fig. \ref{fig_2user_IC} cooperate by exchanging CSI and messages such that they effectively belong to the same transmitter as per Fig. \ref{fig_2user_BC}. The two-user IC then becomes a two-user MISO broadcast channel (BC) with $M=2$ transmit antennas where the base station transmits to two active receiving users whose channels are given by $\mathbf{h}_1$ and $\mathbf{h}_2$, respectively. In the example of Fig. \ref{fig_2user_BC}, $\mathbf{h}_1=\left[\begin{array} {cc} h_{\mathrm{d}} & h_{\mathrm{c}}\end{array}\right]$ and $\mathbf{h}_2=\left[\begin{array} {cc} h_{\mathrm{c}} & h_{\mathrm{d}}\end{array}\right]$. Considering a transmit signal vector $\mathbf{x}$ spanning across the transmit antennas, the received signal at user-$k$ can be written as 
\begin{equation}
    y_k=\mathbf{h}_k\mathbf{x}+n_k, 
\end{equation}
where $n_k\sim\mathcal{CN}(0,\sigma_{n,k}^2)$.
Such a multi-antenna BC is a basic building block of modern downlink communication systems. Assuming perfect CSIT and CSIR, the capacity of this channel is known and is achieved by Dirty Paper Coding (DPC) \cite{DPC1983,capacityRegion2006HW,ZFandZFDPC2003}. DPC is nevertheless complex to implement due to its inherent nonlinear encoding/precoding mechanism. RSMA has appeared in the past few years as an appealing strategy to achieve close performance to DPC while maintaining the low complexity of linear precoding \cite{mao2017rate}. In the presence of imperfect CSIT, the capacity and the capacity-achieving strategy are unknown. It is nevertheless known that RS plays a central role to achieve the optimal DoF (and generalized DoF) \cite{Davoodi_gdof_2018,RS2016hamdi,enrico2017bruno,hamdi2019spawc,Davoodi2021DoF,chenxi2017bruno}, and RSMA can outperform DPC \cite{mao2019beyondDPC}. In the sequel, we delineate progressively the key design principles and schemes of RSMA in the two-user downlink scenario.
\subsubsection{MISO RSMA}\label{MISO_RSMA_2user} 
\begin{figure}
	\centering
\includegraphics[width=0.45\textwidth]{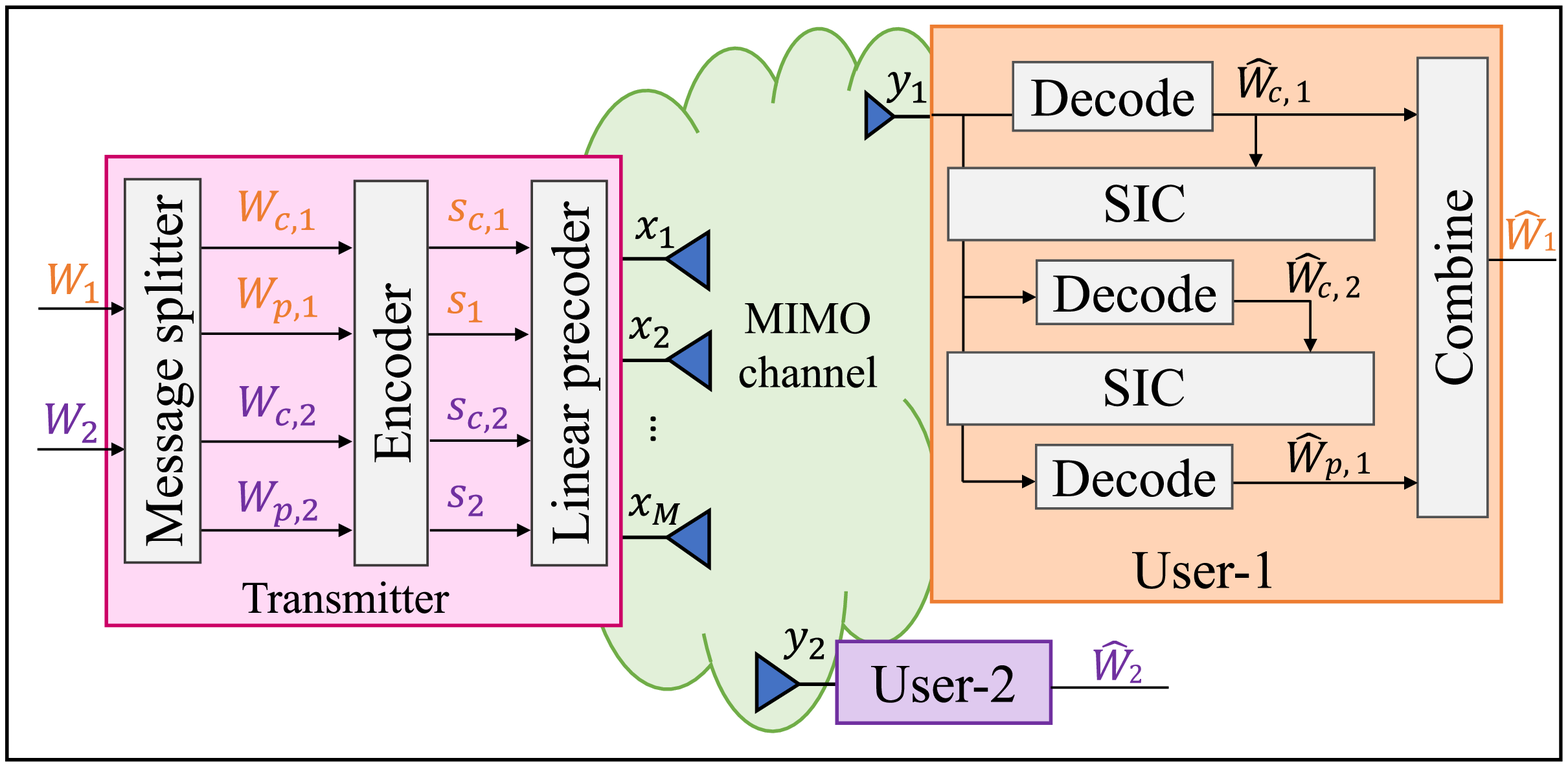}%
	\caption{Two-user downlink MISO RSMA without message combiner. }
	\label{fig_2user_RSMA_nocombiner}
\end{figure}

Inspired by the IC, the first MISO RSMA architecture is obtained by splitting the two messages $W_1$ and $W_2$ into common and private parts and encoding each part into a corresponding stream such that $W_{\mathrm{c},k}\rightarrow s_{\mathrm{c},k}$ and $W_{\mathrm{p},k}\rightarrow s_{\mathrm{p},k}$, $k=1,2$.  Given the presence of multiple antennas, precoding/beamforming can be performed across the two antennas and the transmit signal model can be written as
\begin{equation}\label{RS_BC_1}
\mathbf{x}= \sum_{k=1,2} \mathbf{p}_{\mathrm{c},k} s_{\mathrm{c},k} + \mathbf{p}_k s_{k},
\end{equation}
where $\mathbf{p}_{\mathrm{c},k}$ is the precoder of common stream $k$ and $\mathbf{p}_k$ is the precoder of private stream $k$. 
Defining $\mathbf{s}=[s_{\mathrm{c},1},s_1,s_{\mathrm{c},2},s_2]^T$ and assuming that $\mathbb{E}[\mathbf{s}\mathbf{s}^H]=\mathbf{I}$, the average transmit sum power constraint\footnote{We assume a sum power constraint but a per-antenna power constraint could also be considered.} at the transmitter is $\sum_{k=1,2} P_{\mathrm{c},k}+P_{k}\leq P$ with $P_{\mathrm{c},k}=\left\|\mathbf{p}_{\mathrm{c},k}\right\|^2$ and $P_k=\left\|\mathbf{p}_{k}\right\|^2$ the power allocated to common stream $k$ and private stream $k$, respectively. Uniquely, since $s_{\mathrm{c},1}$ and $s_{\mathrm{c},2}$ are transmitted from the same transmitter and are decoded by both users in this downlink setting, one can choose $\mathbf{p}_{\mathrm{c},1}=\mathbf{p}_{\mathrm{c},2}=\mathbf{p}_{\mathrm{c}}$ so that we only have a single common precoder and $\mathbf{x}=  \mathbf{p}_{\mathrm{c}} \left(s_{\mathrm{c},1}+s_{\mathrm{c},2}\right) +\sum_{k=1,2} \mathbf{p}_k s_{k}$. User-$k$ can now decode using SIC or joint decoding common streams $s_{\mathrm{c},1}$ and $s_{\mathrm{c},2}$ and then private stream $s_{k}$. Using SIC, this architecture would require each receiver $k$ to be equipped with two SIC layers so as to decode three streams $s_{\mathrm{c},j}$, $s_{\mathrm{c},k}$, and $s_{k}$, $j \neq k$. In summary, from two messages, this architecture creates four streams, and two SIC are required to recover the original messages. This RSMA architecture is illustrated\footnote{Only the receiver of user-1 is detailed in Fig. \ref{fig_2user_RSMA_nocombiner}. The receiver of user-2 follows the same principle as that of user-1 where two SIC layers are needed to decode and cancel the common streams before retrieving user-2's private message.} in Fig.\ \ref{fig_2user_RSMA_nocombiner}.

\par
\begin{figure}
	\centering
\includegraphics[width=0.45\textwidth]{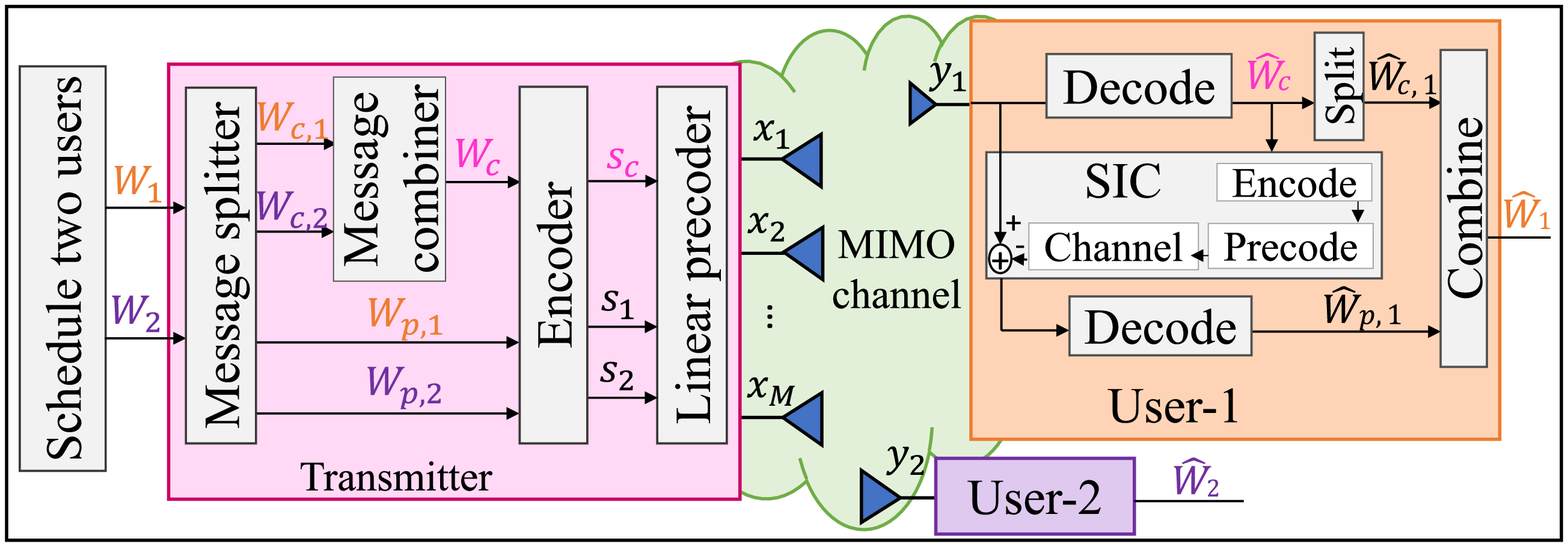}%
	\caption{Two-user downlink MISO RSMA with message combiner (1-layer RS).}
	\label{fig_2user_RSMA_1layerRS}
\end{figure}

 A second MISO RSMA architecture is obtained by noting that an additional benefit of the downlink is that instead of encoding each common part into a common stream $W_{\mathrm{c},k}\rightarrow s_{\mathrm{c},k}$, we can first combine the common parts into a common message $W_{\mathrm{c}}=\left\{W_{\mathrm{c},1},W_{\mathrm{c},2}\right\}$ that is then encoded into a single common stream $W_{\mathrm{c}}\rightarrow s_{\mathrm{c}}$ such that 
\begin{equation}\label{RS_BC_2}
\mathbf{x}= \mathbf{p}_{\mathrm{c}} s_{\mathrm{c}} + \sum_{k=1,2} \mathbf{p}_k s_{k}.
\end{equation}
By defining $\mathbf{s}=[s_{\mathrm{c}},s_1,s_2]^T$ and assuming that $\mathbb{E}[\mathbf{s}\mathbf{s}^H]=\mathbf{I}$, the average transmit (sum) power constraint at the transmitter is $P_{\mathrm{c}}+\sum_{k=1,2}P_{k}\leq P$ with $P_{\mathrm{c}}=\left\|\mathbf{p}_{\mathrm{c},k}\right\|^2$ the power allocated to the unique common stream.
The received signal at user-$k$, $k=1,2$, $j\neq k$, is written as
\begin{equation}\label{MISO_BC_received_signal}
    y_k=\mathbf{h}_k\mathbf{p}_{\mathrm{c}} s_{\mathrm{c}}+\mathbf{h}_k\mathbf{p}_k s_{k}+\mathbf{h}_k\mathbf{p}_j s_{j}+n_k.
\end{equation}
This is the so-called \textit{1-layer RS} architecture of RSMA because it relies on a single common stream and therefore a single SIC layer at each receiver, hence simplifying the encoding complexity (three streams used instead of four) and the decoding complexity (one SIC layer instead of two) compared to \eqref{RS_BC_1} and \eqref{RS_IC_eq}. Indeed both users decode the single common stream $s_{\mathrm{c}}$ into $\widehat{W}_{\mathrm{c}}$ by treating the interference from all private streams as noise. Each user-$k$ then retrieves the estimate $\widehat{W}_{\mathrm{c},k}$ from $\widehat{W}_{\mathrm{c}}$. Using SIC, $\widehat{W}_{\mathrm{c}}$ is then re-encoded, precoded, and subtracted from the received signal such that user-$k$ decodes its private stream $s_k$ into $\widehat{W}_{\mathrm{p},k}$ by treating the remaining interference from the other private streams as noise. Finally user-$k$ recombines $\widehat{W}_{\mathrm{c},k}$ and  $\widehat{W}_{\mathrm{p},k}$ into the message $\widehat{W}_{k}$ which is the same as the original message $W_{k}$ if no decoding error occurs. This architecture is illustrated in Fig.\ \ref{fig_2user_RSMA_1layerRS}. 

\subsubsection{Revisiting the Interference Regimes}

It is worth relating \eqref{MISO_BC_received_signal} to the lessons learned in Section \ref{section_interference}. In \eqref{MISO_BC_received_signal}, after precoding, the two-user MISO BC can effectively be seen as a two-user IC where user-$k$ decodes a common stream $s_{\mathrm{c}}$ and a private stream $s_k$ subject to interference from private stream $s_j$, $j\neq k$. The interference regime experienced by user-$k$ in this effective two-user IC is therefore determined by the strength of the precoded channels $\mathbf{h}_k\mathbf{p}_k$ and $\mathbf{h}_k\mathbf{p}_j$ (with $\mathbf{h}_k\mathbf{p}_k$ and $\mathbf{h}_k\mathbf{p}_j$ taking the role of $h_{\mathrm{d}}$ and $h_{\mathrm{c}}$, respectively):
\begin{itemize}
    \item $|\mathbf{h}_k\mathbf{p}_j|\ll|\mathbf{h}_k\mathbf{p}_k|$: very weak interference regime and interference should be treated as noise, i.e., $ \mathbf{p}_{\mathrm{c}} =0$ and only private streams are used. Decoding interference would perform badly. This scenario can typically occur when the channels are close to being orthogonal.  
    \item $|\mathbf{h}_k\mathbf{p}_j|\leq|\mathbf{h}_k\mathbf{p}_k|$: weak interference regime and non-zero power should be allocated to all common and private streams. This would typically occur whenever the channels are neither orthogonal nor aligned.
    \item $|\mathbf{h}_k\mathbf{p}_j|\geq|\mathbf{h}_k\mathbf{p}_k|$: strong interference regime and interference should be decoded. Treat interference as noise would perform badly. This scenario can typically occur when the channels are aligned.
\end{itemize}
This shows that depending on the propagation conditions in multi-antenna settings (e.g., angle between user channel directions), the interference regime can change from very weak to strong. Consequently, downlink MA schemes therefore need the ability to softly evolve from the extreme of decode interference to treat interference as noise. RSMA has the flexibility to cope with all those interference regimes and propagation conditions. 
\par It will now appear clear in the sequel how existing MA schemes such as SDMA and NOMA are specifically tailored for one specific interference management strategy (e.g., treat interference as noise, decode interference), one specific interference regime (e.g., very weak, strong) and one type of channel conditions (e.g., orthogonal, aligned), and how RSMA can unify them all. 

\subsubsection{Unifying OMA, SDMA, NOMA, and Multicasting}

\begin{figure}
	\centering
\includegraphics[width=0.45\textwidth]{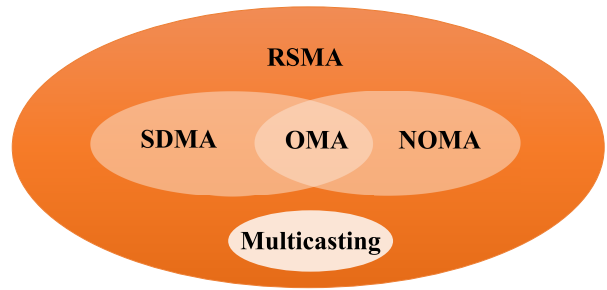}%
	\caption{The relationship between OMA, NOMA, SDMA, physical-layer multicasting, and RSMA in the downlink two-user case. Each set illustrates the optimization space of the corresponding communication strategy. The optimization space of RSMA is larger such that SDMA, NOMA, and physical-layer multicasting are just subsets.}
	\label{Fig_2userbubble}
\end{figure}

\begin{figure}
	\centering
\includegraphics[width=0.5\textwidth]{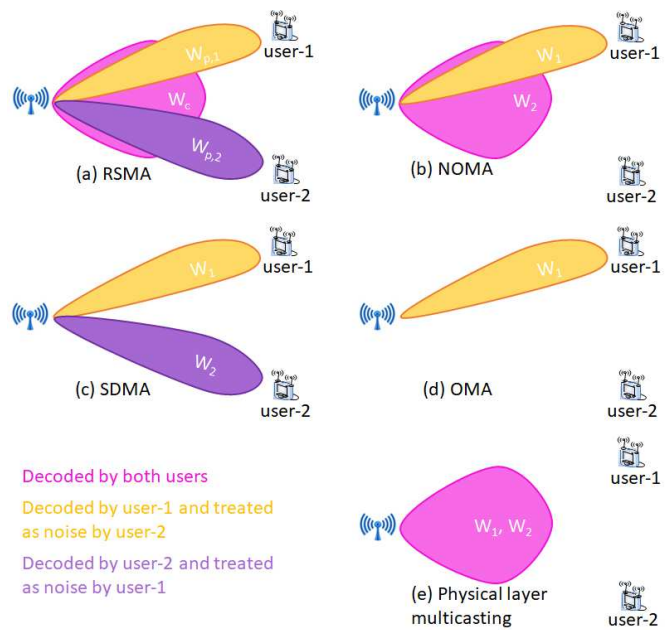}%
	\caption{A beam representation of RSMA and its sub-schemes SDMA, NOMA, OMA, and physical-layer multicasting.}
	\label{MA_beams_fig}
\end{figure}

In the simple two-user case, OMA, SDMA, NOMA, and physical-layer multicasting are particular instances of RSMA, as illustrated by Fig. \ref{Fig_2userbubble}, the message to stream mapping in Table \ref{fig_mapping_BC} \cite{bruno2019wcl}, and Fig. \ref{MA_beams_fig}. 
\par \textit{SDMA} is a special case of RSMA by forcing $ \mathbf{p}_{\mathrm{c}} =0$.
In this way, $W_k$ is directly encoded into $s_k$ and the system model writes as $\mathbf{x}= \sum_{k=1,2} \mathbf{p}_k s_{k}$. By doing so, each stream is decoded by its intended user by treating any residual interference from the other stream as noise. Recalling Tables \ref{fig_mapping} and \ref{tab:lessonFromIT}, from an interference management strategy, SDMA is reminiscent of and builds upon the \textit{Treat Interference as Noise} strategy which would be efficient only if the residual multi-user interference $\left|\mathbf{{h}}_{k}\mathbf{{p}}_{j}\right|$ is sufficiently weak as in orthogonal channels \cite{bruno2019wcl}.
\par \textit{NOMA} is a special case of RSMA by forcing the encoding of message $W_2$ entirely into $s_{\mathrm{c}}$ (i.e., $W_{\mathrm{c}}=W_2$) and $W_1$ into $s_1$ while turning off $s_2$ ($P_2=0$). By doing so, user-1 fully decodes the message of user-2 (and therefore interference created by user-2 stream) and 1-layer RS system model becomes the NOMA system model $\mathbf{x}= \mathbf{p}_{\mathrm{c}} s_{\mathrm{c}} + \mathbf{p}_1 s_{1}$. Note that NOMA also utilizes a common stream since the message of one of the two users, namely $W_2$ in this example, is decoded by both users. Connecting back to Tables \ref{fig_mapping} and \ref{tab:lessonFromIT}, from an interference management strategy, NOMA is reminiscent of and builds upon the \textit{Decode Interference} strategy which would be efficient only if the multi-user interference level is sufficiently strong as in aligned channels \cite{bruno2019wcl}. 

\par \textit{OMA} is a special case of RSMA by forcing only one user to be scheduled, e.g., user-1 (i.e., $\left \| \mathbf{p}_{\mathrm{c}} \right \|^2=\left \| \mathbf{p}_2 \right \|^2=0$).  
\par \textit{Physical-layer multicasting} is a special case of RSMA obtained when messages $W_1, W_2$ are both encoded into $s_{\mathrm{c}}$ (i.e., $W_{\mathrm{c}}=\{W_1,W_2\}$) and the private streams are turned off  ($P_1=P_2=0$). 

\begin{table}
	\caption{Messages-to-streams\! mapping\! in\! the two-user\! MISO\! BC \cite{bruno2019wcl}.}
	\centering
\begin{tabular}{|p{.2\textwidth}|p{.6\textwidth}|p{2.3cm}|p{2.5cm}|}
\hline
\multicolumn{1}{|c|}{}             & \multicolumn{1}{c|}{$s_1$}     & \multicolumn{1}{c|}{$s_2$}     & \multicolumn{1}{c|}{$s_{\mathrm{c}}$}             \\ \hline
\multicolumn{1}{|c|}{ SDMA}         & \multicolumn{1}{c|}{\color{blue} $\,\,\quad \quad W_1\quad\quad\,\,$}     & \multicolumn{1}{c|}{\color{blue} $W_2$}     & \multicolumn{1}{c|}{\color{magenta} --}                \\ \hline
\multicolumn{1}{|c|}{ NOMA}         & \multicolumn{1}{c|}{\color{blue} $W_1$}     & \multicolumn{1}{c|}{\color{blue} --}        & \multicolumn{1}{c|}{\color{magenta} $W_2$}             \\ \hline
\multicolumn{1}{|c|}{ OMA}          & \multicolumn{1}{c|}{\color{blue} $W_1$}     & \multicolumn{1}{c|}{\color{blue} --}        & \multicolumn{1}{c|}{\color{magenta} --}                \\ \hline
\multicolumn{1}{|c|}{Multicasting} & \multicolumn{1}{c|}{\color{blue} --}        & \multicolumn{1}{c|}{\color{blue} --}        & \multicolumn{1}{c|}{\color{magenta} $W_1,W_2$}         \\ \hline
\multicolumn{1}{|c|}{RSMA}           & \multicolumn{1}{c|}{\color{blue} $W_{\mathrm{p},1}$} & \multicolumn{1}{c|}{\color{blue} $W_{\mathrm{p},2}$} & \multicolumn{1}{c|}{\color{magenta} $W_{\mathrm{c},1},W_{\mathrm{c},2}$} \\ \hline
 \multicolumn{1}{c}{}  & \multicolumn{2}{c}{\color{blue} decoded by its intended user and}            &  \multicolumn{1}{c}{\color{magenta} decoded by}                         \\
 \multicolumn{1}{c}{}  &\multicolumn{2}{c}{\color{blue} treated as noise by the other user}          & \multicolumn{1}{c}{ \color{magenta} both users   }                         
\end{tabular}
	\label{fig_mapping_BC}
\end{table}

\begin{table}[t!]
\centering
\caption{Comparison of preferable residual multi-user interference levels of different MA schemes.}
\label{tab:lessonFromIT_2}
\begin{tabular}{|c|c|c|c|}
\hline 
\textbf{Interference levels}
 & \textbf{Very weak} & \textbf{Weak} & \textbf{(Very) Strong} \\ \hline \hline
\textbf{SDMA} & $\surd$       & $\times$          & $\times$       \\ \hline
\textbf{NOMA} & $\times$       & $\times$         & $\surd$       \\ \hline
\textbf{RSMA} & $\surd$       & $\surd$         & $\surd$       \\ \hline
\end{tabular}
\vspace{0.1cm}

{Notations: $\surd$: Suited. $\times$: Not well suited. }
\end{table}

\par Drawing an analogy with Table \ref{tab:lessonFromIT}, we can conclude that MA schemes operate in some preferred residual multi-user interference regimes as shown in Table \ref{tab:lessonFromIT_2}. The above clearly shows how three different non-orthogonal approaches to MA designs, namely SDMA, NOMA, and RSMA, fundamentally differ based on how multi-user interference is managed. NOMA is such that at least one user is forced to fully decode the message(s) of other co-scheduled user. SDMA and RSMA do not follow this approach since they both do not force a user to fully decode the messages of another co-scheduled user. SDMA actually treats any residual interference as noise, and RSMA is built upon the principle of splitting the messages so as to partially treat interference as noise and partially decode the remaining interference. Consequently, this difference in managing interference has deep consequences on the universality of MA schemes, with RSMA being a superset of SDMA and NOMA as per Fig. \ref{Fig_2userbubble}, but also on the performance of those MA schemes as a function of the propagation conditions \cite{bruno2019wcl}.
\par RSMA can also be seen as a smart combination of transmit-side and receive-side interference cancellation strategy where the contribution of the common stream is adjusted according to the level of interference that needs to be canceled by the receiver, therefore departing from the transmit-side only and receive-side only interference cancellation strategies of SDMA (and DPC) and NOMA, respectively. This is further summarized in Table \ref{tab:transmit_receive}.

\begin{table}[t!]
\centering
\caption{Comparison of different MA schemes in terms of transmit-side vs receive-side interference cancellation.}
\label{tab:transmit_receive}
\begin{tabular}{|c|c|c|c|}
\hline 
\textbf{Interf. cancel.}
 & \textbf{transmit-side} & \textbf{receive-side} & \textbf{both sides} \\ \hline \hline
\textbf{SDMA/DPC} & $\surd$       & $\times$         & $\times$       \\ \hline
\textbf{NOMA} & $\times$       & $\surd$         &       $\times$ \\ \hline
\textbf{RSMA} & $\times$       & $\times$         & $\surd$       \\ \hline
\end{tabular}
\vspace{0.1cm}

{Notations: $\surd$: Relevant. $\times$: Not relevant. }
\end{table}

\begin{example}\label{example_RS}
To further illustrate the split of the messages and the flexibility of RSMA, let us  imagine that the message of user-1 $W_1=\left(a_1\:a_2\:a_3\:a_4\right)\in\mathcal{W}_1=\left\{0000,0001,0010,\ldots,1111\right\}$, where $\left|\mathcal{W}_1\right|=16$. Similarly, the message of user-2 is $W_2=\left(b_1\:b_2\:b_3\right)\in\mathcal{W}_2=\left\{000,001,010,\ldots,111\right\}$, where $\left|\mathcal{W}_2\right|=8$. In SDMA, $W_1$ would be encoded into $s_1$ and $W_2$ into $s_2$. Assuming uncoded transmission for simplicity, $s_1$ and $s_2$ would then be a 16-QAM symbol and a 8-PSK symbol, respectively. In NOMA, $W_1$ would be encoded into $s_1$ and $W_2$ into $s_{\mathrm{c}}$, also using a 16-QAM symbol and a 8-PSK symbol, respectively. In RS, we split user-1's message in, e.g., $W_{\mathrm{c},1}=\left(a_1\:a_2\right)$, $W_{\mathrm{p},1}=\left(a_3\:a_4\right)$, and user-2's message in, e.g., $W_{\mathrm{c},2}=\left(b_1\right)$, $W_{\mathrm{p},2}=\left(b_2\:b_3\right)$. The common message is then constructed as $W_{\mathrm{c}}=\left(W_{\mathrm{c},1}\:W_{\mathrm{c},2}\right)=\left(a_1\:a_2\:b_1\right)$, which is then encoded into $s_{\mathrm{c}}$ using a 8-PSK symbol. $W_{\mathrm{p},1}$ and $W_{\mathrm{p},2}$ are encoded into $s_1$ and $s_2$ using QPSK symbols, respectively. The interested reader is referred to \cite{Maosurvey} for more examples.
\end{example}



\subsubsection{Rate Analysis} Under the assumption of Gaussian signaling and infinite blocklength, and perfect CSIT and CSIR, the instantaneous rates for decoding the common and private streams at user-$k$ are given as
\begin{equation}
\label{eq:rate1RS}
\begin{aligned}
R_{\mathrm{c},k}&=\log_{2}\left(1+\frac{\left|\mathbf{{h}}_{k}\mathbf{{p}}_{\mathrm{c}}\right|^{2}}{\left|\mathbf{{h}}_{k}\mathbf{{p}}_{1}\right|^{2}+\left|\mathbf{{h}}_{k}\mathbf{{p}}_{2}\right|^{2}+1}\right),
\\R_{k}&=\log_{2}\left(1+\frac{\left|\mathbf{{h}}_{k}\mathbf{{p}}_{k}\right|^{2}}{\left|\mathbf{{h}}_{k}\mathbf{{p}}_{j}\right|^{2}+1}\right),
\end{aligned}
\end{equation}
where the noise variance was normalized $\sigma_{n,k}^2=1$ without loss of generality.
To ensure that $s_{\mathrm{c}}$ is successfully decoded by both users, its rate cannot exceed
\begin{equation}
\label{eq:commonstream1}
R_{\mathrm{c}}=\min\left\{R_{\mathrm{c},1},R_{\mathrm{c},2}\right\}.
\end{equation}
As $s_{\mathrm{c}}$ contains sub-messages $W_{\mathrm{c},1},W_{\mathrm{c},2}$ of the two users, the rate distribution of $R_{\mathrm{c}}$ among the users is adapted to the amount of sub-messages that each user contributed. Let  $C_k$ denote the portion of rate $R_{\mathrm{c}}$  allocated to user-$k$ for $W_{\mathrm{c},k}$. Then, we have
\begin{equation}
\label{eq:commonstream2}
C_1+C_2=R_{\mathrm{c}}.
\end{equation}
The overall achievable rate of user-$k$ is
\begin{equation}
\label{eq: R_tot}
R_{k,\mathrm{tot}}=C_{k}+R_{k}.
\end{equation}
Here again, the terminology rate-splitting appears clearly in \eqref{eq: R_tot} where we note that the rate of each user is split into two parts, namely, the rate of $s_k$ (i.e., the private rate) and part of the rate of $s_{\mathrm{c}}$  (i.e., the common rate).

\par Common metrics to design the systems include 1) weighted sum rate (WSR) $u_1 R_{1,\mathrm{tot}}+u_2 R_{2,\mathrm{tot}}$ where $u_1$ and $u_2$ are weights set to account for fairness among users (for instance when conducting proportional fair scheduling); 2) max-min fair (MMF) that aims at maximizing $\min_{k=1,2} R_{k,\mathrm{tot}}$; 3) energy efficiency (EE) $\frac{R_{1,\mathrm{tot}}+R_{2,\mathrm{tot}}}{\frac{1}{\eta}(P_1+P_2+P_{\mathrm{c}})+P_{\mathrm{cir}}}$ where $\eta\in(0,1]$ and $P_{\mathrm{cir}}$ are respectively the power amplifier efficiency and the circuit power
consumption. All three metrics could also be subject to a QoS constraint $R_{k,\mathrm{tot}}\geq R_k^{th}$ with $R_k^{th}$ a minimum rate threshold to be achieved by user-$k$.

\begin{remark} We can wonder what happens if $M=1$, namely downlink SISO. RSMA strategy \eqref{RS_BC_2} yields
\begin{equation}
x= \sqrt{P_c} s_{\mathrm{c}} + \sqrt{P_1} s_{1}+\sqrt{P_2} s_{2}
\end{equation}
and the common and private rates write as
\begin{equation}
\begin{aligned}
R_{\mathrm{c},k}&=\log_{2}\left(1+\frac{\left|h_{k}\right|^{2}P_{\mathrm{c}}}{\left|h_{k}\right|^{2}P_1+\left|h_{k}\right|^{2}P_2+1}\right),
\\R_{k}&=\log_{2}\left(1+\frac{\left|h_{k}\right|^{2}P_k}{\left|h_{k}\right|^{2}P_j+1}\right),
\end{aligned}
\end{equation}
for $k=1,2$. Without loss of generality, we consider $\left|h_{2}\right|\leq \left|h_{1}\right|$ so that 
\begin{equation}
R_{\mathrm{c}}=\log_{2}\left(1+\frac{\left|h_{2}\right|^{2}P_{\mathrm{c}}}{\left|h_{2}\right|^{2}P_1+\left|h_{2}\right|^{2}P_2+1}\right).
\end{equation}
Since the capacity of the Gaussian SISO BC with perfect CSIT and CSIR is obtained by performing superposition coding (SC) with SIC (SISO NOMA), optimizing $P_1,P_2,P_c$ to maximize the WSR would lead to choosing $P_2=0$ so that the message of the weaker user (user-2) $W_2$ is entirely encoded in $s_{\mathrm{c}}$ and SISO RSMA becomes SISO NOMA $x= \sqrt{P_c} s_{\mathrm{c}} + \sqrt{P_1} s_{1}$ and its achievable rates are
\begin{align}
R_{1}&=\log_{2}\left(1+\left|h_{1}\right|^{2}P_1\right),\\
R_{\mathrm{c}}&=\log_{2}\left(1+\frac{\left|h_{2}\right|^{2}P_{\mathrm{c}}}{\left|h_{2}\right|^{2}P_1+1}\right).
\end{align}
This shows how RSMA can act as NOMA in Gaussian SISO downlink. 
\end{remark}

\begin{remark}
\label{rem:non-degraded}
The capacity region of a Gaussian SISO BC is known for both cases with fixed and fading channels when there are perfect CSIT and CSIR. However, the problem is open in general when there is imperfect CSIT and only limited cases are known \cite{Tse2012}. The fading SISO BC with only perfect CSIR but imperfect CSIT lacks the degraded structure in general for arbitrary fading distributions. In these cases, SC and SIC (and therefore NOMA) can not achieve the capacity region. In \cite[Theorem 1]{Lin2020}, the sufficient condition that the second channel is stochastically larger than or equal to the first channel, i.e., $H_2 \geq_{st} H_1$ \footnote{Stochastic order $H_2 \geq_{st} H_1$ means that $\mathbb{P}(H_2\geq s) \geq \mathbb{P}(H_1\geq s)$ for all $s$ from the support of $H_1,H_2$.}, is derived for which the capacity region is given by 
\begin{eqnarray}
    R_1 & \leq & I(V;Y_1|H_1), \nonumber \\
    R_2 & \leq & I(X;Y_2|V,H_2) \label{eq:rr-ipcsit}, 
\end{eqnarray}
with Markov chain $V - X - Y_2 - Y_1$. Even though the capacity expressions in (\ref{eq:rr-ipcsit}) look familiar to the rates achievable with SC and SIC (and NOMA), the optimal coding and decoding strategy to achieve (\ref{eq:rr-ipcsit}) remains unknown because the same marginal property of the BC is applied to transform the joint distribution of the underlying channel to fully coupled. 
\end{remark}

\subsubsection{Precoder Design and Power Allocation} Recall from Section \ref{subsection_Interf_regimes} that a qualitative and insightful way of allocating power to the common stream is so that the interference level caused by the private stream has roughly the same level as the other receiver's noise level, i.e., $\left(1-t\right)P\left|h_{\mathrm{c}}\right|^2 \approx 1$. In the downlink scenario, looking at \eqref{eq:rate1RS}, we can think in a similar way and choose the power allocation $P_1$ and $P_2$ to the two private streams such that $\left|\mathbf{{h}}_{2}\mathbf{{p}}_{1}\right|^{2} \approx 1$ and $\left|\mathbf{{h}}_{1}\mathbf{{p}}_{2}\right|^{2} \approx 1$, i.e., instead of allocating all the transmit power $P$ to the private streams (as done in SDMA), we allocate a fraction of the total available power to them such that the signal-to-interference-plus-noise ratios (SINRs) of the private streams are not interference limited. If we were allocating a higher power to the private streams with  $\left|\mathbf{{h}}_{k}\mathbf{{p}}_{j}\right|^{2} \gg  1$, the SINR $\frac{\left|\mathbf{{h}}_{k}\mathbf{{p}}_{k}\right|^{2}}{\left|\mathbf{{h}}_{k}\mathbf{{p}}_{j}\right|^{2}+1}\approx \frac{\left|\mathbf{{h}}_{k}\mathbf{{p}}_{k}\right|^{2}}{\left|\mathbf{{h}}_{k}\mathbf{{p}}_{j}\right|^{2}}$ would saturate and the private rates would not increase further. Hence, the key is for the private streams not to enter the interference limited regime, and allocate any remaining power to the common stream whose SINR will keep increasing linearly as $P_{\mathrm{c}}$ increases. Doing so, $R_1+R_2$ would be roughly equivalent to the sum-rate achieved by SDMA, but the common rate $R_{\mathrm{c}}$ would provide an additional rate increase over what SDMA can offer.

\par The above is particularly insightful when the CSIT is imperfect and the transmitter only has an estimate $\hat{\mathbf{h}}_k$ of user-$k$ channel $\mathbf{h}=\hat{\mathbf{h}}_k+\tilde{\mathbf{h}}_k$ with $\tilde{\mathbf{h}}_k$ the channel acquisition error. In such setting, residual multi-user interference scaling as $\big|\tilde{\mathbf{{h}}}_{k}\mathbf{{p}}_{j}\big|^{2}$ would be unavoidable regardless of how the precoder $\mathbf{{p}}_{j}$ is designed. Following such an approach, we can demonstrate that RSMA is information theoretically optimal from the DoF perspective in the presence of imperfect CSIT \cite{RS2016hamdi,enrico2017bruno,DoF2013SYang} while SDMA and NOMA are not \cite{bruno2021MISONOMA}. Because of its inherent robustness to imperfect CSIT, RSMA can also afford smaller feedback overhead than conventional SDMA \cite{chenxi2015finitefeedback}.

\par More systematic methods to design the precoder and power allocation can be obtained using either closed form low complexity suboptimal techniques or using optimization techniques detailed in \cite{Maosurvey}. Extensive results have demonstrated that the power allocation to the common stream vs private stream depends on a number of factors, including the angle between user channels \cite{mao2017rate,bruno2019wcl}, the disparity of channel strengths \cite{mao2017rate,bruno2019wcl}, the objective function to maximize \cite{RS2016hamdi,hamdi2016robust}, the quality of CSIT \cite{RS2016hamdi,hamdi2016robust}, the QoS, the network load, etc.

\subsubsection{Cooperative RSMA} 
\begin{figure}
	\centering
\includegraphics[width=0.45\textwidth]{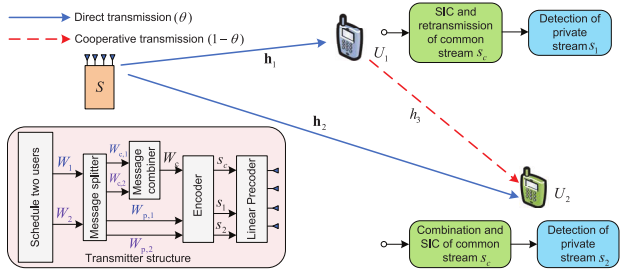}%
	\caption{Cooperative downlink RSMA with user relaying \cite{jian2019crs}.}
	\label{Fig_CRS}
\end{figure}

An interesting extension is achieved by considering that the transmitter can opportunistically ask one of the users to act as a relay. This is known as a cooperative relay BC in information theory \cite{CRS2007TIT,Liang2007RBC,Dabora2006}. Specifically, in our setting, the common stream is decoded by both users and the transmitter can ask the relay user to forward the decoded common message/stream to the other user to efficiently cope with a wide range of propagation conditions (disparity of user channel strengths and directions) and compensate for the performance degradation due to deep fading \cite{jian2019crs,mao2019maxmin}, as illustrated in Fig. \ref{Fig_CRS}. The parameter $0<\theta\leq 1$ refers to the time split between the direct transmission phase and the relaying phase, i.e.,\ $\theta=1$ indicates that no time is allocated to relaying, hence boiling down to the conventional downlink transmission of Fig. \ref{fig_2user_BC}.

The user relaying feature of cooperative RSMA enables to enlarge the pool of possible schemes within the RSMA framework as shown by the messages-to-streams mapping in the two-user cooperative MISO downlink of Table \ref{fig_mapping_coop_BC}. We note how the mapping of Table \ref{fig_mapping_BC} has been extended to include cooperation where conventional decode and forward (DF) and cooperative NOMA (C-NOMA) are particular instances of cooperative RSMA (C-RSMA).

\begin{table}
	\caption{Messages-to-streams\! mapping\! in\! the two-user\! cooperative MISO downlink \cite{jian2019crs}.}
	\centering
\begin{tabular}{|p{.2\textwidth}|p{.6\textwidth}|p{2.3cm}|p{2.5cm}|p{0.8cm}|}
\hline
\multicolumn{1}{|c|}{}             & \multicolumn{1}{c|}{$s_1$}     & \multicolumn{1}{c|}{$s_2$}     & \multicolumn{1}{c|}{$s_{\mathrm{c}}$} & $\theta$            \\ \hline
\multicolumn{1}{|c|}{ SDMA}         & \multicolumn{1}{c|}{\color{blue} $\,\,\quad \quad W_1\quad\quad\,\,$}     & \multicolumn{1}{c|}{\color{blue} $W_2$}     & \multicolumn{1}{c|}{\color{red} --}   & $\theta=1$               \\ \hline
\multicolumn{1}{|c|}{ NOMA}         & \multicolumn{1}{c|}{\color{blue} $W_1$}     & \multicolumn{1}{c|}{\color{blue} --}        & \multicolumn{1}{c|}{\color{red} $W_2$}    & $\theta=1$          \\ \hline
\multicolumn{1}{|c|}{ OMA}          & \multicolumn{1}{c|}{\color{blue} $W_1$}     & \multicolumn{1}{c|}{\color{blue} --}        & \multicolumn{1}{c|}{\color{red} --}     & $\theta=1$            \\ \hline
\multicolumn{1}{|c|}{Multic.} & \multicolumn{1}{c|}{\color{blue} --}        & \multicolumn{1}{c|}{\color{blue} --}        & \multicolumn{1}{c|}{\color{red} $W_1,W_2$}    & $\theta=1$      \\ \hline
\multicolumn{1}{|c|}{RSMA}           & \multicolumn{1}{c|}{\color{blue} $W_{\mathrm{p},1}$} & \multicolumn{1}{c|}{\color{blue} $W_{\mathrm{p},2}$} & \multicolumn{1}{c|}{\color{red} $W_{\mathrm{c},1},W_{\mathrm{c},2}$} & $\theta=1$ \\ \hline
\multicolumn{1}{|c|}{ C-NOMA}         & \multicolumn{1}{c|}{\color{blue} $W_1$}     & \multicolumn{1}{c|}{\color{blue} --}        & \multicolumn{1}{c|}{\color{red} $W_2$}    & $\theta<1$          \\ \hline
\multicolumn{1}{|c|}{ DF}          & \multicolumn{1}{c|}{\color{blue} --}     & \multicolumn{1}{c|}{\color{blue} --}        & \multicolumn{1}{c|}{\color{red} $W_2$}     & $\theta<1$            \\ \hline
\multicolumn{1}{|c|}{C-RSMA}           & \multicolumn{1}{c|}{\color{blue} $W_{\mathrm{p},1}$} & \multicolumn{1}{c|}{\color{blue} $W_{\mathrm{p},2}$} & \multicolumn{1}{c|}{\color{red} $W_{\mathrm{c},1},W_{\mathrm{c},2}$} & $\theta<1$ \\ \hline
 \multicolumn{1}{c}{}  & \multicolumn{2}{c}{\color{blue} decoded by its intended user and}            &  \multicolumn{1}{c}{\color{red} decoded by}                         \\
 \multicolumn{1}{c}{}  &\multicolumn{2}{c}{\color{blue} treated as noise by the other user}          & \multicolumn{1}{c}{ \color{red} both users   }    
\end{tabular}
\vspace{0.1cm}

{Notations: $\theta=1$ refers to non-cooperative schemes as in Table \ref{fig_mapping_BC}. $\theta<1$ refers to cooperative RSMA schemes. }
	\label{fig_mapping_coop_BC}
\end{table}

\subsubsection{Space-Time / Space-Frequency RSMA}\label{ST_RSMA_section}
Most RSMA literature deal with a single channel use (be it in time or frequency). This means the signal model as expressed in \eqref{RS_BC_2} is applied in a given time slot or given frequency subband. For instance, in the non-alternating CSIT pattern of Table \ref{tab:CSIT_patterns}, \eqref{RS_BC_2} would be applied separately on subbands A and B, i.e., each subband determining the split of the message, power allocation and precoders based on the available quality of CSIT in that subband (good on subband A and bad on subband B). One could also apply the same strategy to the alternating CSIT pattern of Table \ref{tab:CSIT_patterns}. However, because of the alternating feature, simply doing the spatial domain RSMA transmission \eqref{RS_BC_2} would not be as efficient as doing an RSMA transmission across the two subbands. In other words, RSMA can benefit from a multi-channel transmission in space-time (ST) or in space-frequency (SF) depending on the CSIT pattern \cite{chenxi2015finitefeedback,Tandon_2013,Elia_2013}. Such a multi-channel RSMA transmission is particularly helpful in the presence of alternating CSIT, i.e., whenever the CSIT changes across time or frequency in an alternating user-specific manner. The alternating CSIT pattern of Table \ref{tab:CSIT_patterns} is a typical (and practical) scenario where the transmitter wants to serve two users but the CSIT of user-1 (resp. user-2) is better on time/frequency B (resp. A) and worse on time/frequency A (resp. B). In such scenario, an ST/SF RSMA scheme can further increase the DoF over conventional RSMA \cite{Tandon_2013,Elia_2013}. 

\begin{table}[t!]
\centering
\caption{Two different CSIT patterns.}
\label{tab:CSIT_patterns}
\begin{tabular}{|c|c|c|}
\hline 
 & User-1       &  User-2        \\ \hline
\textbf{subband A} & $\surd$       &  $\surd$               \\ \hline
\textbf{subband B} & $\times$       & $\times$             \\ \hline
\end{tabular}
\\
\vspace{0.1cm}
{Non-alternating CSIT}\\

\vspace{0.5cm}
\begin{tabular}{|c|c|c|}
\hline 
 & User-1       &  User-2        \\ \hline
\textbf{subband A} & $\times$       &  $\surd$               \\ \hline
\textbf{subband B} & $\surd$       & $\times$             \\ \hline
\end{tabular}
\\
\vspace{0.1cm}
{Alternating CSIT}\\

\vspace{0.5cm}
{Notations:$\surd$: Good CSIT quality. $\times$: Poor CSIT quality.}

\end{table}

Compared to the RSMA scheme of \eqref{RS_BC_2}, ST/SF-RSMA scheme transmits an additional common stream (obtained from a further split of the messages), i.e., $s_0$, across the two channel uses. Specifically, considering the alternating CSIT pattern of Table \ref{tab:CSIT_patterns}, the transmitted signals in subbands A and B write can be expressed as follows:
\begin{align}
    \mathbf{x}^{(A)}&= \mathbf{p}_{\mathrm{0}}^{(A)} s_{\mathrm{0}}+\mathbf{p}_{\mathrm{c}}^{(A)} s_{\mathrm{c}}^{(A)} + \sum_{k=1,2} \mathbf{p}_k^{(A)} s_{k}^{(A)},\\
    \mathbf{x}^{(B)}&= \mathbf{p}_{\mathrm{0}}^{(B)} s_{\mathrm{0}}+\mathbf{p}_{\mathrm{c}}^{(B)} s_{\mathrm{c}}^{(B)} + \sum_{k=1,2} \mathbf{p}_k^{(B)} s_{k}^{(B)},
\end{align}
where the superscript $^{(i)}$ refers to the subband. We note the addition of the new common stream $s_0$ that has been repeated across the two channel uses (precoded by $\mathbf{p}_{\mathrm{0}}^{(i)}$ in channel use $i=1,2$). If the CSIT quality becomes non-alternating, the common stream $s_{\mathrm{0}}$ becomes useless (zero power is allocated to $s_{\mathrm{0}}$) and SF-RSMA boils down to \eqref{RS_BC_2} in each subband. The receiver at both users is more complicated since two common streams have to be decoded at each user before decoding the respective private stream.

The decoding works as follows. Let us focus on user-1 for simplicity. First, user-1 decodes $s_{\mathrm{c}}^{(A)}$ and $s_{\mathrm{0}}$ sequentially using subband A observation by treating the private streams as noise. Secondly, after removing $s_{\mathrm{0}}$ from subband B observation, user-1 recovers $s_{\mathrm{c}}^{(B)}$ by treating the private streams as noise. Thirdly, by removing all common streams, user-1 decodes $s_{1}^{(A)}$ and $s_{1}^{(B)}$ by treating the other private stream as noise.

More details on the CSIT pattern conditions needed for the optimality of separate RSMA in each subband can be found from \cite{Joudeh_separability}. Another interesting use of ST-RSMA, where RS is combined with space-time block coding to address the lack of knowledge of the channel phase information at the transmitter, has been developed in \cite{Mosquera_2021}. 

\subsubsection{MIMO RSMA}
Let us now consider a two-user MIMO downlink with $M$ transmit antennas and $N$ receive antennas at each user. The transmitter wants to transmit two $N$-dimensional vectors of messages $\mathbf{w}_1$ and $\mathbf{w}_2$ to user-1 and user-2, respectively. To that end, for each user, each of those $N$ messages is split into a common part and a private part. Common parts of both users are combined and encoded into a $N$-dimensional vector of common streams $\mathbf{s}_{\mathrm{c}}$ and private parts are encoded into two $N$-dimensional vectors of private streams $\mathbf{s}_k$. $M \times N$ precoding are performed on the common and private stream vectors such that the transmit signal is written as
\begin{equation}\label{2user_mimo}
    \mathbf{x}=\mathbf{P}_{\mathrm{c}} \mathbf{s}_{\mathrm{c}} + \sum_{k=1,2} \mathbf{P}_k \mathbf{s}_{k}.
\end{equation}
\par
\begin{figure}
	\centering
\includegraphics[width=0.48\textwidth]{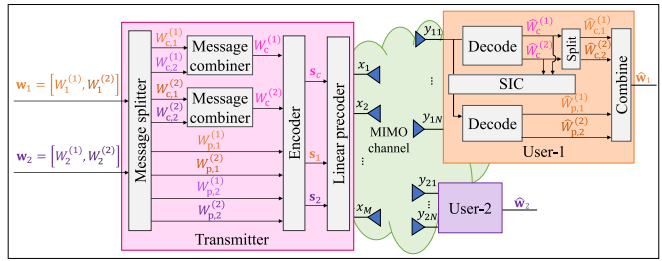}%
	\caption{Downlink MIMO RSMA \cite{anup2021MIMO}.}
	\label{Fig_MIMOBC_architecture}
\end{figure}

 Let us consider an example with $N=2$ to illustrate the universality of the framework. We have four messages to transmit (two for each user) $\mathbf{w}_1=[W_1^{(1)},W_1^{(2)}]$ and $\mathbf{w}_2=[W_2^{(1)},W_2^{(2)}]$, where $W_i^{(j)}$ refers to the $j$th message of user-$i$, $i=1,2$, $j=1,2$. We split the four messages into common and private parts such that 
\begin{align}
W_1^{(1)}&=\left\{W_{\mathrm{c},1}^{(1)},W_{\mathrm{p},1}^{(1)}\right\},\\
W_1^{(2)}&=\left\{W_{\mathrm{c},1}^{(2)},W_{\mathrm{p},1}^{(2)}\right\},\\
W_2^{(1)}&=\left\{W_{\mathrm{c},2}^{(1)},W_{\mathrm{p},2}^{(1)}\right\},\\
W_2^{(2)}&=\left\{W_{\mathrm{c},2}^{(2)},W_{\mathrm{p},2}^{(2)}\right\}.
\end{align}
Common parts are then combined into $W_{\mathrm{c}}^{(1)}$ and $W_{\mathrm{c}}^{(2)}$ and then respectively encoded into common stream $s_{\mathrm{c}}^{(1)}$ and $s_{\mathrm{c}}^{(2)}$
\begin{align}
W_{\mathrm{c}}^{(1)}&=\left\{W_{\mathrm{c},1}^{(1)},W_{\mathrm{c},2}^{(1)}\right\}\rightarrow s_{\mathrm{c}}^{(1)},\\
W_{\mathrm{c}}^{(2)}&=\left\{W_{\mathrm{c},1}^{(2)},W_{\mathrm{c},2}^{(2)}\right\}\rightarrow s_{\mathrm{c}}^{(2)},
\end{align}
so as to form the 2-dimensional common stream vector $\mathbf{s}_{\mathrm{c}}=\big[\begin{array}{cc} s_{\mathrm{c}}^{(1)}&s_{\mathrm{c}}^{(2)}\end{array}\big]^T$. Private parts are encoded into private streams
\begin{align}
W_{\mathrm{p},1}^{(1)}&\rightarrow s_1^{(1)}, \hspace{0.2cm} W_{\mathrm{p},1}^{(2)}\rightarrow s_1^{(2)},\\
W_{\mathrm{p},2}^{(1)}&\rightarrow s_2^{(1)}, \hspace{0.2cm} W_{\mathrm{p},2}^{(2)}\rightarrow s_2^{(2)},
\end{align}
so as to create two 2-dimensional private stream vectors $\mathbf{s}_{1}=\big[\begin{array}{cc} s_{1}^{(1)}&s_{1}^{(2)}\end{array}\big]^T$ and $\mathbf{s}_{2}=\big[\begin{array}{cc} s_{2}^{(1)}&s_{2}^{(2)}\end{array}\big]^T$.
At the receivers, both users decode the vector of common streams first using the two receive antennas, perform SIC, and then decode their respective private stream vector while treating the co-scheduled user's private stream vector as noise. Fig. \ref{Fig_MIMOBC_architecture} illustrates the MIMO RSMA architecture.

\begin{table*}
	\caption{Messages-to-streams mapping in the two-user MIMO BC.}
	\centering
\begin{tabular}{|p{.2\textwidth}|p{.2\textwidth}|p{.2\textwidth}|p{.2\textwidth}|p{.2\textwidth}|p{.2\textwidth}|p{.2\textwidth}|p{.2\textwidth}|}
\hline
\multicolumn{1}{|c|}{}             & \multicolumn{2}{c|}{$\mathbf{s}_1$}     & \multicolumn{2}{c|}{$\mathbf{s}_2$}     &
\multicolumn{2}{c|}{$\mathbf{s}_{\mathrm{c}}$}             \\ \hline
\multicolumn{1}{|c|}{}             & \multicolumn{1}{c|}{$s_1^{(1)}$}     & \multicolumn{1}{c|}{$s_1^{(2)}$}     &\multicolumn{1}{c|}{$s_2^{(1)}$}     & \multicolumn{1}{c|}{$s_2^{(2)}$}     &\multicolumn{1}{c|}{$s_{\mathrm{c}}^{(1)}$}     & \multicolumn{1}{c|}{$s_{\mathrm{c}}^{(2)}$}      \\ \hline
\multicolumn{1}{|c|}{ MU-MIMO}         & \multicolumn{1}{c|}{\color{blue} $\,\,\quad \quad W_1^{(1)}\quad\quad\,\,$}& \multicolumn{1}{c|}{\color{blue} $\,\,\quad \quad W_1^{(2)}\quad\quad\,\,$}     & \multicolumn{1}{c|}{\color{blue} $\,\,\quad \quad W_2^{(1)} \quad\quad\,\,$}   & \multicolumn{1}{c|}{\color{blue} $\,\,\quad \quad W_2^{(2)}\quad\quad\,\,$}  & \multicolumn{1}{c|}{\color{red} --}   & \multicolumn{1}{c|}{\color{red} --}              \\ \hline
\multicolumn{1}{|c|}{ NOMA}         & \multicolumn{1}{c|}{\color{blue} $W_1^{(1)}$}& \multicolumn{1}{c|}{\color{blue} $W_1^{(2)}$}     & \multicolumn{1}{c|}{\color{blue} --}  & \multicolumn{1}{c|}{\color{blue} --}        & \multicolumn{1}{c|}{\color{red} $\,\,\quad \quad W_2^{(1)}\,\,\quad \quad$}    & \multicolumn{1}{c|}{\color{red} $\,\,\quad \quad W_2^{(2)}\,\,\quad \quad$}          \\ \hline
\multicolumn{1}{|c|}{ OMA}         & \multicolumn{1}{c|}{\color{blue} $W_1^{(1)}$}& \multicolumn{1}{c|}{\color{blue} $W_1^{(2)}$}     & \multicolumn{1}{c|}{\color{blue} --}   & \multicolumn{1}{c|}{\color{blue} --}       & \multicolumn{1}{c|}{\color{red} --}      & \multicolumn{1}{c|}{\color{red} --}           \\ \hline
\multicolumn{1}{|c|}{Multicasting} & \multicolumn{1}{c|}{\color{blue} --}    & \multicolumn{1}{c|}{\color{blue} --}& \multicolumn{1}{c|}{\color{blue} --}    & \multicolumn{1}{c|}{\color{blue} --}        & \multicolumn{1}{c|}{\color{red} $W_1^{(1)},W_2^{(1)}$} & \multicolumn{1}{c|}{\color{red} $W_1^{(2)},W_2^{(2)}$}        \\ \hline
\multicolumn{1}{|c|}{RSMA}           & \multicolumn{1}{c|}{\color{blue} $W_{\mathrm{p},1}^{(1)}$}& \multicolumn{1}{c|}{\color{blue} $W_{\mathrm{p},1}^{(2)}$} & \multicolumn{1}{c|}{\color{blue} $W_{\mathrm{p},2}^{(1)}$}& \multicolumn{1}{c|}{\color{blue} $W_{\mathrm{p},2}^{(2)}$} & \multicolumn{1}{c|}{\color{red} $W_{\mathrm{c},1}^{(1)},W_{\mathrm{c},2}^{(1)}$} & \multicolumn{1}{c|}{\color{red} $W_{\mathrm{c},1}^{(2)},W_{\mathrm{c},2}^{(2)}$}\\ \hline
\multicolumn{1}{|c|}{subscheme 1} & \multicolumn{1}{c|}{\color{blue} $W_1^{(1)}$}    & \multicolumn{1}{c|}{\color{blue} --}& \multicolumn{1}{c|}{\color{blue} $W_2^{(1)}$}    & \multicolumn{1}{c|}{\color{blue} --}        & \multicolumn{1}{c|}{\color{red} $W_1^{(2)}$} & \multicolumn{1}{c|}{\color{red} $W_2^{(2)}$}        \\ \hline
\multicolumn{1}{|c|}{subscheme 2} & \multicolumn{1}{c|}{\color{blue} --}    & \multicolumn{1}{c|}{\color{blue} $W_1^{(2)}$}& \multicolumn{1}{c|}{\color{blue} --}    & \multicolumn{1}{c|}{\color{blue} $W_2^{(2)}$}        & \multicolumn{1}{c|}{\color{red} $W_1^{(1)}$} & \multicolumn{1}{c|}{\color{red} $W_2^{(1)}$}        \\ \hline
\multicolumn{1}{|c|}{subscheme 3} & \multicolumn{1}{c|}{\color{blue} $W_1^{(1)}$}    & \multicolumn{1}{c|}{\color{blue} $W_1^{(2)}$}& \multicolumn{1}{c|}{\color{blue} $W_2^{(1)}$}    & \multicolumn{1}{c|}{\color{blue} --}        & \multicolumn{1}{c|}{\color{red} --} & \multicolumn{1}{c|}{\color{red} $W_2^{(2)}$}        \\ \hline
 \multicolumn{1}{c}{}  & \multicolumn{4}{c}{\color{blue} decoded by its intended user and treated as noise by the other user}            &  \multicolumn{2}{c}{\color{red} decoded by both users}                         \\
 
\end{tabular}
	\label{fig_mapping_MIMO_BC}
\end{table*}

\par The mapping of messages to streams of Table \ref{fig_mapping_BC} is now further extended to account for MIMO RSMA in Table \ref{fig_mapping_MIMO_BC}. RSMA again is a superset of all schemes. We note that SDMA is replaced by MU-MIMO where a vector of private streams is transmitted to each user. NOMA in the form of having one user (user-1) decoding all the messages of the other user (user-2) is illustrated, along with OMA. Additionally, other subschemes of RSMA are listed for the sake of illustration. For instance, in subscheme 1, both users decode the 2nd messages and, in subscheme 2, both users decode the 1st messages, i.e., each user decodes one message of the other user. Subscheme 3 is when both users decode the 2nd message of user-2. 
Note that the MIMO NOMA scheme proposed in \cite{schober2021MUMIMONOMA1,schober2021MUMIMONOMA2} is also a subscheme of MIMO RSMA, and it switches among MU-MIMO, NOMA, or the three additional subschemes of RSMA when different numbers of transmit antennas and receive antennas are considered.
\par Optimization of MIMO RSMA with both perfect and imperfect CSIT can be performed as in \cite{anup2021MIMO}. Similar to the MISO case, RSMA outperforms MU-MIMO and NOMA in the MIMO case. Note that RSMA is information theoretically optimal from a DoF perspective in MIMO BC with imperfect CSIT \cite{chenxi2017bruno,Davoodi2021DoF}. In the asymmetric case where the number of receive antennas is not the same at each user, RSMA with a multi-channel transmission (similar to ST/SF RSMA) needs to be used to achieve optimality. The reader is referred to \cite{chenxi2017bruno} for more details on such a ST RSMA scheme for asymmetric MIMO BC. 

\subsection{Two-User Uplink Rate-Splitting Multiple Access}\label{section_2user_ul}
In the uplink, we consider two single-antenna users simultaneously transmitting their messages to a receiver equipped with $M$ antennas.
When $M=1$, such two-user MAC can be considered as a special case of the two-user IC in Fig. \ref{fig_2user_IC} where the two receivers are colocated and cooperatively decode the messages of the two users.
\subsubsection{Two-User Architectures}  Inspired by the RSMA design in IC, the uplink RSMA architecture has been proposed in \cite{Rimo1996} by splitting the message of one user into two parts. Without loss of generality, we assume that user-1's message $W_1$ is split into $W_{1,1}$ and $W_{1,2}$. By independently encoding the two parts into  $s_{1,1}, s_{1,2}$,  respectively allocating transmit power $P_{1,1}, P_{1,2}$, and superposing the two streams, the transmit signal at user-1 is given by
\begin{equation}
    x_1=
\sqrt{P_{1,1}}s_{1,1}+\sqrt{P_{1,2}}s_{1,2}.
\end{equation}
At user-2, the message $W_2$ is directly encoded into $s_2$. By allocating certain power $P_2$, the transmit signal at user-2 is $x_2=\sqrt{P_{2}}s_{2}$.
\par
\begin{figure}
	\centering
\includegraphics[width=0.45\textwidth]{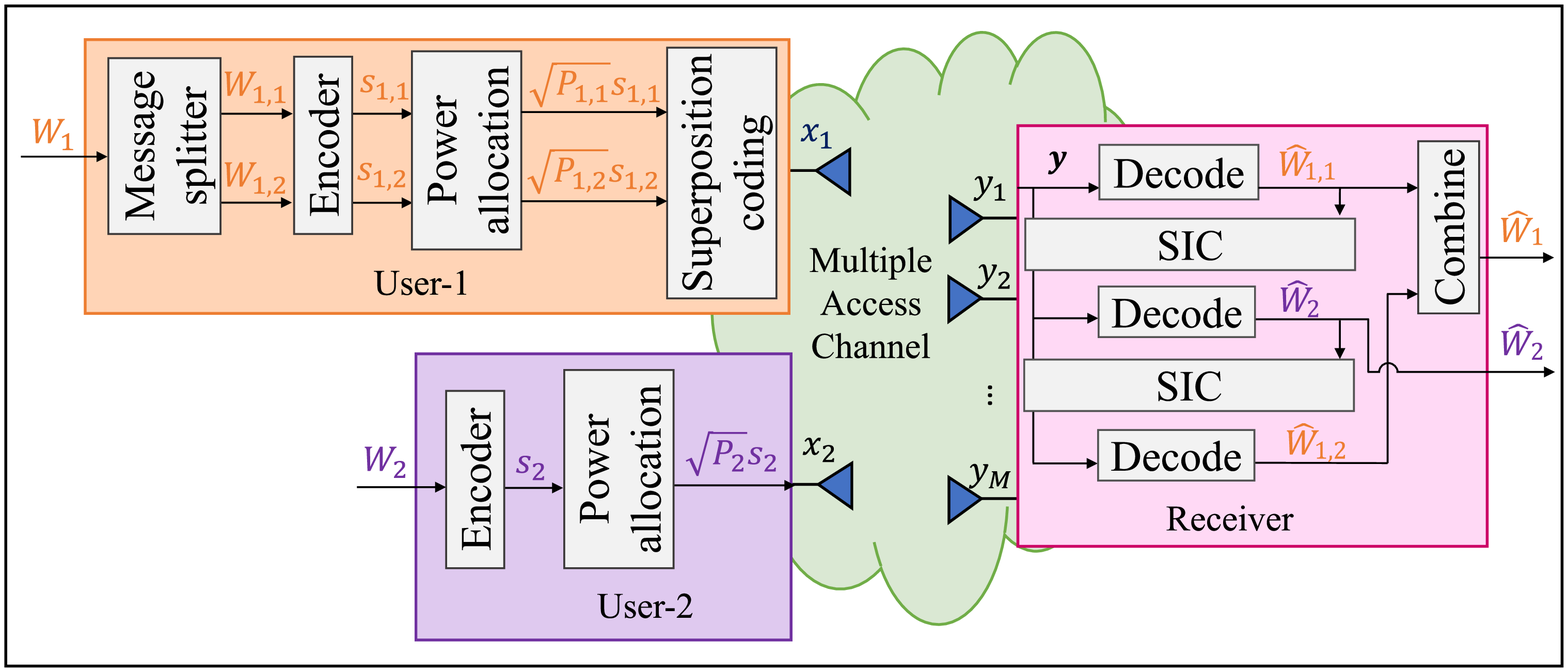}%
	\caption{Two-user uplink SIMO RSMA.}
	\label{Fig_2ULRSMA_architecture}
\end{figure}

 The signal received at the receiver is
\begin{equation}
\begin{aligned}
     \mathbf{y}&=\mathbf{h}_1x_1+\mathbf{h}_2x_2+\mathbf{n}\\
     &=\sqrt{P_{1,1}}\mathbf{h}_1s_{1,1}+\sqrt{P_{1,2}}\mathbf{h}_1s_{1,2}+\sqrt{P_{2}}\mathbf{h}_2s_{2}+\mathbf{n},
\end{aligned}
\end{equation}
where 
$\mathbf{h}_1,\mathbf{h}_2\in \mathbb{C}^{M\times1}$ are the channel vectors and $\mathbf{n}\sim \mathcal{CN}(\mathbf{0},\mathbf{I}_M)$ is the AWGN vector.
The receiver can employ SIC or joint decoding to decode the three streams $s_{1,1}, s_{1,2}, s_2$. If using SIC, two layers of SIC are required at the receiver.
In summary, two-user uplink RSMA  creates two virtual users from the user that splits its message into two parts and in total three streams are sent to the receiver. Two SIC layers are required to recover the original messages of the two users. This uplink RSMA architecture is  illustrated in Fig. \ref{Fig_2ULRSMA_architecture}.

\begin{figure}
	\centering
\includegraphics[width=0.35\textwidth]{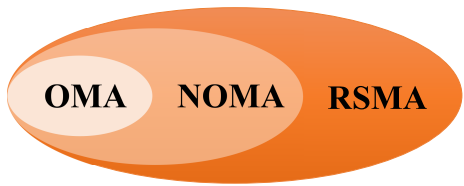}%
	\caption{The relationship between OMA, NOMA, and RSMA in the uplink two-user case. Each set illustrates the optimization space of the corresponding communication strategy. The optimization space of RSMA is larger than OMA and NOMA.}
	\label{Fig_ULRelation}
\end{figure}
\begin{table}[!t]
	\caption{Messages-to-streams\! mapping\! in\! the two-user\! multiple\! access\! channel.}
	\centering
\begin{tabular}{|p{.2\textwidth}|p{2.5cm}|p{2.5cm}|p{2.5cm}|}
\hline
 \multicolumn{1}{|c|}{}  &\multicolumn{2}{c|}{User-1}          & \multicolumn{1}{c|}{User-2} 
\\\hline
\multicolumn{1}{|c|}{}             & \multicolumn{1}{c|}{$s_{1,1}$}     & \multicolumn{1}{c|}{$s_{1,2}$}     & \multicolumn{1}{c|}{$s_{2}$}             \\ \hline
\multicolumn{1}{|c|}{ NOMA}         & \multicolumn{1}{c|}{ $W_1$}     & \multicolumn{1}{c|}{ --}     & \multicolumn{1}{c|}{$W_{2}$}                \\ \hline
\multicolumn{1}{|c|}{ OMA}          & \multicolumn{1}{c|}{$W_1$}     & \multicolumn{1}{c|}{--}        & \multicolumn{1}{c|}{--}                \\ \hline
\multicolumn{1}{|c|}{RSMA}           & \multicolumn{1}{c|}{$W_{1,1}$} & \multicolumn{1}{c|}{ $W_{1,2}$} & \multicolumn{1}{c|}{$W_{2}$} \\ \hline
\end{tabular}
	\label{fig_mapping_MAC}
\end{table}
\subsubsection{Unifying OMA and NOMA} 
In the two-user MAC, both OMA and NOMA are subschemes of RSMA, as per Fig. \ref{Fig_ULRelation}. The corresponding messages to streams mapping is illustrated in Table \ref{fig_mapping_MAC}.
\par \textit{NOMA} is a  subscheme of RSMA by forcing the encoding of message $W_1$ entirely into $s_{\mathrm{1,1}}$  and $W_2$ into $s_2$ while turning off $s_{1,2}$ ($P_{1,2}=0$). 
There is no message splitting at user-1 and one layer of SIC is required at the receiver to decode and remove the entire message of one user (i.e., user-1) before decoding the message of the other user (i.e., user-2). Note that, though we here refer to this strategy as (uplink) NOMA, this strategy is nothing else than the traditional SIC at the receiver to achieve the corner points of the MAC capacity region.
\par \textit{OMA} is also a subscheme of RSMA by forcing one user to be scheduled, i.e., $W_1$ is directly encoded into $s_{1,1}$ while $W_2$ is turned off.  

\subsubsection{Uplink MIMO RSMA}
An extension to the two-user uplink SIMO RSMA in Fig. \ref{Fig_2ULRSMA_architecture} is the two-user uplink MIMO RSMA with $N$ transmit antennas at each user and $M$ receive antennas at the receiver.
In this case, each user-$k$ transmits a $N$-dimensional vector of messages $\mathbf{w}_k=[W_k^{(1)},\ldots,W_k^{(N)}]$ to the receiver. At user-1, each of its $N$ messages $W_1^{(j)}, j\in\{1,\ldots,N\}$ is split into two parts, $W_{1,1}^{(j)}$ and $W_{1,2}^{(j)}$, and encoded independently into $s_{1,1}^{(j)}$ and $s_{1,2}^{(j)}$. The two $N$-dimensional vectors $\mathbf{s}_{1,1}=[s_{1,1}^{(1)},\ldots,s_{1,1}^{(N)}]^T$ and $\mathbf{s}_{1,2}=[s_{1,2}^{(1)},\ldots,s_{1,2}^{(N)}]^T$ are linearly precoded by the precoders $\mathbf{P}_{1,1}, \mathbf{P}_{1,2} \in \mathbb{C}^{M\times N}$ and superposed such that the transmit signal at user-$1$ is
\begin{equation}
    \mathbf{x}_1=\mathbf{P}_{1,1}\mathbf{s}_{1,1}+\mathbf{P}_{1,2}\mathbf{s}_{1,2}.
\end{equation}
At user-2, the message vector $\mathbf{w}_2$ is encoded into a $N$-dimensional stream vector $\mathbf{s}_2$ and linearly precoded by $\mathbf{P}_{2}$. The transmit signal at user-2 is  $\mathbf{x}=\mathbf{P}_{2}\mathbf{s}_{2}$.
\par
\begin{figure}
	\centering
\includegraphics[width=0.5\textwidth]{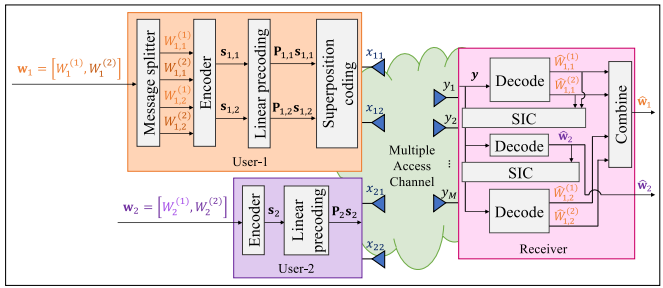}%
	\caption{Two-user uplink MIMO RSMA.}
	\label{Fig_ULMIMORSMA}
\end{figure}

 
An example of the two-user uplink MIMO RSMA when $N=2$ is illustrated in Fig. \ref{Fig_ULMIMORSMA}. Each message in the message vector $\mathbf{w}_1$ of user-1 is split into two parts as
\begin{equation}
\begin{aligned}
        W_1^{(1)}&=\left\{W_{1,1}^{(1)},W_{1,2}^{(1)}\right\},\\
W_1^{(2)}&=\left\{W_{1,1}^{(2)},W_{1,2}^{(2)}\right\}.
\end{aligned}
\end{equation}
The four submessages are respectively encoded into four streams and create two 2-dimensional stream vectors $\mathbf{s}_{1,1}=\big[\begin{array}{cc} s_{1,1}^{(1)}&s_{1,1}^{(2)}\end{array}\big]^T$ and $\mathbf{s}_{1,2}=\big[\begin{array}{cc} s_{1,2}^{(1)}&s_{1,2}^{(2)}\end{array}\big]^T$. The two stream vectors are linearly precoded and superposed to form the transmit signal $\mathbf{x}_1$. 
At user-2, the two messages in $\mathbf{w}_2$ are directly encoded into two streams in $\mathbf{s}_2=\big[\begin{array}{cc} s_{2}^{(1)}&s_{2}^{(2)}\end{array}\big]^T$ and linearly precoded to form the transmit signal $\mathbf{x}_2$. 
The mapping of messages to streams for this example is illustrated in Table \ref{fig_mapping_MIMO_MAC}. Again, MIMO NOMA and OMA are subschemes of MIMO RSMA in the uplink. Though we here did not split user-2 messages, other variants of uplink RSMA schemes can also be derived by also splitting $\mathbf{w}_2$.
\begin{table}
	\caption{Messages-to-streams\! mapping\! in\! the two-user\! MIMO \!multiple\! access\! channel.}
	\centering
\begin{tabular}{|p{.2\textwidth}|p{2.5cm}|p{2.5cm}|p{2.5cm}|p{2.5cm}|p{2.5cm}|p{2.5cm}|p{2.5cm}|}
\hline
\multicolumn{1}{|c|}{}             & \multicolumn{4}{c|}{User-1}       &
\multicolumn{2}{c|}{User-2}             \\ \hline
\multicolumn{1}{|c|}{}             & \multicolumn{2}{c|}{$\mathbf{s}_{1,1}$}     & \multicolumn{2}{c|}{$\mathbf{s}_{1,2}$}     &
\multicolumn{2}{c|}{$\mathbf{s}_{2}$}             \\ \hline
\multicolumn{1}{|c|}{}             & \multicolumn{1}{c|}{$s_{1,1}^{(1)}$}     & \multicolumn{1}{c|}{$s_{1,1}^{(2)}$}     &\multicolumn{1}{c|}{$s_{1,2}^{(1)}$}     & \multicolumn{1}{c|}{$s_{1,2}^{(2)}$}     &\multicolumn{1}{c|}{$s_{2}^{(1)}$}     & \multicolumn{1}{c|}{$s_{2}^{(2)}$}      \\ \hline
\multicolumn{1}{|c|}{NOMA}         & \multicolumn{1}{c|}{$W_1^{(1)}$}& \multicolumn{1}{c|}{$W_1^{(2)}$}     & \multicolumn{1}{c|}{--}  & \multicolumn{1}{c|}{--}        & \multicolumn{1}{c|}{$W_2^{(1)}$}    & \multicolumn{1}{c|}{$W_2^{(2)}$}          \\ \hline
\multicolumn{1}{|c|}{ OMA}         & \multicolumn{1}{c|}{$W_1^{(1)}$}& \multicolumn{1}{c|}{$W_1^{(2)}$}     & \multicolumn{1}{c|}{--}   & \multicolumn{1}{c|}{--}       & \multicolumn{1}{c|}{--}      & \multicolumn{1}{c|}{--}           \\ \hline
\multicolumn{1}{|c|}{RSMA}           & \multicolumn{1}{c|}{$W_{1,1}^{(1)}$}& \multicolumn{1}{c|}{$W_{1,1}^{(2)}$} & \multicolumn{1}{c|}{$W_{1,2}^{(1)}$}& \multicolumn{1}{c|}{$W_{1,2}^{(2)}$}  & \multicolumn{1}{c|}{$W_2^{(1)}$}    & \multicolumn{1}{c|}{$W_2^{(2)}$} \\ \hline
\end{tabular}
	\label{fig_mapping_MIMO_MAC}
\end{table}

\subsection{Lessons Learned}

\noindent\fbox{%
    \parbox{0.97\columnwidth}{%
       \begin{itemize}
           \item OMA, SDMA, and NOMA are respectively based on the orthogonalization, treat interference as noise, and decode interference strategy. This limits the application of each of those MA schemes to a specific interference regime.
            \item RSMA relies on the RS interference strategy and is efficient in all interference regimes. 
            \item RSMA bridges, unifies, and generalizes OMA, SDMA and NOMA.  
            \item RSMA is a superset of OMA, SDMA and NOMA, and can specialize to each of them depending on how messages are mapped to streams.
            \item RSMA is applicable to both downlink and uplink, to SISO, MISO and MIMO settings.
            \item RSMA can be extended to the cooperative relay setting and to space-time/frequency transmission.
       \end{itemize}
    }%
}

\section{$K$-User RSMA}\label{section_Kuser}

The schemes developed for two-user in the previous sections can all be extended to $K$-user and conclusions derived for two-user hold for $K$-user as well. Nevertheless, the $K$-user scenario also opens the door to other RSMA schemes with a variable number of SIC layers. In this section, the state-of-the-art $K$-user RSMA schemes are delineated in the downlink, uplink, and multi-cell MIMO settings where each node is equipped with multiple antennas.

\subsection{Downlink}
\label{sec:DL}
We first consider a downlink symmetric MIMO BC where a transmitter equipped with $M$ antennas serves $K$ users, each equipped with $N$ receive antennas. The users are indexed by $\mathcal{K}=\{1,\ldots,K\}$. Without loss of generality, we assume a $Q_k$-dimensional vector of messages is transmitted to user-$k$, i.e., $\mathbf{w}_k=[W_k^{(1)},\ldots,W_k^{(Q_k)}]^T$, where $Q_k\leq \min(M,N)$.
Depending on the different RSMA schemes adopted, user messages $\{\mathbf{w}_k, k\in\mathcal{K}\}$ are split, combined, and encoded into different stream vectors, which are then mapped to the transmit antennas using linear or non-linear precoders and forms the transmit signal $\mathbf{x}$.
The signal is transmitted through a MIMO BC and the receive signal at user-$k$ is given by 
$\mathbf{y}_k=\mathbf{H}_k^H\mathbf{x}+\mathbf{n}_k$, where  $\mathbf{H}_k\in\mathbb{C}^{M\times N}$ is the channel matrix between the base station and user-$k$,  $\mathbf{n}_k\sim \mathcal{CN}(\mathbf{0},\sigma_{n,k}^2\mathbf{I}_N)$ is the AWGN vector at user-$k$. 
Without loss of generality, the noise variances of all users are assumed to be equal to 1, i.e., $\sigma_{n,k}^2=1, \forall k\in\mathcal{K}$.
Next, we detail the transceiver architectures for different RSMA schemes in such symmetric MIMO BC.
\subsubsection{1-layer RS}
\label{sec:1RS}
\begin{figure}
	\centering
\includegraphics[width=0.5\textwidth]{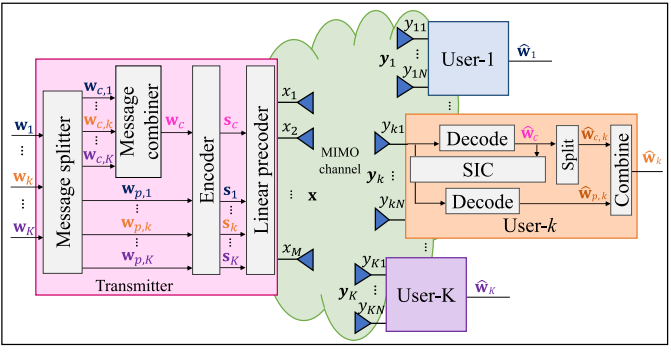}%
	\caption{$K$-user downlink 1-layer MIMO RSMA.}
	\label{Fig_DL1LMIMORSMA}
\end{figure}

The system model of 1-layer RS for MIMO BC is first proposed in \cite{anup2021MIMO}. 
As illustrated in Fig. \ref{Fig_DL1LMIMORSMA}, each  message $W_k^{(i)}$  in $\mathbf{w}_k$ for  user-$k$ is split into one common sub-message and one private sub-message as $W_k^{(i)}=\{W_{\mathrm{c},k}^{(i)},W_{\mathrm{p},k}^{(i)}\}, \forall i\in\{1,\ldots,Q_k\}$, resulting in one common message vector $\mathbf{w}_{\mathrm{c},k}=[W_{\mathrm{c},k}^{(1)},\ldots,W_{\mathrm{c},k}^{(Q_k)}]^T$ and one private message vector $\mathbf{w}_{\mathrm{p},k}=[W_{\mathrm{p},k}^{(1)},\ldots,W_{\mathrm{p},k}^{(Q_k)}]^T$ for user-$k$.
The common messages vectors  of all users $\{\mathbf{w}_{\mathrm{c},1},\ldots,\mathbf{w}_{\mathrm{c},K}\}$ are combined into $Q_{\mathrm{c}}$ ($Q_{\mathrm{c}}\leq \min(M,N)$) common messages $\mathbf{w}_{\mathrm{c}}\in\mathbb{C}^{Q_{\mathrm{c}}\times1}$ and encoded into one common stream vector  $\mathbf{s}_{\mathrm{c}}=[s_{\mathrm{c}}^{(1)},\ldots,s_{\mathrm{c}}^{(Q_{\mathrm{c}})}]^T$, which is decoded by all users. 
The private message vector $\mathbf{w}_{\mathrm{p},k}$ of user-$k$ is independently encoded into the private stream vector  $\mathbf{s}_{k}=[s_{k}^{(1)},\ldots,s_{k}^{(Q_k)}]^T$ to be decoded by user-$k$ only.
The $K+1$ data stream vectors $\mathbf{s}_{\mathrm{c}},\mathbf{s}_1,\ldots,\mathbf{s}_K$ are linearly precoded by the precoders $\mathbf{P}_{\mathrm{c}},\mathbf{P}_1,\ldots,\mathbf{P}_K$ and superposed at the transmitter. The resulting transmit signal is
\begin{equation}
    \mathbf{x}=\mathbf{P}_{\mathrm{c}}\mathbf{s}_{\mathrm{c}}+\sum_{k=1}^{K}\mathbf{P}_{k}\mathbf{s}_{k}.
\end{equation}
\par
The signal received at user-$k$ is
\begin{equation}
     \mathbf{y}
     =\mathbf{H}_k^H\mathbf{P}_{\mathrm{c}}\mathbf{s}_{\mathrm{c}}+\sum_{j=1}^{K}\mathbf{H}_k^H\mathbf{P}_{j}\mathbf{s}_{j}+\mathbf{n}_k.
\end{equation}
User-$k$ employs SIC or joint decoding to decode two stream vectors $\mathbf{s}_{\mathrm{c}}$ and $\mathbf{s}_k$. Only one single layer of SIC is required at each user if employing SIC to decode the data streams. 
Assuming Gaussian signalling, the instantaneous rates of decoding the common and private stream vectors based on SIC are given by
\begin{equation}
    \begin{aligned}
    &R_{\mathrm{c},k}=\log_2\det\left(\mathbf{I}+\mathbf{P}_{\mathrm{c}}^H\mathbf{H}_k\left(\mathbf{R}_{\mathrm{c},k}\right)^{-1}\mathbf{H}_k^H\mathbf{P}_{\mathrm{c}}\right),\\
    &R_{k}=\log_2\det\left(\mathbf{I}+\mathbf{P}_k^H\mathbf{H}_k\left(\mathbf{R}_{k}\right)^{-1}\mathbf{H}_k^H\mathbf{P}_k\right),
    \end{aligned}
\end{equation}
where $\mathbf{R}_{\mathrm{c},k}$ and $\mathbf{R}_{k}$ are noise plus interference covariance matrices defined as
\begin{equation}
\begin{aligned}
    &\mathbf{R}_{\mathrm{c},k}=\mathbf{I}+\sum_{j=1}^K\mathbf{H}_k^H\mathbf{P}_{j}\mathbf{P}_{j}^H\mathbf{H}_k,\\
    &\mathbf{R}_{k}=\mathbf{I}+\sum_{j=1,j\neq k}^K\mathbf{H}_k^H\mathbf{P}_{j}\mathbf{P}_{j}^H\mathbf{H}_k.
\end{aligned}
\end{equation}
To guarantee the common stream vector is successfully decoded by all users, it is achievable rate is $R_{\mathrm{c}}=\min_k\{R_{\mathrm{c},1},\ldots,R_{\mathrm{c},K}\}$.
As $\mathbf{s}_{\mathrm{c}}$ contains sub-messages of $K$ users, we have $\sum_{k\in\mathcal{K}}C_k=R_{\mathrm{c}}$, where $C_k$ denotes the portion of $R_{\mathrm{c}}$ allocated to user-$k$ for the transmission of $\mathbf{w}_{{\mathrm{c}},k}$. 
The overall achievable rate of user-$k$ is $R_{k,\mathrm{tot}}=C_k+R_{k}$.

\subsubsection{Hierarchical RS}
Hierarchical RS (HRS) is proposed in \cite{Minbo2016MassiveMIMO} for MISO BC where the transmitter is equipped with multiple antennas and each receiver is equipped with a single antenna. In this subsection, we extend the system model to MIMO BC where each node has multiple antennas. 
In HRS, users are clustered into $G$ separate groups indexed by $\mathcal{G}=\{1,\ldots,G\}$ according to the similarity of their channel covariance matrices.  Each group-$g$ contains a subset of users $\mathcal{K}_g$ such that $\bigcup _{g\in\mathcal{G}}\mathcal{K}_g=\mathcal{K}$. 
In contrast to 1-layer RS where each user message is split into two parts, in HRS, each user message $W_k^{(i)}, i\in\{1,\ldots, Q_k\}$ in $\mathbf{w}_k$ for user-$k$ (assuming user-$k$ is in group $g$) is split into three sub-messages as $W_k^{(i)}=\{W_{\mathcal{K},k}^{(i)},W_{\mathcal{K}_g,k}^{(i)},W_{k,k}^{(i)}\}$, resulting in three sub-message vectors, namely, an inter-group common message vector $\mathbf{w}_{k}^{\mathcal{K}}=\{W_{\mathcal{K},k}^{(1)},\ldots,W_{\mathcal{K},k}^{(Q_k)}\}$,
an inner-group common message vector $\mathbf{w}_{k}^{\mathcal{K}_g}=\{W_{\mathcal{K}_g,k}^{(1)},\ldots,W_{\mathcal{K}_g,k}^{(Q_k)}\}$, and a private message vector $\mathbf{w}_{k}^k=\{W_{k,k}^{(1)},\ldots,W_{k,k}^{(Q_k)}\}$. 
The inter-group common messages of all users  $\{\mathbf{w}_{1}^{\mathcal{K}},\ldots,\mathbf{w}_{K}^{\mathcal{K}}\}$ are combined into $Q_{\mathrm{c}}$ ($Q_{\mathrm{c}}\leq \min(M,N)$) common messages $\mathbf{w}_{\mathcal{K}}\in\mathbb{C}^{Q_{\mathrm{c}}\times1}$ and encoded into one common stream vector  $\mathbf{s}_{\mathcal{K}}=[s_{\mathcal{K}}^{(1)},\ldots,s_{\mathcal{K}}^{(Q_{\mathrm{c}})}]^T$, which is decoded by all users.
For group-$g, g\in\mathcal{G}$, the inner-group common message vectors of  users in group-$g$  $\{\mathbf{w}_{k}^{\mathcal{K}_g}, \forall k\in\mathcal{K}_g\}$ are combined into $Q_{\mathrm{c}g}$ ($Q_{\mathrm{c}g}\leq \min(M,N)$) common messages $\mathbf{w}_{\mathcal{K}_g}\in\mathbb{C}^{Q_{\mathrm{c}g}\times1}$ and encoded into one common stream vector  $\mathbf{s}_{\mathcal{K}_g}=[s_{\mathcal{K}_g}^{(1)},\ldots,s_{\mathcal{K}_g}^{(Q_{\mathrm{c}g})}]^T$ to be decoded by all users in group-$g$. 
The private message vector $\mathbf{w}_{k}^k$  of user-$k$ is independently encoded into the private stream vector  $\mathbf{s}_{k}=[s_{k}^{(1)},\ldots,s_{k}^{(Q_k)}]^T$ to be decoded by user-$k$ only.
The overall $K+G+1$ data stream vectors $\mathbf{s}_{\mathcal{K}}, \mathbf{s}_{\mathcal{K}_1},\ldots,\mathbf{s}_{\mathcal{K}_G},\mathbf{s}_{1},\ldots,\mathbf{s}_{K}$ are linearly precoded by the corresponding precoders $\mathbf{P}_{\mathcal{K}},\mathbf{P}_{\mathcal{K}_1},\ldots,\mathbf{P}_{\mathcal{K}_G},\mathbf{P}_{1},\ldots,\mathbf{P}_{K}$. The resulting transmit signal is
\begin{equation}
    \mathbf{x}=\mathbf{P}_{\mathcal{K}}\mathbf{s}_{\mathcal{K}}+\sum_{g=1}^{G}\mathbf{P}_{\mathcal{K}_g}\mathbf{s}_{\mathcal{K}_g}+\sum_{k=1}^{K}\mathbf{P}_{k}\mathbf{s}_{k}.
\end{equation}
\par
The signal received at user-$k$ is
\begin{equation}
     \mathbf{y}
     =\mathbf{H}_k^H\mathbf{P}_{\mathcal{K}}\mathbf{s}_{\mathcal{K}}+\sum_{g=1}^{G}\mathbf{H}_k^H\mathbf{P}_{\mathcal{K}_g}\mathbf{s}_{\mathcal{K}_g}+\sum_{j=1}^{K}\mathbf{H}_k^H\mathbf{P}_{j}\mathbf{s}_{j}+\mathbf{n}_k.
\end{equation}
In contrast to 1-layer RS where each user only decodes two streams, in HRS, each user-$k$ ($k\in\mathcal{K}_g$) employs SICs or joint decoding to decode three stream vectors $\mathbf{s}_\mathcal{K}$, $\mathbf{s}_{\mathcal{K}_g}$,  and $\mathbf{s}_k$. Two layers of SIC are required at each user if employing SIC. 
Assuming Gaussian signalling, the instantaneous rates of decoding $\mathbf{s}_\mathcal{K}$, $\mathbf{s}_{\mathcal{K}_g}$,  and $\mathbf{s}_k$ based on SIC are given by
\begin{equation}
    \begin{aligned}
    &R_{\mathcal{K},k}=\log_2\det\left(\mathbf{I}+\mathbf{P}_\mathcal{K}^H\mathbf{H}_k\left(\mathbf{R}_{\mathcal{K},k}\right)^{-1}\mathbf{H}_k^H\mathbf{P}_\mathcal{K}\right),\\
    &R_{{\mathcal{K}_g},k}=\log_2\det\left(\mathbf{I}+\mathbf{P}_{{\mathcal{K}_g}}^H\mathbf{H}_k\left(\mathbf{R}_{{\mathcal{K}_g},k}\right)^{-1}\mathbf{H}_k^H\mathbf{P}_{{\mathcal{K}_g}}\right),\\
    &R_{k}=\log_2\det\left(\mathbf{I}+\mathbf{P}_{k}^H\mathbf{H}_k\left(\mathbf{R}_{k}\right)^{-1}\mathbf{H}_k^H\mathbf{P}_{k}\right),\\
    \end{aligned}
\end{equation}
where the noise plus interference covariance matrices $\mathbf{R}_{\mathcal{K},k}$, $\mathbf{R}_{{\mathcal{K}_g},k}$,  and $\mathbf{R}_{k}$ are defined as
\begin{equation}
\begin{aligned}
    &\mathbf{R}_{\mathcal{K},k}=\mathbf{I}+\sum_{g=1}^{G}\mathbf{H}_k^H\mathbf{P}_{\mathcal{K}_g}\mathbf{P}_{\mathcal{K}_g}^H\mathbf{H}_k+\sum_{j=1}^K\mathbf{H}_k^H\mathbf{P}_{j}\mathbf{P}_{j}^H\mathbf{H}_k,\\
    &\mathbf{R}_{{\mathcal{K}_g},k}=\mathbf{I}+\sum_{i=1, i\neq g}^{G}\mathbf{H}_k^H\mathbf{P}_{\mathcal{K}_i}\mathbf{P}_{\mathcal{K}_i}^H\mathbf{H}_k+\sum_{j=1}^K\mathbf{H}_k^H\mathbf{P}_{j}\mathbf{P}_{j}^H\mathbf{H}_k,\\
    &\mathbf{R}_{k}=\mathbf{I}+\sum_{i=1, i\neq g}^{G}\mathbf{H}_k^H\mathbf{P}_{\mathcal{K}_i}\mathbf{P}_{\mathcal{K}_i}^H\mathbf{H}_k+\sum_{j=1,j\neq k}^K\mathbf{H}_k^H\mathbf{P}_{j}\mathbf{P}_{j}^H\mathbf{H}_k.
\end{aligned}
\end{equation}
To guarantee the successful decoding of the inter-group and inner-group common stream vectors, the achievable rates follow $R_{\mathcal{K}}=\min_k\{R_{{\mathcal{K}},k},k\in\mathcal{K}\}$ and $R_{\mathcal{K}_g}=\min_k\{R_{{\mathcal{K}_g},k}, k\in{\mathcal{K}_g}\}$.
As $\mathbf{s}_{\mathcal{K}}$ and $\mathbf{s}_{\mathcal{K}_g}$ contain sub-messages of multiple users, we have $\sum_{k\in\mathcal{K}}C_k^{\mathcal{K}}=R_{\mathcal{K}}$ and $\sum_{k\in\mathcal{K}_g}C_k^{\mathcal{K}_g}=R_{\mathcal{K}_g}$, where $C_k^{\mathcal{K}}$ and $C_k^{\mathcal{K}_g}$ respectively denote the portion of $R_{\mathcal{K}}$ and $R_{\mathcal{K}_g}$ allocated to user-$k$ for the transmission of $\mathbf{w}_{k}^{{\mathcal{K}}}$ and $\mathbf{w}_{k}^{{\mathcal{K}_g}}$. 
The overall achievable rate of user-$k$ is $R_{k,\mathrm{tot}}=C_k^{\mathcal{K}}+C_k^{\mathcal{K}_g}+R_{k}$.
It is easy to observe that HRS is a more general framework than  1-layer RS for MIMO BC. HRS boils down to 1-layer RS when the inner-group common stream vectors $\{\mathbf{s}_{{\mathcal{K}_g}}, \forall g\in{\mathcal{G}}\}$ are turned off, i.e., $\mathbf{P}_{{\mathcal{K}_g}}=\mathbf{0}, \forall g\in{\mathcal{G}}$. In this case, $\mathbf{s}_{{\mathcal{K}}}$ for HRS is equivalent to $\mathbf{s}_{c}$ for 1-layer RS.

\par
\begin{figure}
	\centering
\includegraphics[width=0.5\textwidth]{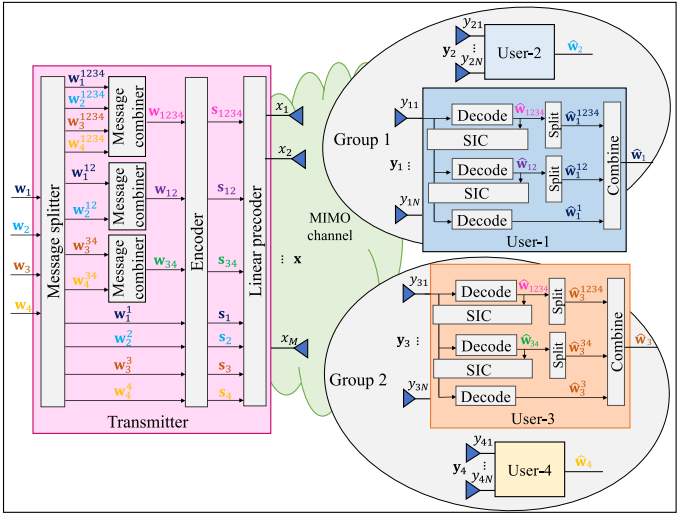}%
	\caption{$4$-user downlink MIMO HRS.}
	\label{Fig_DLMIMOHRS}
\end{figure}

 
In Fig. \ref{Fig_DLMIMOHRS}, a four-user two-group HRS example is illustrated with user-1 and user-2 in group-1, user-3 and user-4 in group-2. The user sets $\mathcal{K}=\{1,2,3,4\}$,  $\mathcal{K}_1=\{1,2\}$, $\mathcal{K}_2=\{3,4\}$ are simply denoted as $1234$, $12$, $34$, respectively. User-1 and user-2 are required to decode the inter-group common stream vector $\mathbf{s}_{1234}$ and the inner-group common stream vector $\mathbf{s}_{12}$ while user-3 and user-4 are required to decode the  $\mathbf{s}_{1234}$ and  $\mathbf{s}_{34}$.

\subsubsection{Generalized RS}
Generalized RS (GRS) is proposed in \cite{mao2017rate} for MISO BC to further enhance the spectral efficiency by splitting and encoding multiple common streams, which are required to be decoded by different subsets of users. It is a generalized transmission framework that embraces 1-layer RS, HRS, linearly precoded MU-MIMO, and MIMO NOMA as special cases. We here extend it to MIMO BC. 
At the transmitter, each user message is split into $2^{K-1}$ parts, i.e., $W_k^{(i)}=\{W_{\mathcal{A}',k}^{(i)}\mid \mathcal{A}'\in\mathcal{K}, k\in\mathcal{A}'\}$. The message vector $\mathbf{w}_k$ is therefore split into $2^{K-1}$ sub-message vectors $\mathbf{w}_{k}^{\mathcal{A}'}, \forall \mathcal{A}'\in\mathcal{K}, k\in\mathcal{A}'\}$ with 
$\mathbf{w}_{k}^{\mathcal{A}'}=\{W_{\mathcal{A}',k}^{(1)},\ldots,W_{\mathcal{A}',k}^{(Q_k)}\}$. 
The sub-message vectors $\{\mathbf{w}_{k}^{\mathcal{A}}\mid k\in\mathcal{A}\}$ for all users in a given subset $\mathcal{A}\in\mathcal{K}$ are combined and encoded into one stream vector $\mathbf{s}_{\mathcal{A}}$ of dimension ${Q_{\mathrm{c}\mathcal{A}}\times1}$ which is only decoded by users in ${\mathcal{A}}$ and treated as noise by the remaining users.
Following the concept of stream order introduced in GRS for MISO BC \cite{mao2017rate}, we denote the stream vectors to be decoded by $l$ number of users as $l$-order stream vectors.
All $l$-order stream vectors $\{\mathbf{s}_{\mathcal{A}'}\mid \mathcal{A}'\in\mathcal{K},  |\mathcal{A}'|=l\}$ are wrapped up into a larger dimensional vector $\mathbf{s}_l\in \mathbb{C}^{\sum_{\mathcal{A}'\in\mathcal{K}, \mathcal |\mathcal{A}'|=l}Q_{\mathrm{c}\mathcal{A'}}\times1}$ and linearly precoded by the precoding matrix $\mathbf{P}_l$ composed by $\{\mathbf{P}_{\mathcal{A}'}\mid \mathcal{A}'\in\mathcal{K}, |\mathcal{A}'|=l\}$. The resulting transmit signal is
\begin{equation}
	\mathbf{{x}}=\sum_{l=1}^{K}\mathbf{{P}}_{{l}}\mathbf{{s}}_{{l}}=\sum_{l=1}^{K}\sum_{\mathcal{A}'\subseteq\mathcal{K},|\mathcal{A}'|=l}\mathbf{{P}}_{\mathcal{A}'}{\mathbf{s}}_{\mathcal{A}'}.
\end{equation}

\par At user-$k$,   $2^{K-1}-1$  layers of SIC or joint decoding is employed to decode the  required streams vectors. 
The set of $l$-order stream vectors required to be decoded at user-$k$ is denoted by $\mathcal{S}_{l,k}=\{\mathbf{s}_{\mathcal{A'}}\mid \mathcal{A}'\subseteq\mathcal{K},|\mathcal{A}'|=l,k\in\mathcal{A}'\}$.
Following the decoding order from the $K$-order stream vector down to the $1$-order stream vector as well as a certain decoding order $\pi_{l,k}$  to decode the $l$-order stream vectors in $\mathcal{S}_{l,k}$ (i.e., $\mathbf{s}_{\pi_{l,k}(i)}$ is decoded before $\mathbf{s}_{\pi_{l,k}(j)}$ if $i<j$), we obtain the rate of decoding $l$-order stream vector $\mathbf{s}_{\pi_{l,k}(i)}$ at user-$k$ under Gaussian signalling and SIC decoding as
\begin{equation}
	\begin{aligned}
		&R_{\pi_{l,k}{(i)},k}=\\
		&\,\,\log_2\det\left(\mathbf{I}+\mathbf{P}_{\pi_{l,k}{(i)}}^H\mathbf{H}_k\left(\mathbf{R}_{\pi_{l,k}{(i)},k}\right)^{-1}\mathbf{H}_k^H\mathbf{P}_{\pi_{l,k}{(i)}}\right),
	\end{aligned}
\end{equation}
where $\mathbf{R}_{\pi_{l,k}{(i)},k}$ is given as
\begin{equation}
	\begin{aligned}
		\mathbf{R}_{\pi_{l,k}{(i)},k}&=\mathbf{I}+\sum_{j>i}\mathbf{{H}}_{k}^{H}\mathbf{{P}}_{\pi_{l,k}(j)}\mathbf{{P}}_{\pi_{l,k}(j)}^H\mathbf{{H}}_{k}\\
		&+\sum_{l'=1}^{l-1}\sum_{j=1}^{|\mathcal{S}_{l',k}|}\mathbf{{H}}_{k}^{H}\mathbf{{P}}_{\pi_{l',k}(j)}\mathbf{{P}}_{\pi_{l',k}(j)}^H\mathbf{{H}}_{k}\\
		&+\sum_{\mathcal{A}'\subseteq\mathcal{K},k\notin\mathcal{A}'}\mathbf{{H}}_{k}^{H}\mathbf{{P}}_{{\mathcal{A}'}}\mathbf{{P}}_{{\mathcal{A}'}}^{H}\mathbf{{H}}_{k}.
	\end{aligned}
\end{equation}
To guarantee the successful decoding of each stream vector $\mathbf{s}_{\mathcal{A}}$, the achievable rates follow $R_{\mathcal{A}}=\min_k\{R_{{\mathcal{A}},k},k\in{\mathcal{A}}\}$.
As $\mathbf{s}_{\mathcal{A}}$ contains sub-messages of users in ${\mathcal{A}}$, we have $\sum_{k\in\mathcal{A}}C_k^{\mathcal{K}}=R_{\mathcal{A}}$, where $C_k^{\mathcal{A}}$ denotes the portion of $R_{\mathcal{A}}$  allocated to user-$k$ for the transmission of $\mathbf{w}_{k}^{{\mathcal{A}}}$. 
The overall achievable rate of user-$k$ is $R_{k,\mathrm{tot}}=\sum_{k\in \mathcal{A}'}C_{k}^{\mathcal{A}'}+R_{k,k}$.
GRS is a more general framework than 1-layer RS and HRS for MIMO BC. It boils down to 1-layer RS when only the $K$-order stream vector $\mathbf{s}_{\mathcal{K}}$ and the $1$-order stream vectors $\mathbf{s}_k, k\in\mathcal{K}$  are active.  $\mathbf{s}_{\mathcal{K}}$ for GRS is equivalent to $\mathbf{s}_{c}$ for 1-layer RS in such case. It reduces to HRS when only $\mathbf{s}_{\mathcal{K}}$,  $\mathbf{s}_k, k\in\mathcal{K}$, and $\mathbf{s}_{\mathcal{K}_g}, g\in\mathcal{G}$  are active. 

\par
\begin{figure}
	\centering
\includegraphics[width=0.5\textwidth]{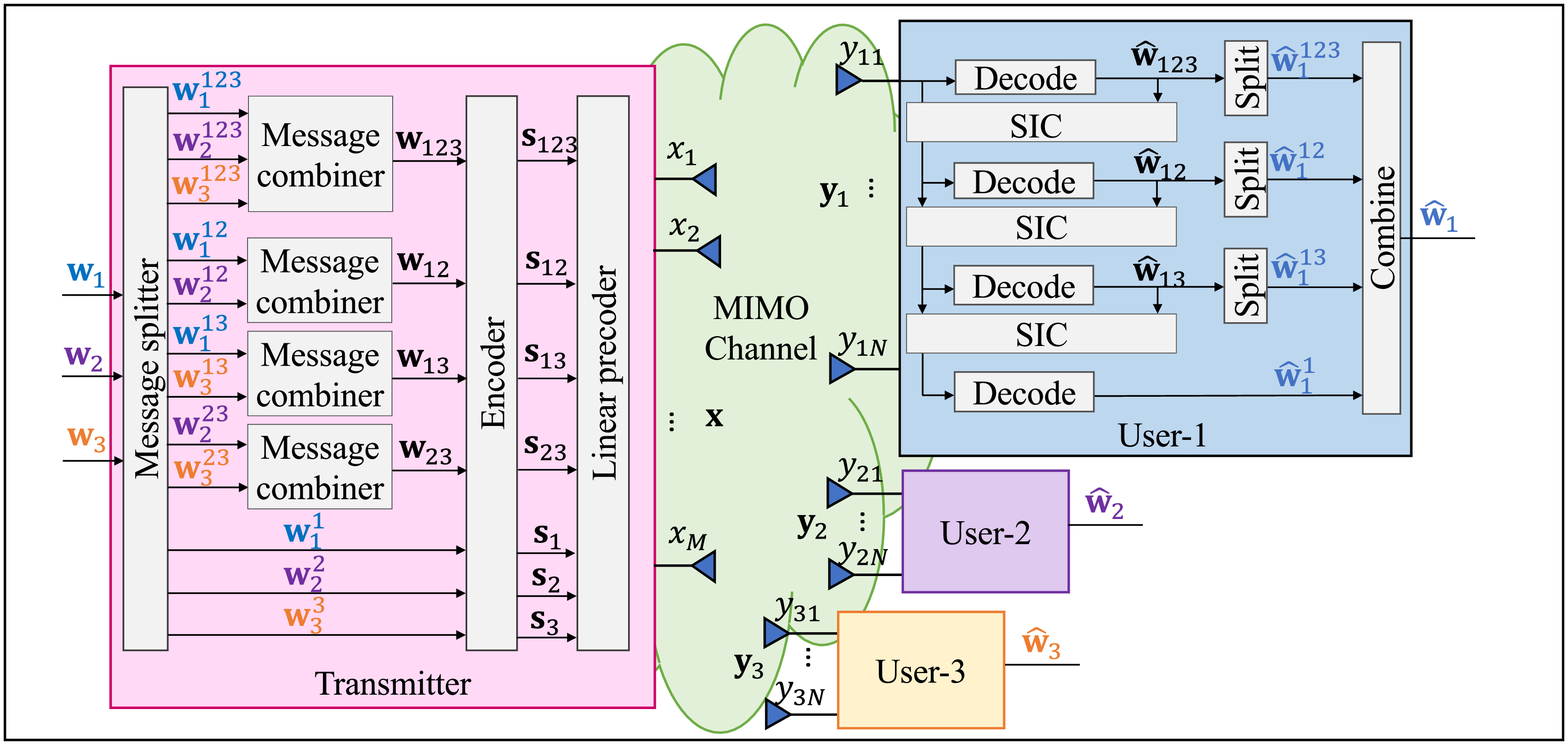}%
	\caption{$3$-user downlink MIMO generalized RS.}
	\label{Fig_DLMIMOGRS}
\end{figure}

 
A toy $3$-user GRS example is illustrated in Fig. \ref{Fig_DLMIMOGRS}. 
The user message vector of each user is split into 4 sub-message vectors. For user-1,  $\mathbf{w}_1$ is split into $\{\mathbf{w}_{1}^{123}$, $\mathbf{w}_{1}^{12}$, $\mathbf{w}_{1}^{13}$, $\mathbf{w}_{1}^{1}\}$, each is required to be decoded by different groups of users. $\{\mathbf{w}_{1}^{123}, \mathbf{w}_{2}^{123}, \mathbf{w}_{3}^{123}\}$ are combined into $\mathbf{w}_{123}$, which is then encoded into the $3$-order stream vector $\mathbf{s}_{123}$. Similarly,  $\{\mathbf{w}_{1}^{12},\mathbf{w}_{2}^{12}\}$ are combined into $\mathbf{w}_{12}$ and encoded into  the $2$-order stream vector $\mathbf{s}_{12}$. Following this method, we obtain in total 7 transmit stream vectors.
3 layers of SIC is adopted at each user to decode the intended $4$ stream vectors. In this example, user-$1$ decodes $\mathbf{s}_{123}$, $\mathbf{s}_{12}$, $\mathbf{s}_{13}$, $\mathbf{s}_{1}$ sequentially with $\mathbf{s}_{\pi_{2,1}{(1)}}=\mathbf{s}_{12}$ and $\mathbf{s}_{\pi_{2,1}{(2)}}=\mathbf{s}_{13}$.

\subsubsection{Dirty paper coded RS}
Dirty paper coded RS (DPCRS) is a non-linearly precoded RS framework built upon DPC.  It is proposed in \cite{mao2019beyondDPC} for MISO BC and is extended to MIMO BC in this subsection. 
We first illustrate the transmission architecture of the simplest DPCRS model 1-layer Dirty Paper Coded RS (1-DPCRS), and then briefly discuss its extension to  multi-layer dirty paper coded rate-splitting (M-DPCRS).
\par
\begin{figure}
	\centering
\includegraphics[width=0.5\textwidth]{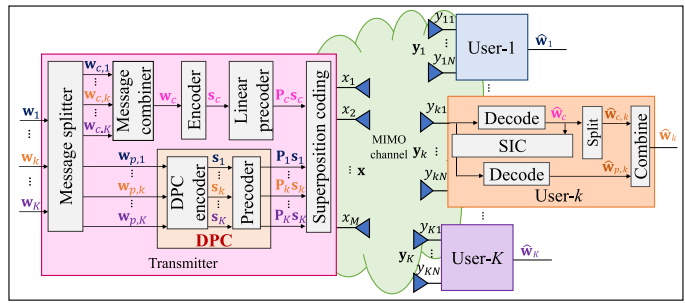}%
	\caption{$K$-user downlink 1-layer MIMO DPCRS.}
	\label{Fig_DL1LMIMODPCRS}
\end{figure}

 
Fig. \ref{Fig_DL1LMIMODPCRS} illustrates the proposed 1-layer MIMO DPCRS. It is an extension of 1-layer RS in Section \ref{sec:1RS} by enabling dirty paper encoded private streams.
Following 1-layer RS, the $K$ user messages are split and  combined into one common message vector $\mathbf{w}_{\mathrm{c}}$ and $K$ private message vectors $\mathbf{w}_{\mathrm{p},1},\ldots,\mathbf{w}_{\mathrm{p},K}$. $\mathbf{w}_{\mathrm{c}}$ is encoded into a common stream vector $\mathbf{s}_{\mathrm{c}}$ using a codebook shared by all users while $\mathbf{w}_{\mathrm{p},1},\ldots,\mathbf{w}_{\mathrm{p},K}$ are encoded and precoded by DPC for a certain encoding order $\pi$ into the private stream vectors  $\mathbf{s}_{\pi(1)},\ldots,\mathbf{s}_{\pi(K)}$.
The resulting transmit signal is
\begin{equation}
	\mathbf{x}={{\mathbf{P}_{\mathrm{c}}\mathbf{s}_{\mathrm{c}}}}+{{\sum_{k=1}^K\mathbf{P}_{\pi(k)}\mathbf{s}_{\pi(k)}}},
\end{equation}
where $\mathbf{P}_{\mathrm{c}}, \mathbf{P}_{\pi(1)}, \ldots, \mathbf{P}_{\pi(K)}$ are the  precoders for the corresponding streams $\mathbf{s}_{\mathrm{c}}, \mathbf{s}_{\pi(1)}, \ldots, \mathbf{s}_{\pi(K)}$.
\par
\begin{figure}
	\centering
\includegraphics[width=0.5\textwidth]{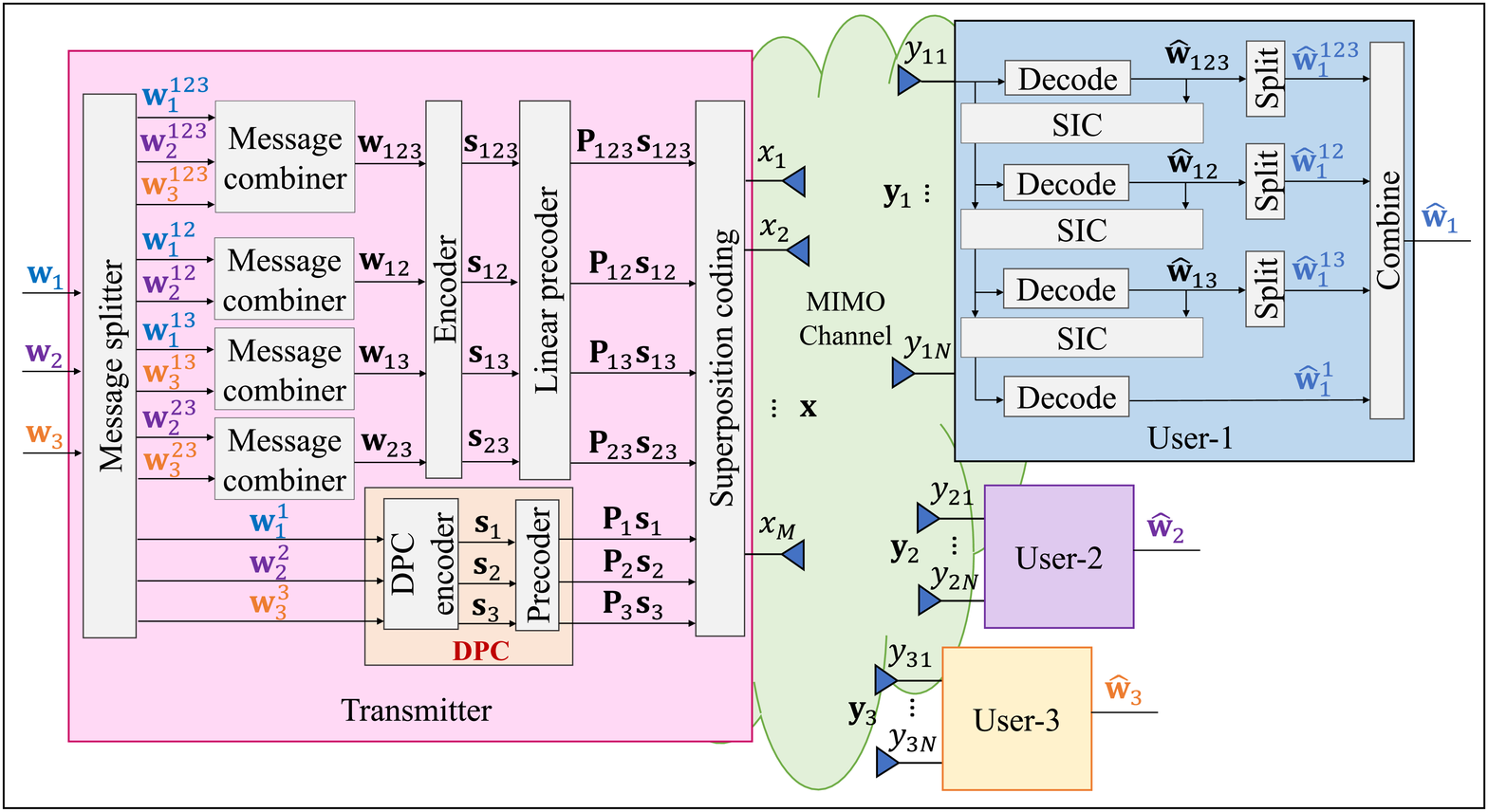}%
	\caption{$3$-user downlink multi-layer MIMO DPCRS.}
	\label{Fig_DLMLMIMODPCRS}
\end{figure}

 
Each user-$\pi(k)$ decodes $\mathbf{s}_{\mathrm{c}}$ and $\mathbf{s}_{\pi(k)}$ based on SIC or joint decoding. 
Assuming Gaussian signalling, the instantaneous rates of decoding the common and private stream vectors based on SIC are given by
\begin{equation}
	\begin{aligned}
		&R_{\mathrm{c},\pi(k)}=\log_2\det\left(\mathbf{I}+\mathbf{P}_{\mathrm{c}}^H\mathbf{H}_{\pi(k)}\left(\mathbf{R}_{\mathrm{c},\pi(k)}\right)^{-1}\mathbf{H}_{\pi(k)}^H\mathbf{P}_{\mathrm{c}}\right),\\
		&R_{\pi(k)}=\log_2\det\left(\mathbf{I}+\mathbf{P}_{\pi(k)}^H\mathbf{H}_{\pi(k)}\left(\mathbf{R}_{\pi(k)}\right)^{-1}\mathbf{H}_{\pi(k)}^H\mathbf{P}_{\pi(k)}\right),
	\end{aligned}
\end{equation}
where $\mathbf{R}_{\mathrm{c},{\pi(k)}}$ and $\mathbf{R}_{\pi(k)}$ are noise plus interference covariance matrices defined as
\begin{equation}
	\begin{aligned}
		&\mathbf{R}_{\mathrm{c},{\pi(k)}}=\mathbf{I}+\sum_{j=1}^K\mathbf{H}_{\pi(k)}^H\mathbf{P}_{\pi(j)}\mathbf{P}_{\pi(j)}^H\mathbf{H}_{\pi(k)},\\
		&\mathbf{R}_{\pi(k)}=\mathbf{I}+\sum_{j>k}^K\mathbf{H}_{\pi(k)}^H\mathbf{P}_{\pi(j)}\mathbf{P}_{\pi(j)}^H\mathbf{H}_{\pi(k)}.
	\end{aligned}
\end{equation}
The calculation of the overall achievable rate for each user follows 1-layer RS, which is not detailed here. 
Besides 1-layer MIMO DPCRS, we can also integrate DPC and different RS schemes to enhance the system performance especially in imperfect CSIT. For instance, Fig. \ref{Fig_DLMLMIMODPCRS} illustrates another DPCRS scheme, namely, multi-layer MIMO DPCRS, which marries DPC and GRS. In contrast to GRS in Fig. \ref{Fig_DLMIMOGRS} which relies on linear precoding to precode all streams,  multi-layer MIMO DPCRS considers dirty paper coded private streams.

\subsection{Uplink}\label{uplink_k_user}
Uplink RSMA is first introduced  in \cite{Rimo1996} for SISO MAC and further extended to SIMO MAC in \cite{Maosurvey}. 
The main benefit of uplink RSMA discovered in existing works is that the capacity region of the Gaussian MAC can be achieved by uplink RSMA without time sharing among users. 
In the following, we introduce a generic uplink RSMA transmission framework for MIMO MAC, which is simply denoted by uplink MIMO RSMA.
\par
\begin{figure}
	\centering
\includegraphics[width=0.5\textwidth]{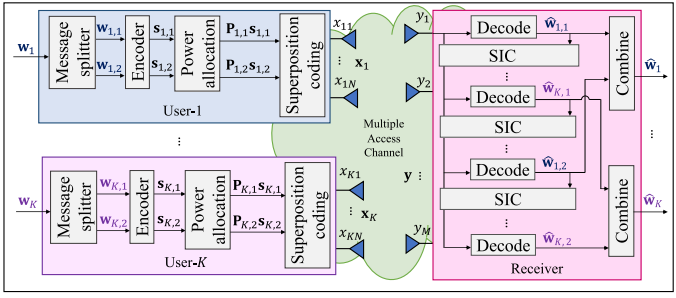}%
	\caption{$K$-user uplink MIMO RSMA.}
	\label{Fig_ULKMIMORSMA}
\end{figure}

Fig. \ref{Fig_ULKMIMORSMA}  illustrates the proposed $K$-user uplink MIMO RSMA based on SC at the transmitters  and SIC at the receiver. 
A $Q_k$-dimensional message vector $\mathbf{w}_k=[W_k^{(1)},\ldots,W_k^{(Q_k)}]^T$ is transmitted from user-$k$ to the receiver, where $Q_k\leq \min(M,N)$. 
At user-$k$, each user message $W_k^{(i)}$ is split into two sub-messages $W_{k,1}^{(i)}$ and $W_{k,2}^{(i)}$, resulting in two sub-message vectors $\mathbf{w}_{k,1}$ and $\mathbf{w}_{k,2}$. 
Note that splitting  $\sum_{k=1}^{K}Q_k-1$ messages is sufficient to achieve the capacity region without time sharing in this setting. For illustration simplicity, we here consider a more general model in which the user messages of all users are split.
At user-$k$, the sub-message vectors $\mathbf{w}_{k,1}$ and $\mathbf{w}_{k,2}$  are independently encoded into $\mathbf{s}_{k,1}$ and $\mathbf{s}_{k,2}$,  precoded by $\mathbf{P}_{k,1}$ and $\mathbf{P}_{k,2}$, and superposed at the transmitter.
The resulting transmit signal is
\begin{equation}
	\mathbf{x}_k=\mathbf{P}_{k,1}\mathbf{s}_{k,1}+\mathbf{P}_{k,2}\mathbf{s}_{k,2}, \forall k\in\mathcal{K}.
\end{equation}

\par The signals of all users are transmitted via MIMO MAC, the receive signal is
\begin{equation}
	\mathbf{y}=\sum_{k\in\mathcal{K}}\mathbf{H}_k\mathbf{x}_k+\mathbf{n},
\end{equation}
where 
$\mathbf{H}_k\in \mathbb{C}^{M\times N}$ is the channel vector between  user-$k$  and the receiver, and $\mathbf{n}\sim \mathcal{CN}(\mathbf{0},\mathbf{I}_M)$ is the AWGN vector.
The receiver then employs  the receive filters $\mathbf{W}_{k,1}, \mathbf{W}_{k,2}, k\in\mathcal{K}$ to detect the stream vectors for all users.
Denote the decoding order of the $2K$ received stream vectors $\{\mathbf{s}_{k,i} \mid k\in\{1,\ldots,K\}, i\in\{1,2\}\}$ by $\pi$ such that $\pi_{k,i}<\pi_{k',i'}$ if  $\mathbf{s}_{k,i}$ is decoded before $\mathbf{s}_{k',i'}$. 
 $2K-1$ layers of SIC is needed to decode all stream vectors.
Assuming Gaussian signaling and infinite blocklength, the rate of decoding $s_{ki}$ at the receiver is given as follows
\begin{equation}
	\begin{aligned}
		&R_{k,i}=\log_2\det\left(\mathbf{I}+\mathbf{P}_{k,i}^H\mathbf{H}_k^H\mathbf{W}_{k,i}^H\left(\mathbf{R}_{k,i}\right)^{-1}\mathbf{W}_{k,i}\mathbf{H}_k\mathbf{P}_{k,i}\right),
	\end{aligned}
\end{equation}
where $\mathbf{R}_{k,i}$ is defined as
\begin{equation}
	\begin{aligned}
		&\mathbf{R}_{k,i}=\mathbf{I}+\sum\limits_{\pi_{k',i'}>\pi_{k,i}}\mathbf{W}_{k,i}\mathbf{H}_{k'}\mathbf{P}_{k',i'}\mathbf{P}_{k',i'}^H\mathbf{H}_{k'}^H\mathbf{W}_{k,i}^H.
	\end{aligned}
\end{equation}
The proposed uplink MIMO RSMA model is generic. It embraces uplink SISO RSMA in \cite{Rimo1996} and uplink SIMO RSMA in \cite{Maosurvey} as two sub-schemes. 
The study of uplink MIMO RSMA is still in its infancy. Its SE/EE performance and applications in different services such as eMBB, URLLC, and mMTC or their hybrid services are worth more investigations.

\subsection{Multi-cell}
In multi-cell networks, the transmission from multiple transmitters to multiple receivers can be categorized into ``coordinated transmission" and ``cooperative transmission" depending on whether the data is shared among the transmitters \cite{clerckx2013mimo}. 
RSMA has been investigated in both transmissions \cite{chenxi2017brunotopology,wonjae2020multicell, mao2018networkmimo, alaa2020cranimperfectCSIT}. It is shown to be a promising strategy to enhance SE by providing a powerful inter-cell and intra-cell interference management capability.
In the following subsections, the transmission models of MIMO RSMA for ``coordinated transmission" and ``cooperative transmission"  are respectively delineated.
\subsubsection{Coordinated transmission} 
\begin{figure}[t!]
	\centering
\subfigure[Coordinated transmission]{
		\centering
		\includegraphics[width=0.22\textwidth]{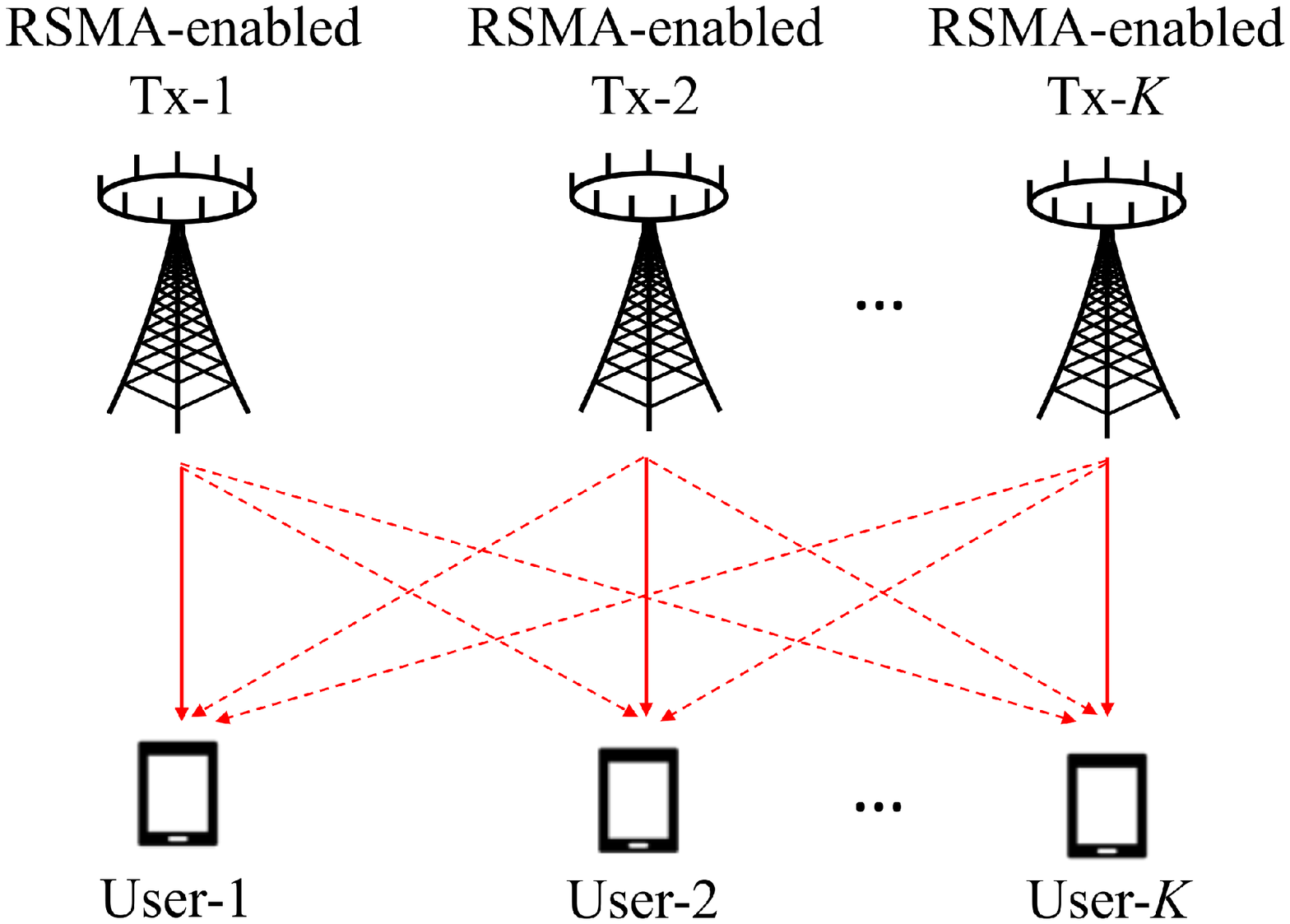}%
		\label{fig:multiCellcoordination}
}
	~
\subfigure[Cooperative transmission]{
		\centering
		\includegraphics[width=0.22\textwidth]{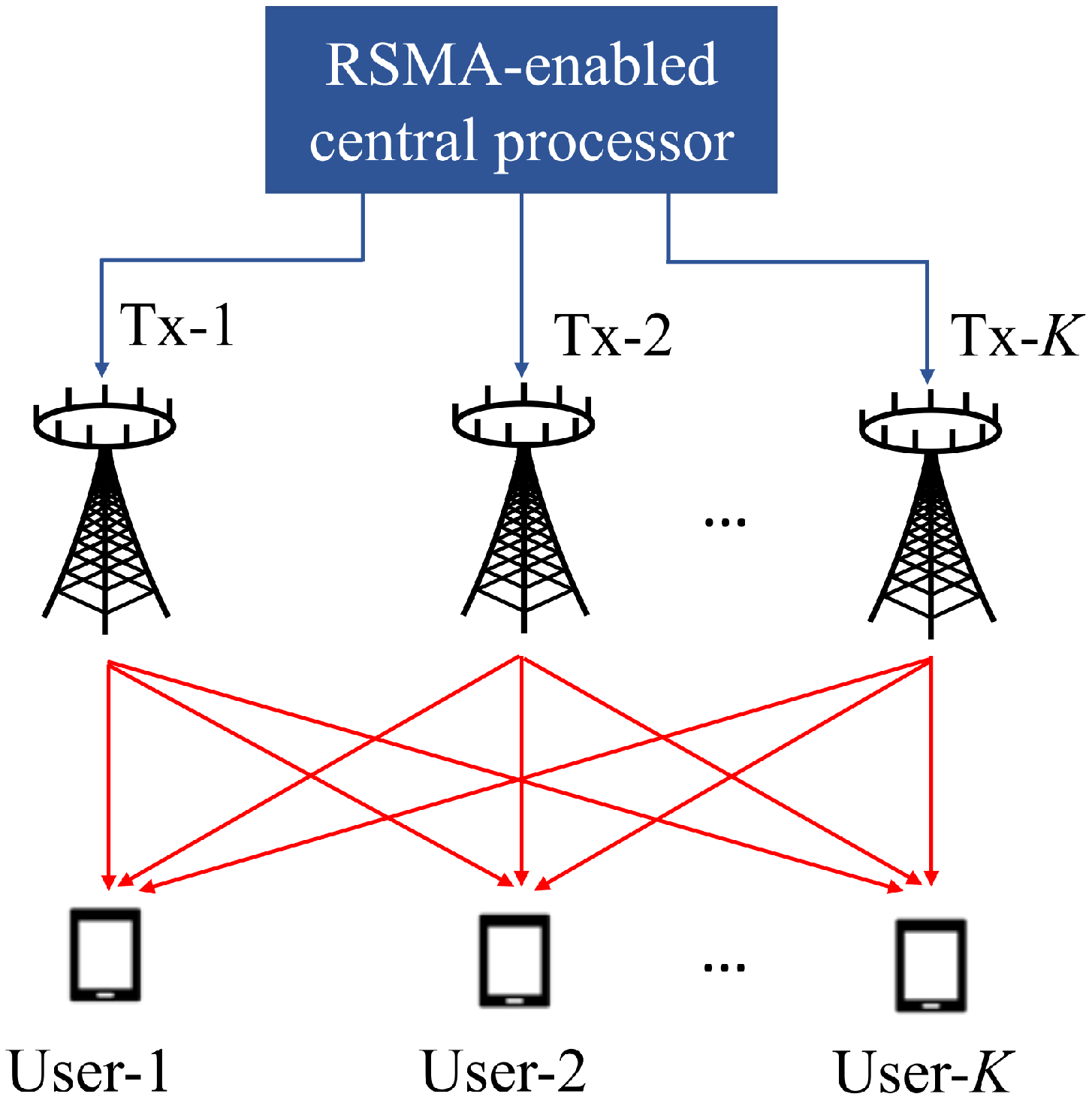}%
			\label{fig:multiCellcooperation}
}
	\caption{Multi-cell RSMA-enabled transmission \cite{Maosurvey}.}
	\label{fig:multiCell}
\end{figure}

A $K$-cell coordinated MIMO is illustrated in Fig.\ref{fig:multiCellcoordination}. 
The message of each user is sent from the serving cell (with one RSMA-enabled transmitter) while the resource allocation and user scheduling are coordinated among cells. 
At each transmitter-$k$, the message vector $\mathbf{w}_k$ for user-$k$ is split into two sub-message vectors $\mathbf{w}_{\mathrm{c},k}$ and $\mathbf{w}_{\mathrm{p},k}$, independently encoded into two stream vectors $\mathbf{s}_{\mathrm{c},k}$ and $\mathbf{s}_{\mathrm{p},k}$, and linearly precoded by the precoding matrices $\mathbf{P}_{\mathrm{c},k}$ and $\mathbf{P}_{\mathrm{p},k}$. The  transmit signal at transmitter-$k$ is given as
\begin{equation}
	\mathbf{{x}}_k=\mathbf{{P}}_{\mathrm{c},k}\mathbf{s}_{\mathrm{c},k}+\mathbf{{P}}_{\mathrm{p},k}\mathbf{s}_{\mathrm{p},k}.
\end{equation}
\par The signal of all transmitters are transmitted to all users. The signal received at user-$k$ is  
\begin{equation}
	\mathbf{y}_k=\sum_{j=1}^K\mathbf{H}_{kj}^H(\mathbf{{P}}_{\mathrm{c},j}\mathbf{s}_{\mathrm{c},j}+\mathbf{P}_{\mathrm{p},j}\mathbf{s}_{\mathrm{p},j})+\mathbf{n}_k,
\end{equation}
where $\mathbf{H}_{kj}^H\in \mathbb{C}^{N\times M}$ is the channel between transmitter-$j$ and user-$k$ and $\mathbf{n}_k\sim \mathcal{CN}(\mathbf{0},\mathbf{I}_N)$ is the AWGN vector at user-$k$. 
Each user-$k$ is required to decode all common stream vectors $\mathbf{s}_{\mathrm{c},1}, \ldots, \mathbf{s}_{\mathrm{c},K}$ and the intended private stream $\mathbf{s}_{\mathrm{p},k}$ based on SIC or joint decoding.
For a certain decoding order $\pi_k$ such that $\mathbf{s}_{\mathrm{c},\pi_k(i)}$ is decoded before $\mathbf{s}_{\mathrm{c},\pi_k(j)}$ at user-$k$ if $i<j$.
Assuming Gaussian signaling and infinite blocklength, the rate of decoding $\mathbf{s}_{\mathrm{c},\pi_k(i)}$ and $\mathbf{s}_{\mathrm{p},k}$ at user-$k$ are obtained as follows
\begin{equation}
	\begin{aligned}
		&R_{k,\pi_k(i)}^{\mathrm{c}}=\\
		&\log_2\det\left(\mathbf{I}+\mathbf{P}_{\mathrm{c},\pi_k(i)}^H\mathbf{H}_{k\pi_k(i)}\left(\mathbf{R}_{\mathrm{c},{\pi_k(i)}}\right)^{-1}\mathbf{H}_{{k\pi_k(i)}}^H\mathbf{P}_{\mathrm{c},\pi_k(i)}\right),\\
		&R_{k}^{\mathrm{p}}=\log_2\det\left(\mathbf{I}+\mathbf{P}_{\mathrm{p},k}^H\mathbf{H}_{kk}\left(\mathbf{R}_{k}\right)^{-1}\mathbf{H}_{kk}^H\mathbf{P}_{\mathrm{p},k}\right).
	\end{aligned}
\end{equation}
where $\mathbf{R}_{\mathrm{c},{\pi_k(i)}}$ and $\mathbf{R}_{k}$ are defined as
\begin{equation}
	\begin{aligned}
		&\mathbf{R}_{\mathrm{c},{\pi_k(i)}}=\mathbf{I}+\sum_{j>i}\mathbf{H}_{k\pi_k(j)}^H\mathbf{P}_{\mathrm{c},\pi_k(j)}\mathbf{P}_{\mathrm{c},\pi_k(j)}^H\mathbf{H}_{k\pi_k(j)}\\
		&\quad\quad\quad\,\,+\sum_{j=1}^K\mathbf{H}_{kj}^H\mathbf{P}_{\mathrm{p},j}\mathbf{P}_{\mathrm{p},j}^H\mathbf{H}_{kj},\\
		&\mathbf{R}_{k}=\mathbf{I}+\sum_{j\in\mathcal{K},j\neq k}\mathbf{H}_{kj}^H\mathbf{P}_{\mathrm{p},j}\mathbf{P}_{\mathrm{p},j}^H\mathbf{H}_{kj}.
	\end{aligned}
\end{equation}
The achievable rate of user-$k$ is
\begin{equation}
	R_{k,\mathrm{tot}}=\min\left\{R_{1,k}^{\mathrm{c}},\ldots,R_{K,k}^{\mathrm{c}}\right\}+R_{k}^{\mathrm{p}}.
\end{equation}

\par 
As discussed in \cite{chenxi2017brunotopology}, 
The performance of coordinated MIMO RSMA could be further boosted by splitting the message of each user into $N>2$ parts.  Each part is decoded by a different group of users with the same instantaneous CSIT quality.
\subsubsection{Cooperative transmission} 
As discussed in \cite{Maosurvey}, cooperative MIMO requires all transmitters to be connected as a virtual giant transmitter to serve all users in the downlink. All downlink transmission frameworks of RSMA discussed in \ref{sec:DL} therefore can be applied for cooperative MIMO.
Compared with single-cell MIMO BC which is subject to a sum-power constraint, multi-cell cooperative MIMO BC requires per-cell transmit power constraint and the fronthaul capacity constraints \cite{alaa2020cranimperfectCSIT}.

\subsection{Numerical Results}
Next, we show numerically the benefits of RSMA in terms of the significant wireless communication performance metrics, namely,  spectral efficiency, energy efficiency, generality, flexibility, robustness, reliability and latency.
\par
Unless stated otherwise, the elements of the channel matrix $\mathbf{H}_k$ are generated as i.i.d. complex Gaussian random variables $\mathcal{CN}(0,\sigma_k^2)$ and the noise variances are normalized. The imperfect CSIT model is  $\mathbf{H}_k=\widehat{\mathbf{H}}_k+\widetilde{\mathbf{H}}_k$, where both the channel estimate $\widehat{\mathbf{H}}_k$ and the channel error $\widetilde{\mathbf{H}}_k$ have i.i.d. complex Gaussian random entries drawn from  $\mathcal{CN}(0,\sigma_k^2-\sigma_{e,k}^2)$ and $\mathcal{CN}(0,\sigma_{e,k}^2)$, respectively.
All results are averaged over 100  channel instances.
\subsubsection{Spectrally and Energy Efficient}
\begin{figure}[t!]
	\centering
\subfigure[$\sigma_1^2=1, \sigma_2^2=1$]{
		\centering
		\includegraphics[width=0.22\textwidth]{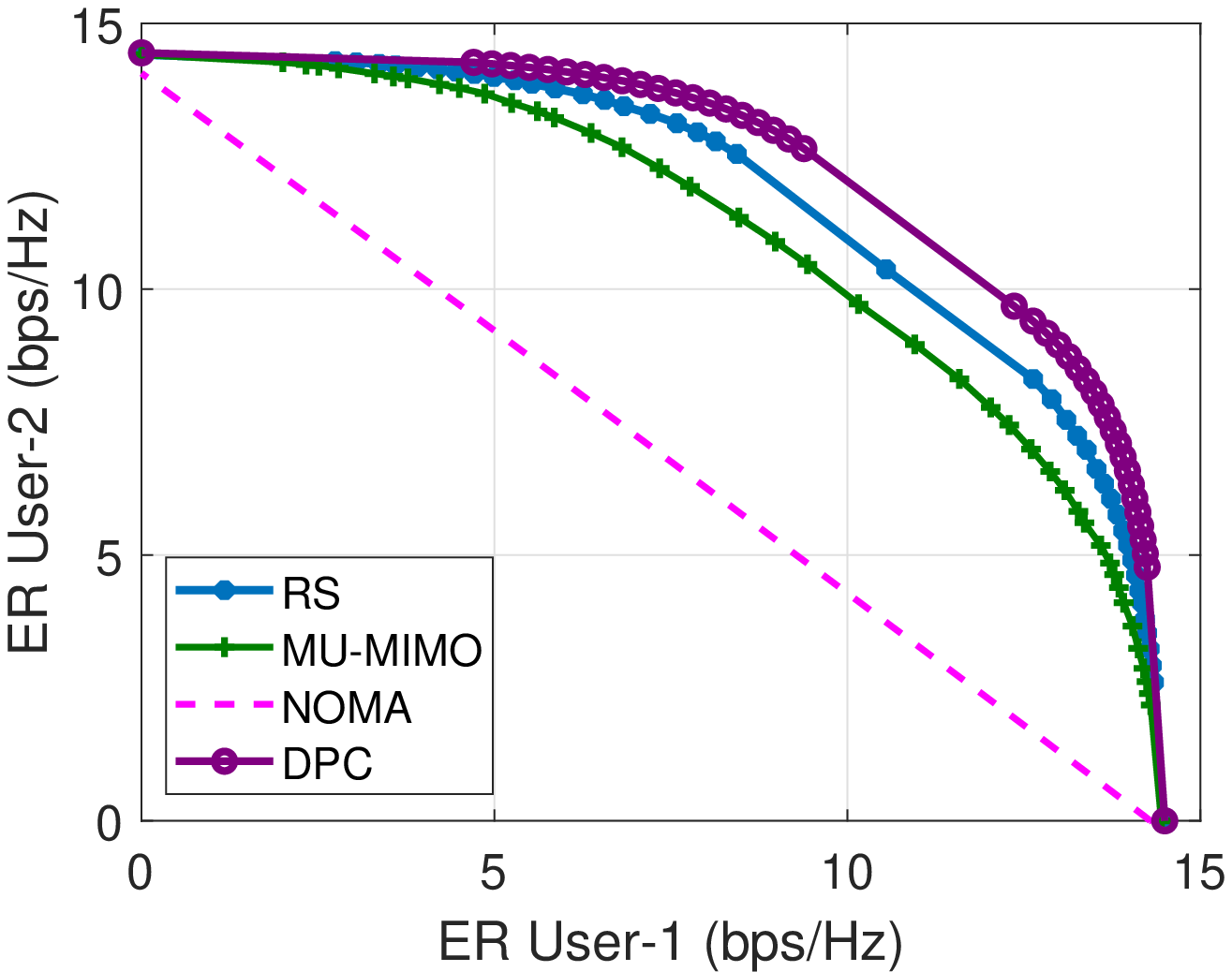}%
		\label{fig:SE1a}
}
~
\subfigure[$\sigma_1^2=1, \sigma_2^2=0.09$]{
		\centering
		\includegraphics[width=0.22\textwidth]{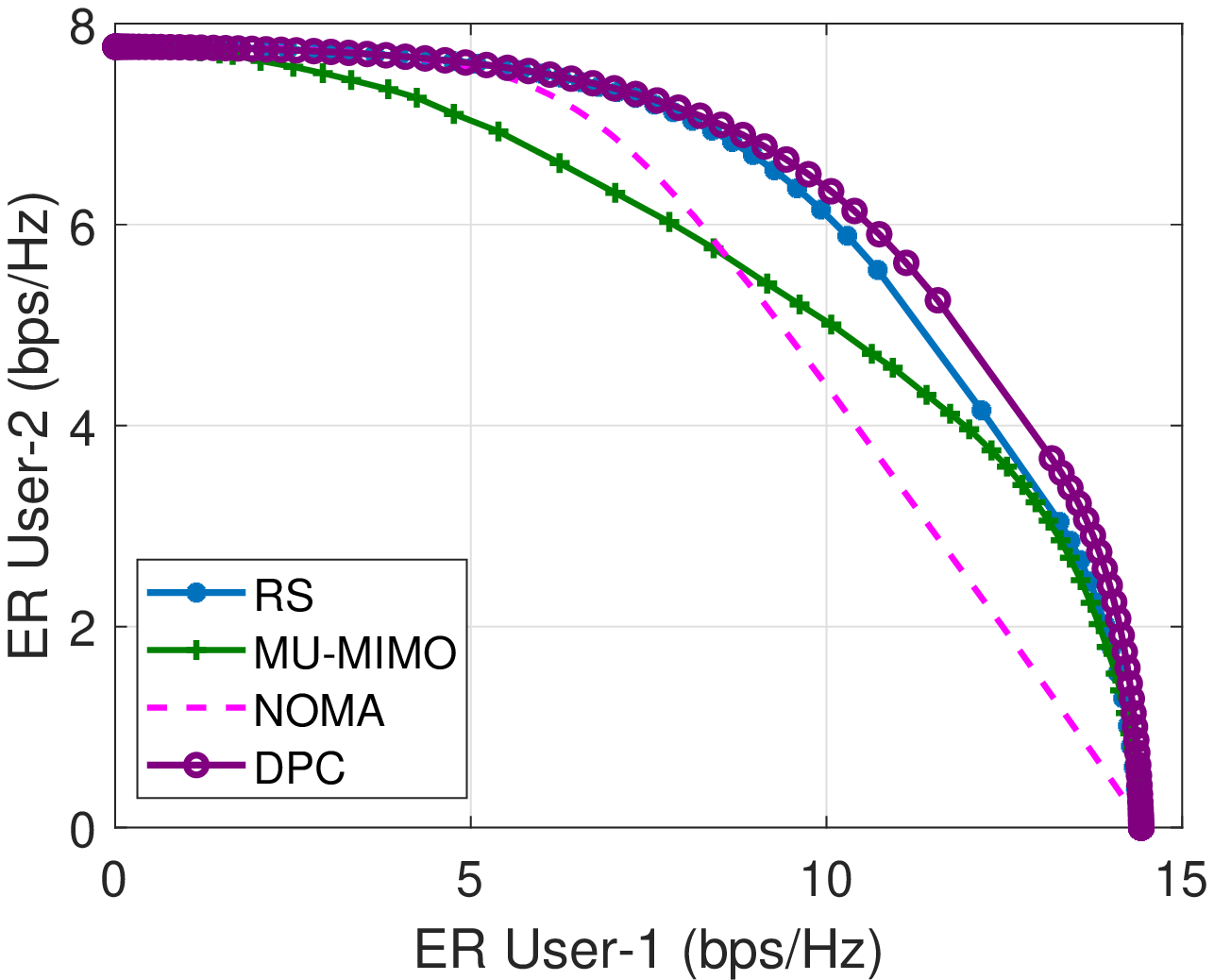}%
			\label{fig:SE1b}
}
	\caption{Ergodic rate region comparison  in MIMO BC with perfect CSIT, SNR = 20 dB, $M=4$, $K=2, N=2$, and $Q_{\mathrm{c}}=Q_1=Q_2=2$ \cite{anup2021MIMO}.}
	\label{fig:SE1}
\end{figure}

Consider a downlink MU-MIMO with one transmitter equipped with $M=4$ antennas  and $K=2$ users, each is equipped with $N=2$ antennas. Each user requires $Q_1=Q_2=2$ streams, which are split, combined, and encoded into a common stream vector of length $Q_{\mathrm{c}}=2$  and two private stream vectors of length 2. 
Readers are referred to \cite{anup2021MIMO} for more details on the simulation setting and the rate region optimization algorithm.
Fig. \ref{fig:SE1} illustrates the ergodic rate region of RS, MU-MIMO, NOMA, and DPC in perfect CSIT with equal channel variances or 10 dB channel variance disparity between the  users. 
Note that in the two-user case, HRS and GRS are equivalent to 1-layer RS, which are therefore simply denoted by ``RS". 
We observe that in both subfigures, the rate region of RS is larger than that of MU-MIMO and NOMA, and is much closer to the capacity region of MIMO BC achieved by DPC.
When the users have equal channel variances as in Fig. \ref{fig:SE1a}, NOMA achieves the worst performance as no channel strength disparity can be leveraged to manage interference. 
Although MU-MIMO is capable of exploiting the full DoF in this setting (i.e., underloaded and perfect CSIT), there is still a rate region gap between RS and MU-MIMO. 
As MU-MIMO always treats interference as noise, it only works well when user channels are (semi-)orthogonal \cite{mao2017rate}. 
When the users have a 10 dB channel variance disparity, the rate region gap between NOMA and RS reduces. Interestingly, the rate region of NOMA and MU-MIMO outperform each other in part while RS always outperforms both. 
Although NOMA can utilize the channel variance disparity to improve its rate region, it incurs a theoretical DoF loss and therefore rate loss due to its interference management principle of forcing one user to fully decode the streams of all users \cite{bruno2021MISONOMA}.
Owing to its powerful interference management capability of partially decoding the interference and partially treating the interference as noise, RS enhances the spectral efficiency in various user deployments. 
\begin{figure}[t!]
	\centering
\subfigure[$\sigma_1^2=1, \sigma_2^2=1$]{
		\centering
		\includegraphics[width=0.22\textwidth]{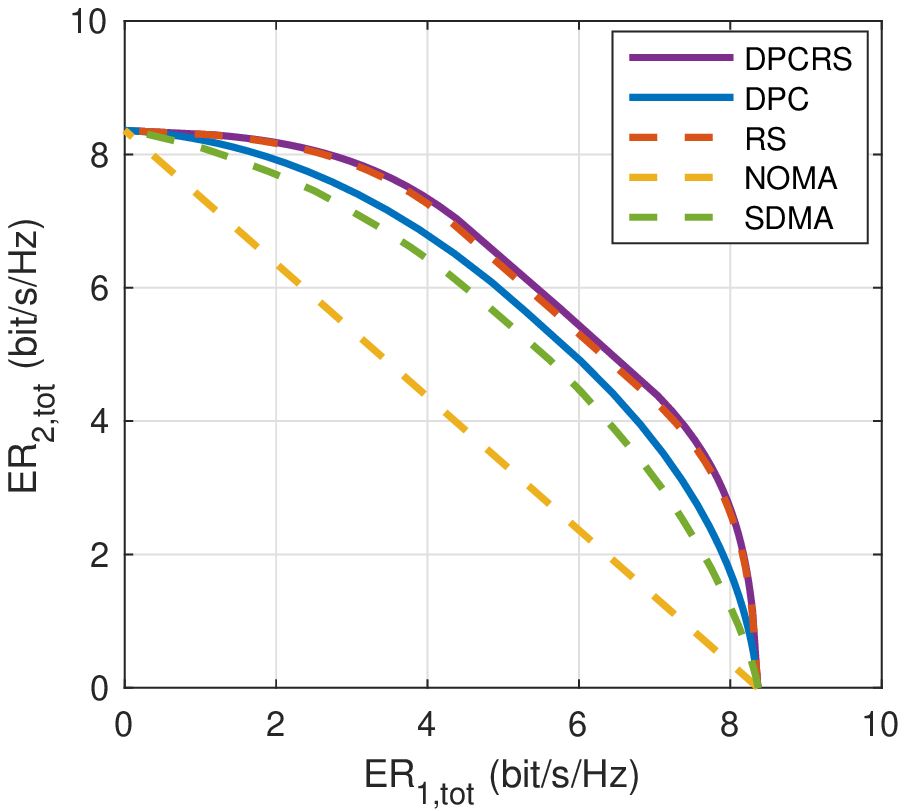}%
		\label{fig:SE2a}
}
	~
\subfigure[$\sigma_1^2=1, \sigma_2^2=0.09$]{
		\centering
		\includegraphics[width=0.22\textwidth]{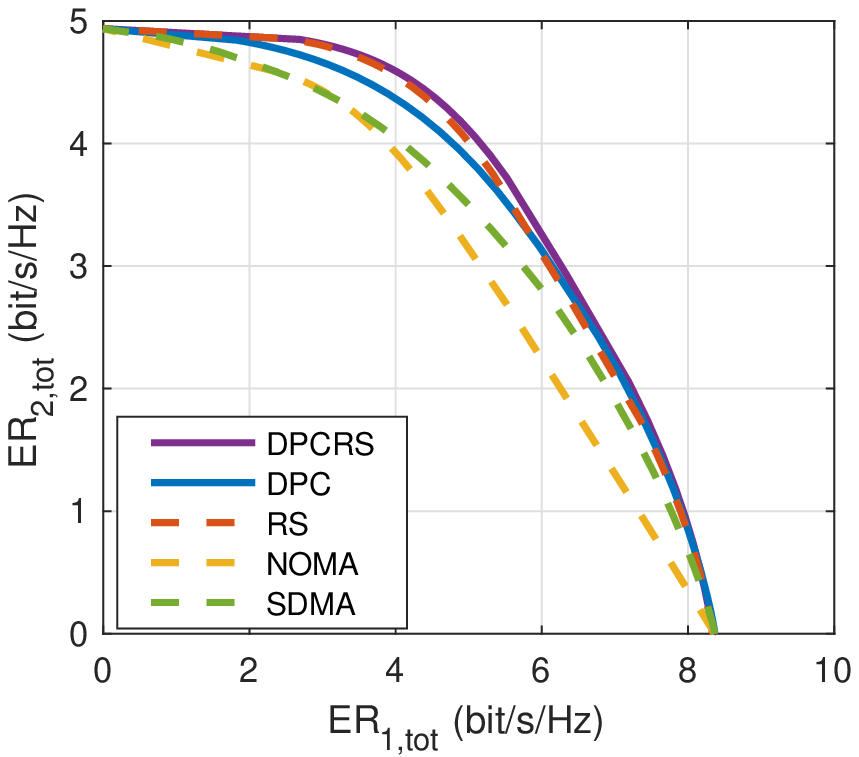}%
			\label{fig:SE2b}
}
	\caption{Ergodic rate region comparison in MISO BC with imperfect CSIT, SNR = 20 dB, $\alpha=0.6$, $M=4, K=2$ \cite{mao2019beyondDPC}.}
	\label{fig:SE2}
\end{figure}
\par 
In Fig. \ref{fig:SE2}, we further compare SE of different strategies in imperfect CSIT. Consider a downlink MU-MISO with $M=4$ and $K=2$. The power of the channel error is defined as $\sigma_{e,k}^2=\sigma_k^2P^{\alpha}$, where $P=\mathrm{tr}(\mathbf{P}\mathbf{P}^H)$ and $\alpha$ is fixed to 0.6.
More details of the simulation setting and the rate region optimization algorithm can be found in \cite{mao2019beyondDPC}.
Besides linearly precoded RS, the ergodic rate region of 1-layer DPCRS is also illustrated. 
In both subfigures, the ergodic rate region of DPC drops significantly as it is sensitive to CSIT uncertainty.
Surprisingly, linearly precoded  RS achieves a larger rate region than DPC while maintaining a much lower transceiver complexity. 
As RS has been proved to achieve the optimal multiplexting gain/DoF of MISO BC with imperfect CSIT \cite{enrico2017bruno}, such gain is  reflected in its spectral efficiency gain over MU-MIMO and NOMA.
DPCRS, which marries the advantages of DPC and RS, achieves the largest achievable rate region in both subfigures at the sacrifice of a higher transceiver complexity than DPC and RS.  
\par
\begin{figure}[t!]
	\centering
\subfigure[$\sigma_1^2=1, \sigma_2^2=1$]{
		\centering
		\includegraphics[width=0.22\textwidth]{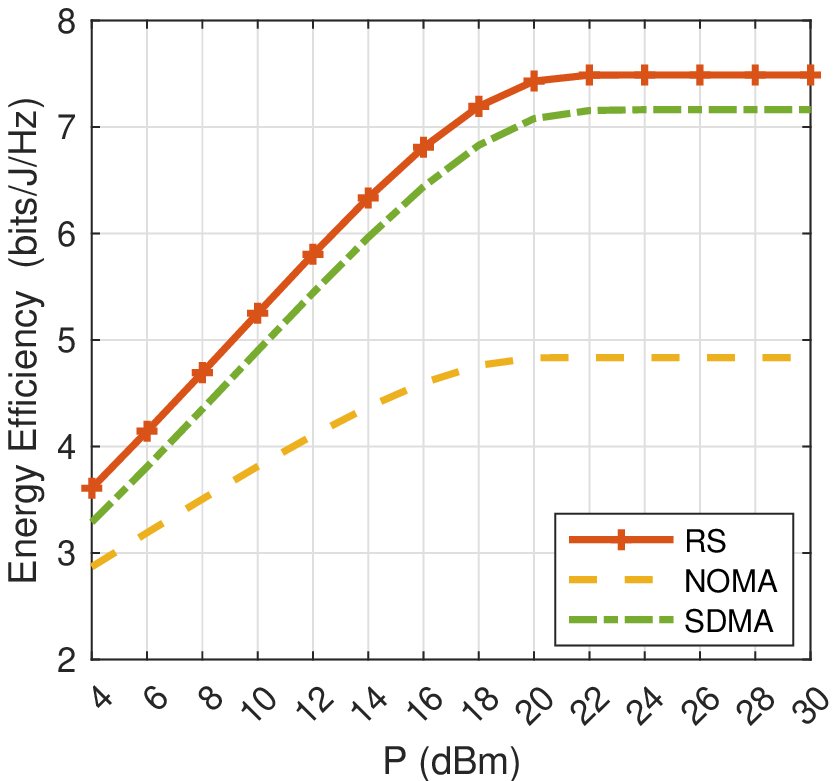}%
		\label{fig:EEa}
}
	~
\subfigure[$\sigma_1^2=1, \sigma_2^2=0.09$]{
		\centering
		\includegraphics[width=0.22\textwidth]{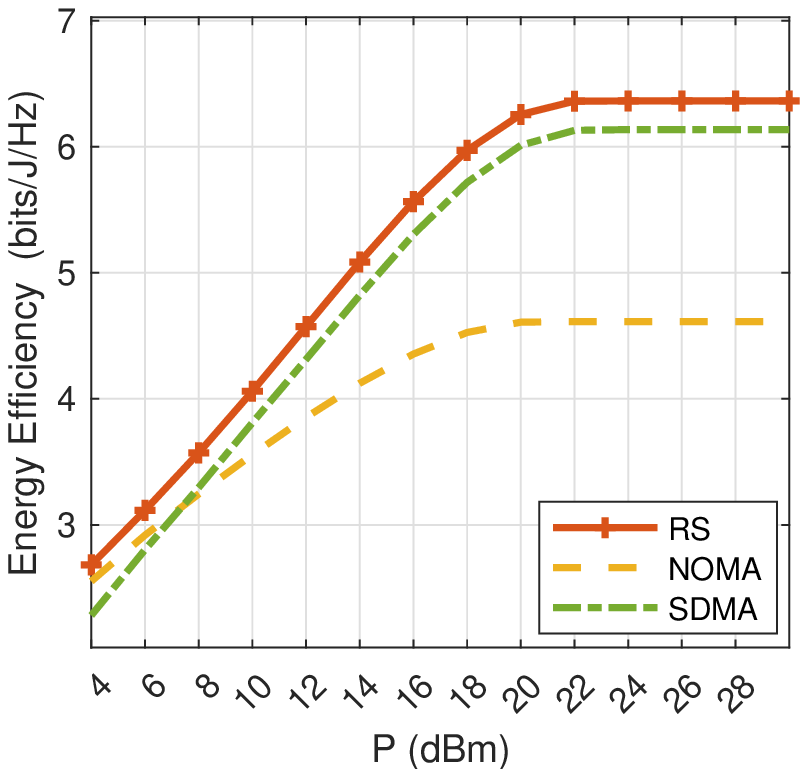}%
			\label{fig:EEb}
}
	\caption{Energy efficiency versus transmit power budget comparison in MISO BC with perfect CSIT.}
	\label{fig:EE}
\end{figure}

 Fig. \ref{fig:EE} compares EE of different strategies for two-user MU-MISO with perfect CSIT, $M=2, N=2$. 
EE is defined as $\frac{\sum_{k=1}^KR_{k,\mathrm{tot}}}{\frac{1}{\eta}\mathrm{tr}(\mathbf{P}\mathbf{P}^H)+P_{\textrm{cir}}}$ \cite{mao2018EE, bho2021globalEE}, where $\eta=0.35$ is the power amplifier efficiency, $P_{\textrm{cir}}=MP_{\textrm{dyn}}+P_{\textrm{sta}}$ is the static circit power consumption with $P_{\textrm{dyn}}=27$ dBm and  $P_{\textrm{sta}}=1$ mW. There is a 1 bit/s/Hz minimum rate threshold for each user.
RS achieves a larger EE than NOMA and SDMA in both subfigures.
Again, NOMA achieves the EE worst performance when there is no channel strength disparity while it slightly outperforms SDMA when there is a 10 dB channel strength disparity and low transmit power constraint.
\par
\begin{figure}[t!]
	\centering
		\includegraphics[width=0.4\textwidth]{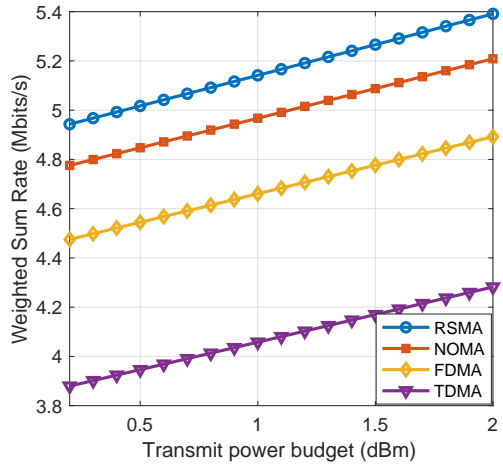}%
		\caption{Sum rate versus the  transmit power budget of each user in SISO MAC with perfect CSIT \cite{zhaohui2020ULRS}.}
	\label{fig:uplinkSE}
\end{figure}

 
Fig. \ref{fig:SE1}--\ref{fig:EE} show the SE and EE gains of RSMA in the downlink for different user deployments and CSIT conditions. In the uplink, the authors in \cite{zhaohui2020ULRS} have also shown that RSMA achieves a higher spectral efficiency than NOMA and other OMA strategies, as illustrated in Fig. \ref{fig:uplinkSE} thanks to its capability of achieving each point at the boundary of the capacity region. 
In contrast,  NOMA can only reach the corner points at the boundary of the capacity region and time sharing is needed to achieve the points along the line segment between the two corner points.

\subsubsection{General and Unified}
\begin{figure}[t!]
	\centering
\subfigure[SNR$=10$ dB]{
		\centering
		\includegraphics[width=0.22\textwidth]{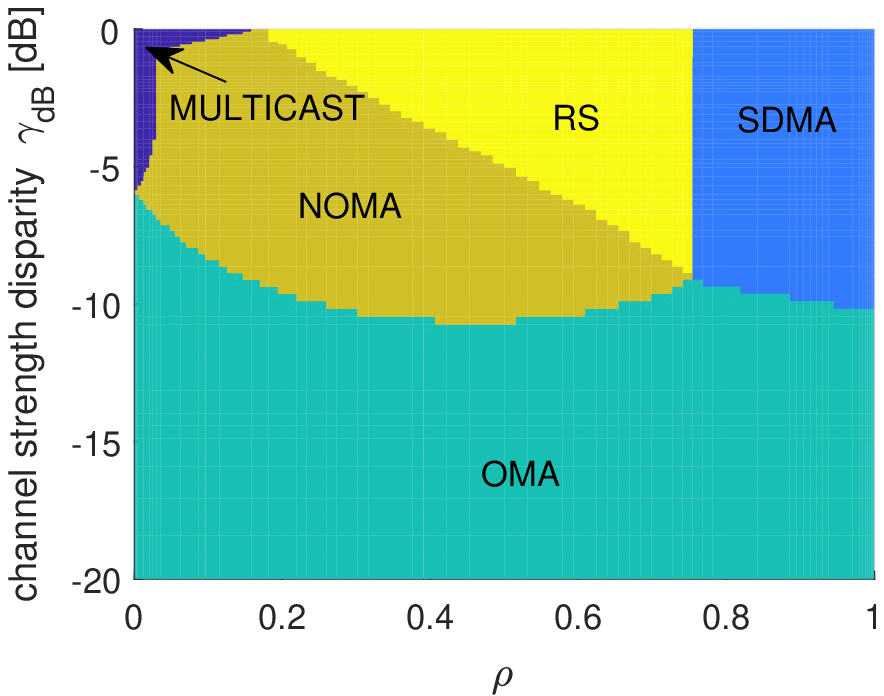}%
		\label{fig:universala}
}
	~
\subfigure[SNR$=30$ dB]{
		\centering
		\includegraphics[width=0.22\textwidth]{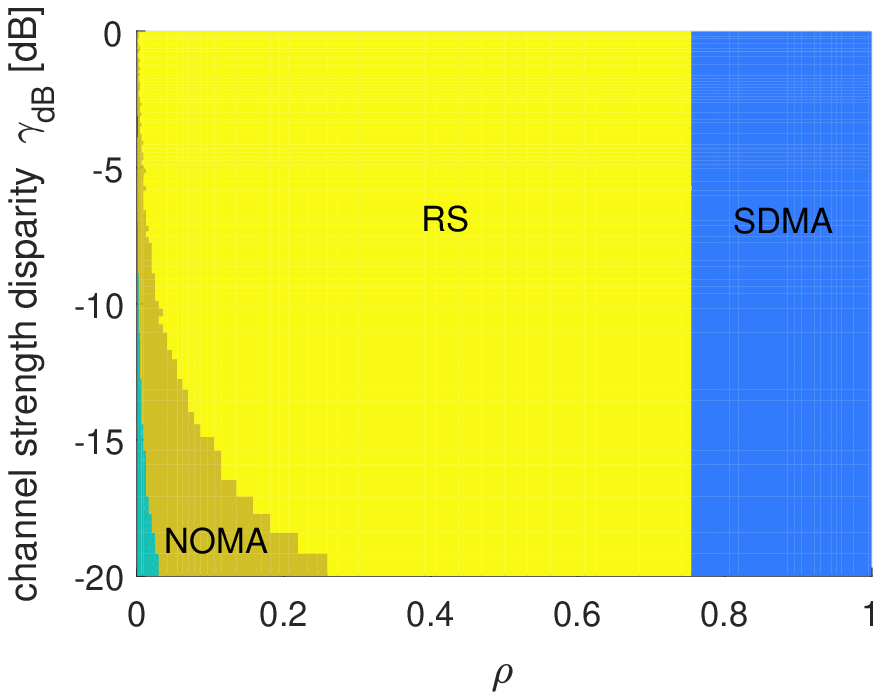}%
			\label{fig:universalb}
}  

	\caption{Operation regions for RS, SDMA, NOMA, OMA and Multicast with perfect CSIT \cite{bruno2019wcl}.}
	\label{fig:universal}
\end{figure}

As widely acknowledged and discussed in the literature of RSMA \cite{mao2017rate, bruno2019wcl,mao2019beyondDPC,onur2021sixG,Maosurvey}, RSMA is a general and universal MA scheme, interference management strategy, and non-orthogonal transmission framework that unifies SDMA, NOMA, OMA, physical-layer multicasting and treats them as sub-schemes or special cases. 
Such universality is reflected 
in the SE and EE gain over existing MA schemes as per Fig. \ref{fig:SE1}--\ref{fig:uplinkSE}, as well as in its operational region illustrated in Fig. \ref{fig:universal}.
In Fig. \ref{fig:universal}, a specific channel realization $\mathbf{h}_1=\frac{1}{\sqrt{2}}[1,1]^H$ and $\mathbf{h}_2=\frac{\gamma}{\sqrt{2}}[1,e^{j\theta}]^H$ is considered. 
The $x$-axis is $\rho=1-\frac{|\mathbf{h}_1^H\mathbf{h}_2|^2}{\|\mathbf{h}_1\|^2\|\mathbf{h}_2\|^2}$, which indicates the channel angle between the two users ($\rho=0$ and $\rho=1$ corresponding to aligned and orthogonal channel directions, respectively). The $y$-axis is the channel disparity between the two users in dB, which is given as $\gamma_{\mathrm{dB}}=20\log_{10}(\gamma)$.
The parameters and the precoder design are detailed in \cite{bruno2019wcl}.
RS automatically reduces to the existing MA schemes in some regimes while it outperforms all existing schemes in the remaining regimes especially when the SNR is high. 
Therefore, instead of using OMA, NOMA, SDMA and optimize them for each propagation deployment, one can use a single and universal RS scheme in wireless communication networks. 

\subsubsection{Flexible}
\begin{figure}
	\centering
	\subfigure[{Overloaded $M=3$}]{
		\centering
		\includegraphics[width=0.22\textwidth]{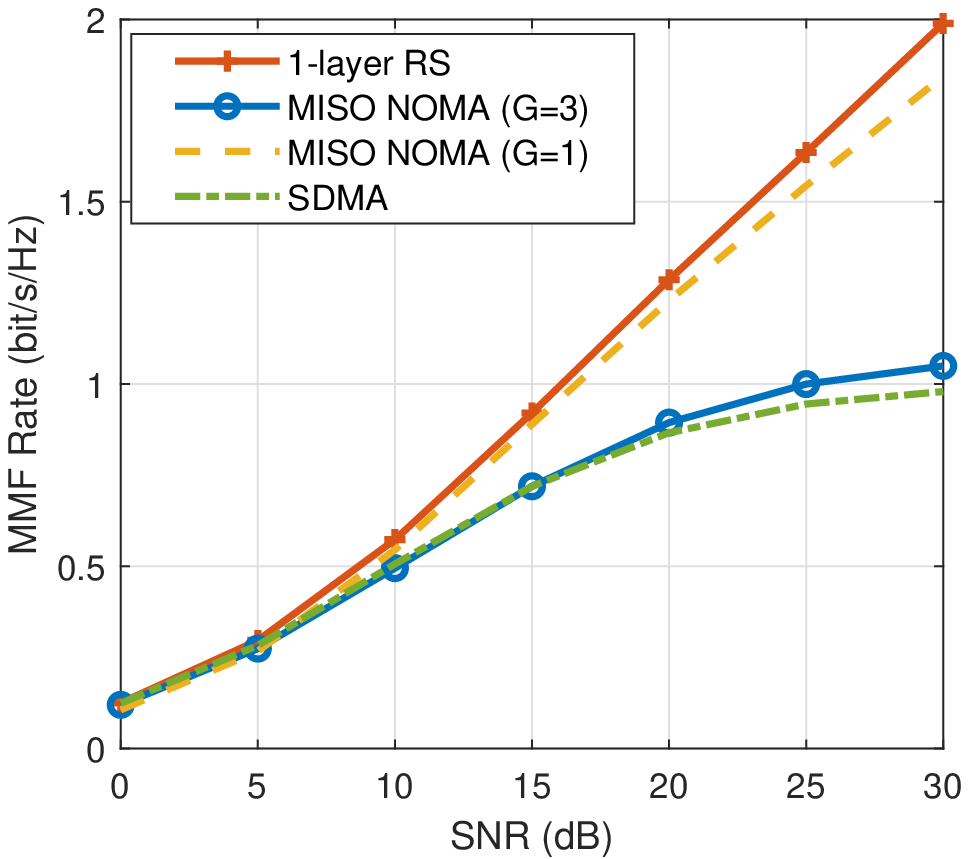}%
	}
	~
    \subfigure[{Overloaded $M=4$}]{
		\centering
		\includegraphics[width=0.22\textwidth]{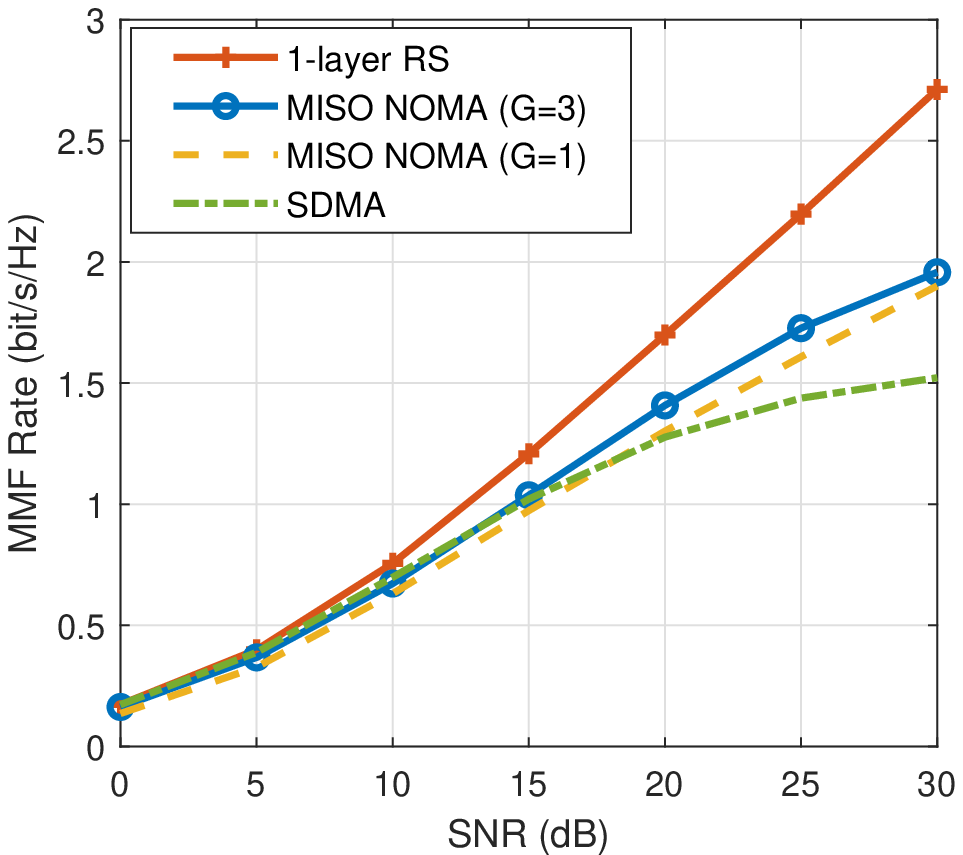}%
		}
		~\\
		\subfigure[{Overloaded $M=5$}]{
		\centering
		\includegraphics[width=0.22\textwidth]{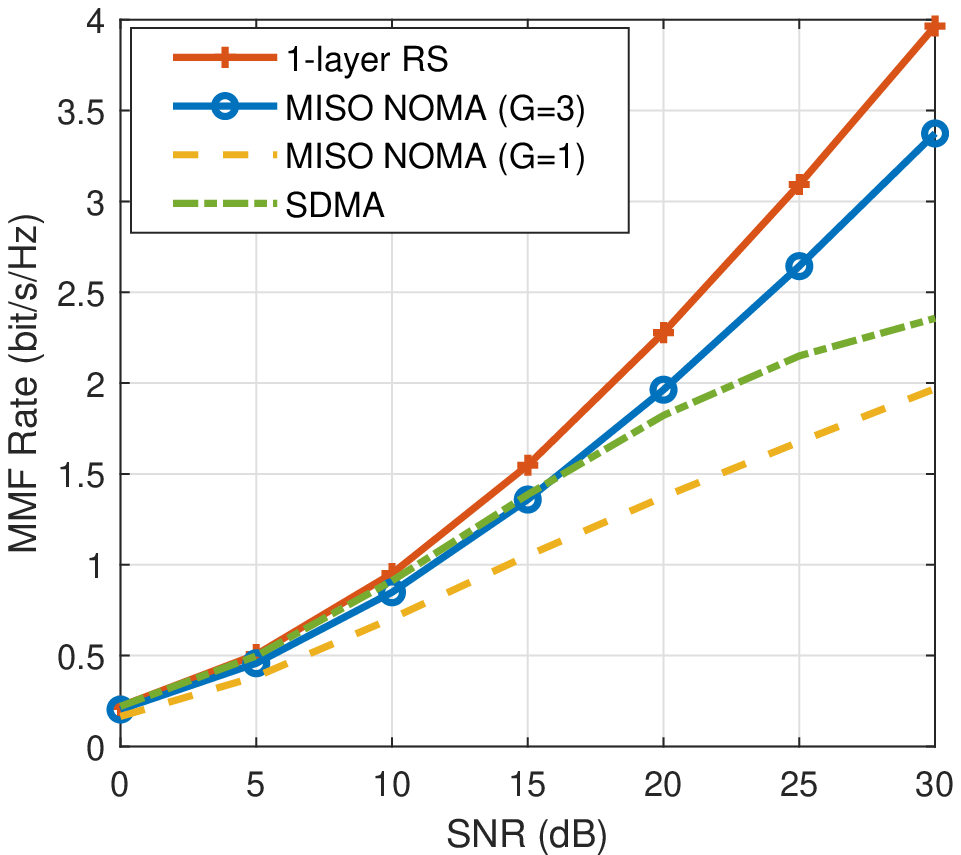}%
	}
	~
    \subfigure[{Underloaded $M=6$}]{
		\centering
		\includegraphics[width=0.22\textwidth]{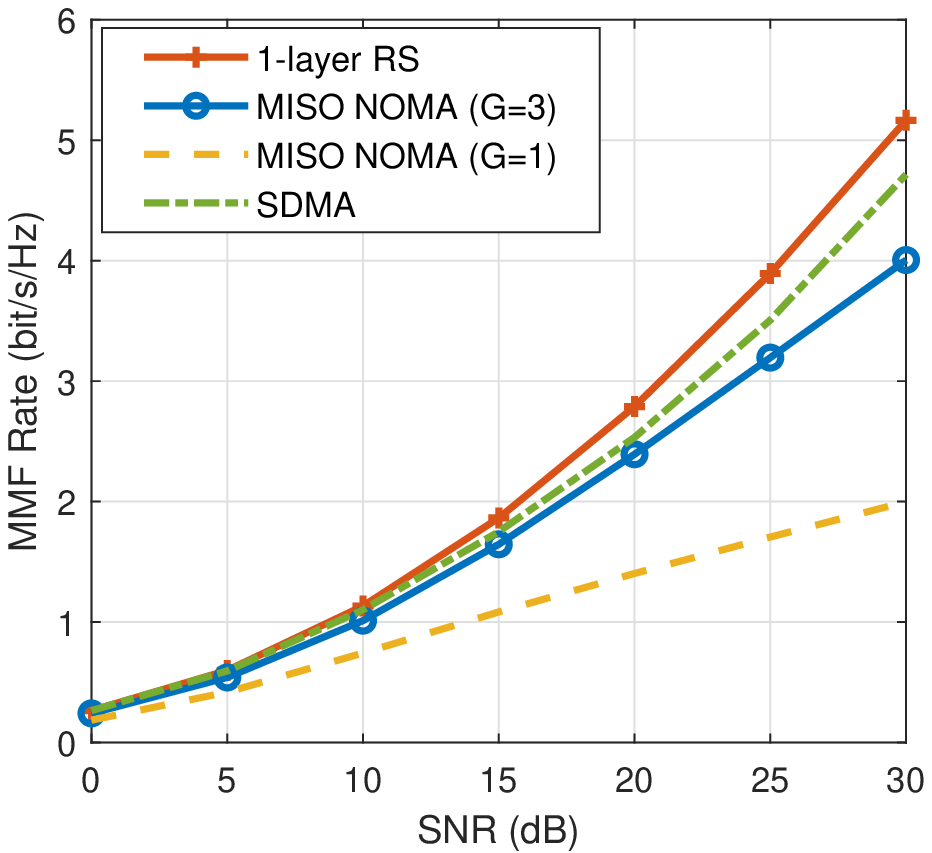}%
		}
	\caption{Max-min rate vs. SNR comparison in MISO BC with imperfect CSIT, $\alpha=0.5$, $K=6$, $\sigma_k^2\in [0.1,1]$ \cite{bruno2021MISONOMA}.}
	\label{fig:flexible}
\end{figure}

Fig. \ref{fig:flexible} illustrates the impact of network load to the system performance for a MISO BC with $K=6$ users and imperfect CSIT with $\alpha=0.5$.  The variances of user channels are randomly selected from $[0.1, 1]$.
Two MISO NOMA schemes are considered as baseline schemes, namely, MISO NOMA with a single user group ($G=1$) which is motivated by SC and SIC in SISO BC, and MISO NOMA with three user groups ($G=3$) which clusters 6 users into 3 user groups and inner-group users are served by SC and SIC.
Readers are referred to \cite{bruno2021MISONOMA} for more details on the parameter settings and baseline schemes of MISO NOMA.
When the number of transmit antennas is $M=3$, the network load is extremely overloaded. 
In this case, MISO NOMA ($G=1$) achieves a better MMF rate than SDMA and MISO NOMA ($G=3$).
However, as the number of transmit antennas increases, the performance of MISO NOMA ($G=1$) decreases due to a significant loss in DoF. MU--LP  outperforms the two MISO NOMA when the network is underloaded ($M=6$). 
In all subfigures, 1-layer RS, which only uses a single layer of SIC at each user, achieves the best MMF rate no matter the network is underloaded or overloaded. 
As per Fig. \ref{fig:SE1}--\ref{fig:flexible}, RSMA is a flexible MA scheme which is suited for different user deployments (diverse channel directions and strength), network loads (underloaded and overloaded), CSIT condition (perfect and imperfect CSIT), and SNR regimes (low, medium, and high SNR regimes).
The root of such flexibility is the powerful interference management ability of RSMA introduced by the common stream (vectors). By  adjusting the sub-messages encapsulated in each common stream as well as the power allocation, RSMA dynamically alters the portion of interference to be pre-canceled at the transmitter and decoded at the receivers. 
\begin{figure}[t!]
	\centering
\subfigure[$M=32, K=8, N=1, Q_c=Q_p=1$]{
		\centering
		\includegraphics[width=0.22\textwidth]{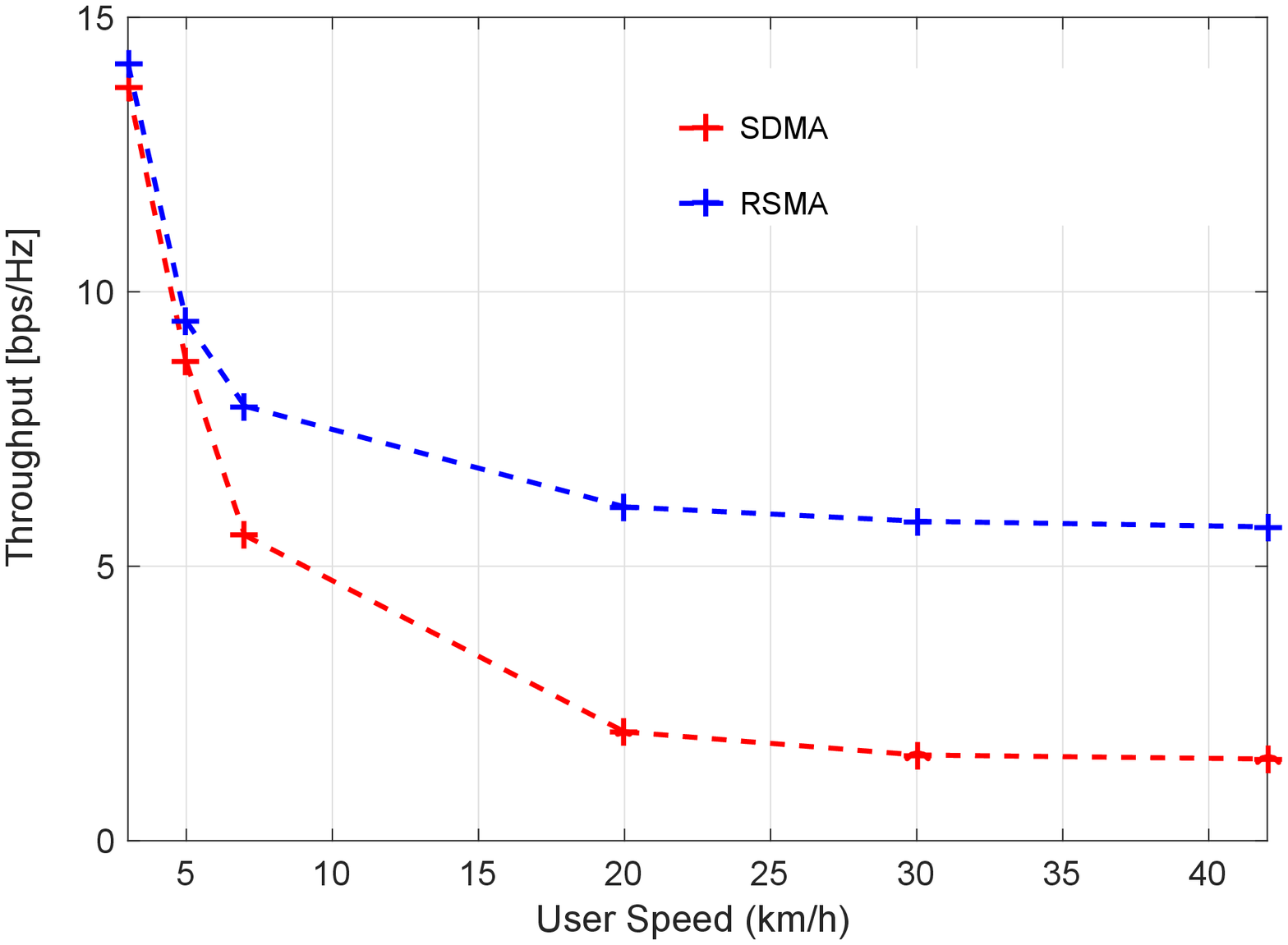}%
		\label{fig:robusta}
}
	~
\subfigure[$M=64, K=8, N_r=4, Q_c=Q_p=3$]{
		\centering
		\includegraphics[width=0.22\textwidth]{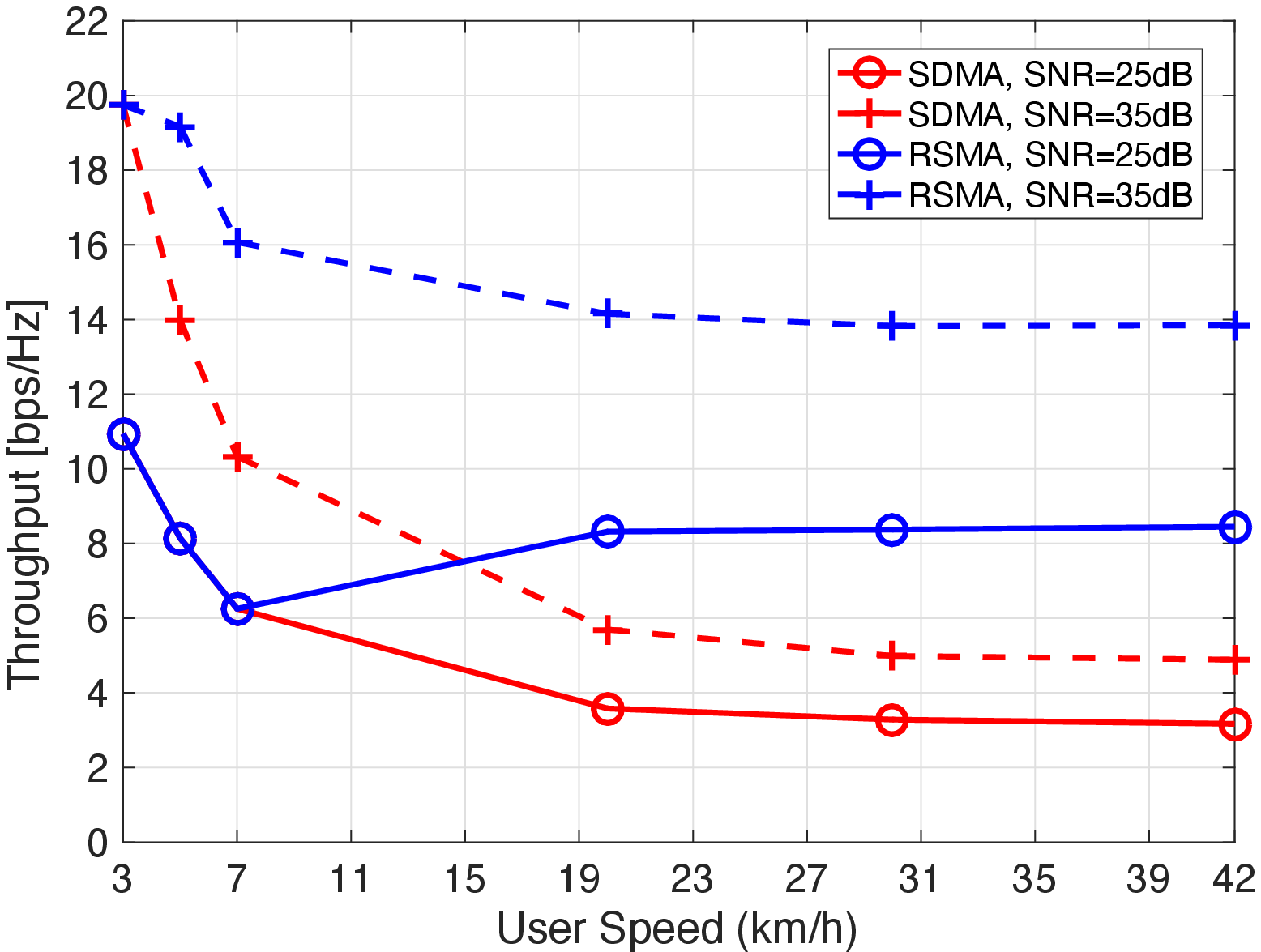}%
			\label{fig:LLSmobility_MIMO}
}
	\caption{Throughput vs. user speed comparison in (massive) MIMO BC with outdated CSIT employing OFDM and a 3GPP channel model, and $10$ ms CSI feedback delay \cite{onur2021mobility}.}
	\label{fig:robust}
\end{figure}

\subsubsection{Robust}
Robustness is one of the most important benefits of RSMA discovered in recent years \cite{enrico2017bruno, RS2016hamdi, hamdi2017bruno,mao2019beyondDPC}. 
The robustness of RSMA is grounded in deep information theoretic results, where RSMA is shown to achieve the optimal DoF in MISO BC when  CSIT is imperfect \cite{enrico2017bruno}. 
Motivated by such discovery, different sources that impair CSIT are studied with RSMA such as pilot contamination \cite{anup2022_massive}, channel estimation errors \cite{RS2016hamdi}, user mobility \cite{onur2021mobility}, etc.
RSMA is shown a significant SE gain over existing MA baselines for all sources of CSIT impairments. 
Taking user mobility as an example in Fig. \ref{fig:robust}, we have shown in \cite{onur2021mobility} that RSMA achieves a significantly higher user speed (i.e., 40 km/h) than SDMA (i.e., 5 km/h) for a given QoS rate constraint (i.e., 8 bps/Hz) thanks to its powerful interference management ability.

\subsubsection{Reliable and Low Latency}
\begin{figure}[t!]
	\centering
\subfigure[Underloaded $M=4, K=2$]{
		\centering
		\includegraphics[width=0.22\textwidth]{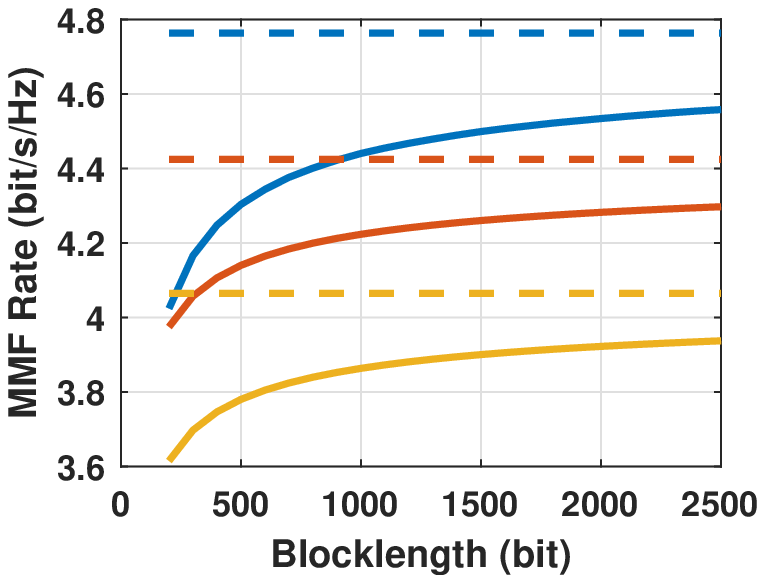}%
}
	~
\subfigure[Overloaded (M=4, K=8)]{
		\centering
		\includegraphics[width=0.22\textwidth]{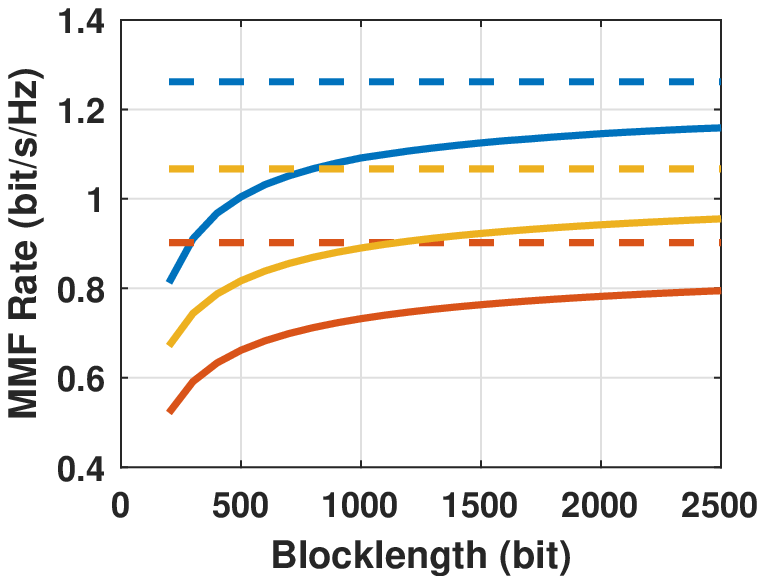}%
		\label{fig:MMF_FBLb}
}  

	\caption{MMF rate versus blocklength (solid lines) of downlink SDMA (red), NOMA (yellow), RSMA (blue) at SNR$=$20dB \cite{yunnuo2022FBL}. Dashed line is the upperbound achieved with infinite blocklength.}
	\label{fig:MMF_FBL}
\end{figure}

: As shown in \cite{yunnuo2021FBL,yunnuo2022FBL} for downlink and in \cite{Jiawei2022} for uplink, RSMA is more reliable than SDMA and NOMA under finite blocklength. This is illustrated in Fig. \ref{fig:MMF_FBL} for underloaded and overloaded downlink settings where we note that RSMA can achieve a given MMF rate at a lower blocklength, therefore enabling lower latency communications, compared to other MA schemes. It is important to recall here again that 1-layer RS scheme of RSMA is used in Fig. \ref{fig:MMF_FBL}, and therefore requires a single SIC. This is in contract with NOMA that achieves lower performance at the expense of the need for 7 SIC layers at the receivers in Fig. \ref{fig:MMF_FBLb}. 

\subsection{Lessons Learned}

\noindent\fbox{%
    \parbox{\columnwidth}{%
       \begin{itemize}
            \item RSMA framework is a superset of OMA, SDMA and NOMA, and can specialize to each of them depending on how messages are mapped to streams. This holds for both uplink and downlink, as well as in SIMO, MISO and MIMO settings.
            \item All instances of RSMA are supersets of SDMA and OMA, but not all, such as 1-layer RS scheme, are supersets of NOMA. This enables RSMA schemes with better performance than NOMA at a much lower receiver complexity (e.g., only one SIC layer).
            \item RSMA can be linearly precoded or non-linearly precoded.
            \item RSMA is applicable to all major settings of a cellular network, namely downlink, uplink, and multicell, for general MIMO deployments (with SISO, SIMO, MISO as special cases).
            \item RSMA is spectrally and energy efficient, general and unified, flexible, robust, reliable and has lower latency.
       \end{itemize}
    }%
}

\section{Numerous Applications for RSMA}\label{section_applications}
\begin{figure*}
	\centering
		\includegraphics[width=0.9\textwidth]{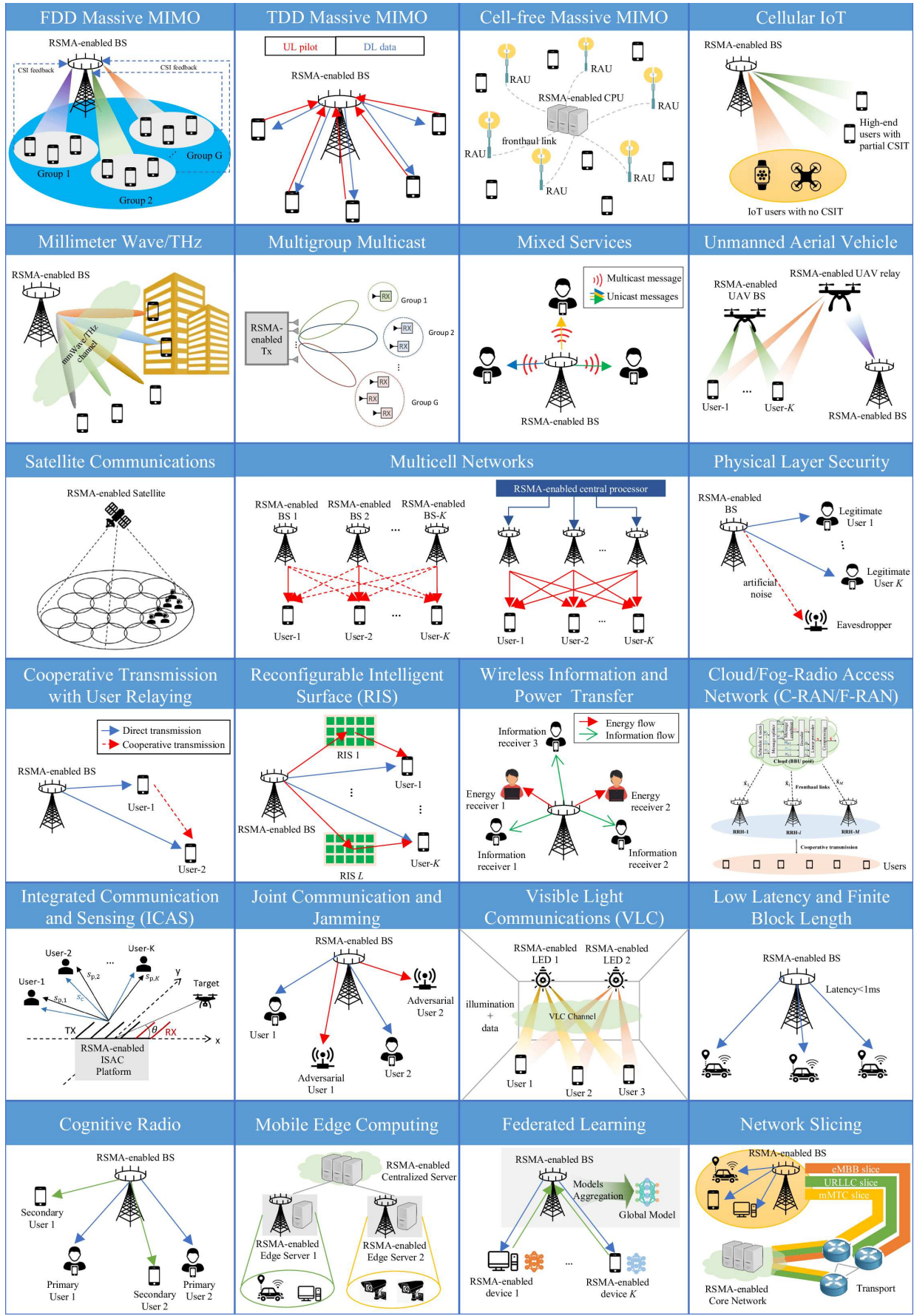}%
		\caption{A subset of promising scenarios and applications of RSMA.}
	\label{apps_scen_RSMA}
\end{figure*}

Given its fundamental communication theoretic principles, and its unique features (efficiency, universality, flexibility, robustness and resilience, reliability and low latency), RSMA finds applications in all modern multi-user scenarios.
We here provide a description of over forty promising scenarios and applications of RSMA and briefly explain how and why RSMA is beneficial\footnote{Other promising scenarios and applications are also discussed in Section \ref{section_asked_questions} on frequently asked questions.}. A subset of those scenarios and applications is illustrated in Fig. \ref{apps_scen_RSMA}. All those applications demonstrate the suitability of RSMA for FeMBB, eURLLC, umMTC, and new wireless services in 6G \cite{onur2021sixG,onur2020sixG}. Importantly, the benefits and applications of RSMA have not only been explored from an academic perspective with the performance superiority over other MA techniques confirmed via stochastic analysis \cite{Qiao_2022,Demarchou_2021_ISWCS,Demarchou_2020_ICC,Tegos_2022} but have also been confirmed using realistic link-level simulations over 5G compliant channel models \cite{Onur2020LLS,hongzhi2020LLS,longfei2021LLS,anup2021MIMO,JRCLoli2022,DLloli2022}.
Interested readers are also encouraged to consult \cite{Maosurvey,CL1_2022,CL2_2022,CL3_2022} for some more descriptions of interesting applications and future works.

\subsubsection{Downlink SISO}  RSMA can be used for both degraded or non-degraded BC. In the degraded BC, RSMA boils down to NOMA (since NOMA is capacity achieving for degraded BC) but RSMA can be used to decrease the receiver complexity and still come close to the capacity region with a reduced number of SIC layers compared to NOMA \cite{mao2017rate}. In the non-degraded BC, RSMA achieves a strictly larger rate region than NOMA (see Remark \ref{rem:non-degraded}).
\subsubsection{Downlink MISO} RSMA can be used in various MISO settings with perfect and imperfect CSIT. The design and precoder optimization of RSMA would depend on the objective value, e.g., WSR, MMF, EE, and the modeling of the CSIT imperfection, e.g., unbounded vs bounded error \cite{RS2016hamdi,mao2017rate,hamdi2016robust,Medra2018SPAWC,bho2021globalEE,bho2022global,Zheng2020JSAC,Park2021WCNC}. One challenge with imperfect CSIT is to make sure that the rate optimized are achievable. 1-layer RS is the most common scheme but other schemes based on multi-layer RSMA can also be used \cite{mao2017rate}. A strong benefit of RSMA is its inherent robustness to imperfect CSIT.
\subsubsection{Downlink MIMO} RSMA can be used in settings with multiple receive antennas at the users. MIMO RSMA enables to transmit vectors of common and private streams, which requires a special design \cite{chenxi2017bruno,bruno2020MUMIMO,anup2021MIMO,Zheng2021TIT,schober2021MUMIMORS,tuan2019,Kaulich_2020}. Various CSIT/CSIR assumptions and objective values can also be considered in the MIMO setting.
\subsubsection{Uplink} Uplink RSMA avoids the need of time sharing to achieve the capacity region, which finds new applications in modern systems subject to latency constraints  \cite{Rimo1996,Grant2001,Jiawei2022,zhaohui2020ULRS,Hongwu2020ULRS,Cao2007distributeRS}. Much attention has been brought to the uplink SISO RSMA, but uplink multi-antenna RSMA is also possible as shown in Section \ref{uplink_k_user}.
\subsubsection{Unifying OMA, SDMA, NOMA, Multicasting} RSMA is universal and general in the sense that it encompasses many different schemes (such as OMA, SDMA, NOMA, physical-layer multicasting) as particular instances \cite{bruno2019wcl,mao2017rate,book2021Mao,bruno2021MISONOMA}. It is interesting to study and understand under what propagation conditions, e.g., disparity of channel strengths and angle between user channels, RSMA boils down to each of those schemes \cite{bruno2019wcl}.
\subsubsection{Statistical CSIT} RSMA is helpful in scenarios where only statistical CSIT (i.e., the distribution
information of the user channels) is available, This is a particular instance of imperfect CSIT, but important in practice as it leads to low feedback overhead. This can be for instance a channel covariance matrix calculated by averaging over frequency and time resources and reported once in a while to the base station. RSMA has been found robust to scenarios where only statistical CSIT is available \cite{minbo2017mmWave,Minbo2016MassiveMIMO,longfei2021wcnc,Amor_2020}
\subsubsection{Quantized Feedback} Quantized feedback is another popular CSI feedback mechanism that relies on a codebook to quantize the channel. Due to quantization error, the CSIT is imperfect and RSMA has been found robust to the quantization error \cite{chenxi2015finitefeedback,Lu2018}. Interestingly the number of feedback bits  needed to achieve a given sum-rate performance is decreased with RSMA compared to conventional SDMA/MU-MIMO, therefore enabling an overhead reduction \cite{chenxi2015finitefeedback,NJindalMIMO2006}.
\subsubsection{Imperfect CSIT and CSIR} Aside imperfect CSIT, for which we know RSMA is robust, imperfect CSIR is also an important problem that has received less attention \cite{wonjae2021imperfectCSIR}. The rate achievability and the SIC error propagation are two important issues to consider when studying RSMA with imperfect CSIR. 
\subsubsection{Frequency  Division  Duplex (FDD) Massive MIMO} One popular strategy in FDD massive MIMO is to rely on two-tier precoders where the first tier precoder is based on channel statistics to cancel inter-cluster interference and the second tier based on instantaneous CSI feedback to manage intra-cluster interference \cite{JSDM,Lau_two_tier}. Unfortunately if the space spanned by the covariance matrices of the clusters overlap or if the channel statistics and/or the CSIT is imperfect, the performance of such approaches degrade and RSMA can be used to mitigate those issues \cite{Minbo2016MassiveMIMO}.
\subsubsection{Time  Division  Duplex (TDD) Massive MIMO} TDD massive MIMO also has its impairments such as pilot contamination due to multiple users sharing the same\footnote{In massive machine-type communications, due to the large number of devices, their sporadic access behaviour and limited coherence interval, active devices have a higher chances to utilize the same pilot for uplink channel estimation.} pilot sequence when performing uplink channel sounding and channel aging due to mobility and latency between the channel sounding phase and the data transmission phase. Those two issues effectively lead to imperfect CSIT that need to be tackled. RSMA has been shown robust to mitigate both problems \cite{onur2021mobility,thomas2020PilotContami,anup2022_massive}
\subsubsection{Hardware Impairments} Another impairment in Massive MIMO comes from the hardware, e.g., phase noise, digital-to-analogue (DAC) and analogue-to-digital (ADC) converters. Phase noise can effectively lead to a scenario where the CSIT is imperfect and RSMA has been shown to be robust against hardware impairments such as phase noise \cite{AP2017bruno}. Quantization errors due to finite resolution DAC and ADC also lead to multi-user interference that can be efficiently and flexibly managed using RSMA \cite{onur2021DAC,onur2022DFRC,Seokjun_2022}.
\subsubsection{Cell-Free Massive MIMO} Cell-free refers to distributed antennas systems, which is subject to the same pilot contamination problem as conventional TDD massive MIMO. The disparity of path loss nevertheless makes the analysis and performance gap between schemes different. RSMA was found robust to pilot contamination in various cell-free topologies \cite{anup2022_cfmassive}.
\subsubsection{Millimeter Wave and TeraHertz Systems} Higher frequency bands are subject to their own set of challenges such as high path loss, blocking and expensive RF chains. RSMA can be used and designed in conjunction with hybrid analogue-digital precoding to outperform conventional SDMA strategies at high frequencies \cite{minbo2017mmWave,Kola2018SPAWC,zhengli2019mmWaveRS}. RSMA in combination with cooperative communications can also be used to further combat the path loss and blocking issues \cite{Junil_2022_THz}.
\subsubsection{Non-Linear Precoding and DPC} RSMA is commonly designed using linear precoding due to its practical use in real systems. Nevertheless, non-linear precoding could also be used to precode the private streams, using for instance DPC, Tomlinson-Harashima precoding, vector perturbation, etc. Recent works have demonstrated the benefits of non-linear precoded RSMA \cite{Flores2018ISWCS,mao2019beyondDPC,Andre2021THP,Andre2021THP_ICASSP}. The combination of RSMA and DPC (so-called DPCRS strategy as per Fig. \ref{Fig_DL1LMIMODPCRS} and Fig. \ref{Fig_DLMLMIMODPCRS}) is particularly suited to imperfect CSIT settings since we know that DPC is capacity achieving with perfect CSIT. As the CSIT quality improves, the power allocation to the common stream decreases, and DPCRS progressively boils down to conventional DPC for multi-antenna BC. It is of interest to investigate how we could further enhance the performance in imperfect CSIT beyond that achieved by DPCRS strategy \cite{mao2019beyondDPC}.

\subsubsection{Cooperative and Relaying Systems} Cooperative communications and RSMA, i.e., cooperative RSMA as in Fig. \ref{Fig_CRS}, form a happy marriage as shown in \cite{jian2019crs,mao2019maxmin} since a user can decode the common stream and its private stream and forward the common stream to help improving the decodability of the common stream at a cell edge user. By doing so, RSMA can efficiently cope with a wide range of propagation conditions (disparity of user channel strengths and directions), compensate for the performance degradation due to large path loss, extend the coverage, and outperforms SDMA, NOMA, and cooperative NOMA. Those observations also hold in the imperfect CSIT setting \cite{Jian_CRS_imperfectCSIT}. RSMA can also find other interesting applications in multi-user relaying systems \cite{Edward2018SPAWC,Salen2018ISWCS,NOMA2017Zheng,Khattab_2022}.
It would be worth further exploring how, in the classical relay channel \cite{Cover_1979}, as well as the cooperative communication works \cite{Sendonaris:2003,Laneman:2004}, RS inherently plays a role in the decode-and-forward protocol and its variations.

\subsubsection{Full Duplex} Self interference is a major problem in full duplex systems. Thanks to its flexible interference management ability, RSMA can increase the range of self-interference over which full duplex outperforms half duplex \cite{Asos2018MultiRelay}.

\subsubsection{Physical Layer Security}\label{PHY_layer_security}
Security at the physical layer has attracted a lot of attention in the past decade \cite{Bassily_2013}. Two interesting scenarios occur for RSMA: first, the eavesdropper is not one of the users served by RSMA (hence, the eavesdropper is not
the intended recipient of any message transmitted by the transmitter, but
only intercepts confidential messages sent to other authorized
users), and second, the eavesdropper is one of the users and can therefore decode the common stream \cite{fuhao2020secrecyRS,Liping2020secrecyCRS}. In the former case, the common stream can be used to manage the interference between users and therefore increase the sum-rate of those users but also to act as an artificial noise (AN) to confuse the eavesdroppers. In the latter case, an eavesdropper could decode a common stream and could therefore reconstruct the original message if it can decode the corresponding private. This leads to a secrecy constraint for the private stream and a tradeoff between sum-rate and secrecy rate \cite{Xia_secrecy_2022,Xia_secrecy_wcnc_2022,Salem_2022}. RSMA achieves better WSR and is more robust to channel errors than SDMA while ensuring all users’ security requirements. In contrast, because the entirety of the message of a user is mapped to the common stream in NOMA, NOMA cannot guarantee all users' secrecy rate constraints \cite{Xia_secrecy_2022}. Other interesting analysis of RSMA in the presence of untrusted users has recently appeared in \cite{Abolpour_2022}.

\subsubsection{Energy Efficient Networks} EE is an increasingly important metric. In general, the superiority of RSMA in terms of spectral efficiency translates into an EE gain over conventional MA baselines such as OMA, SDMA and NOMA \cite{mao2018EE,mao2019TCOM}. Nevertheless, SE maximization and EE maximization are two conflicting objectives in the moderate and high SNR regimes. This calls for the study of the tradeoff between the two criteria. In \cite{Gui2020EESEtradeoff}, the performance of RSMA is shown to be superior to or equal to SDMA and NOMA in terms of SE, EE, and their tradeoff. 

\subsubsection{Reconfigurable Intelligent Surfaces (RIS)} RIS is equipped with a large number of passive elements placed in the environment that can be adjusted so as to provide a passive beamforming gain and make the channel propagation more favourable. The interplay between RIS and RSMA is very attractive  and has been shown very promising by several works \cite{alaa2020IRScran,zhaohui2020IRS,bansal2021IRS,CL3_2022,jolly2021IRS,IRSRS2021TVT,alaa2020IRScran_EE,FU_2021,Fang_2022,deSena_2022,Kangchun_2022}. The advantages of RIS-aided RSMA are higher spectral efficiency, coverage extension and beam control flexibility thanks to the presence of the common stream, robustness to CSI imperfection which is welcome given the channel acquisition challenge in RIS, and lower computational and hardware complexity because RSMA superiority means RSMA can afford operating with less complex RIS architectures while maintaining the same overall performance \cite{CL3_2022}.   

\subsubsection{Multi-Cell Networks} The concept of RS for the 2-user IC can be extended to more than 2 users, or to multiple cells in cellular network context \cite{Bauch_2012}. Furthermore each node in each cell can be equipped with multiple transmit antennas and precoders jointly with the message split need to be optimized across all cells to maximize the objective function accounting for any potential imperfection in the CSIT \cite{WYu2011rs,chenxi2017brunotopology,chenxi2017bruno,Jia2020SEEEtradeoff,wonjae2020multicell,su2021MIMOICRS}. Of particular interest, \cite{chenxi2017brunotopology} showed that in a multi-cell multi-antenna setting with imperfect CSIT, the transmitters should adopt a so called topological RS (TRS) strategy that consists in a multi-layer structure and in transmitting multiple common messages to be decoded by groups of users rather than all users. Though TRS was studied from a communication and information theoretic point of view, it remains to be studied how TRS could be optimized using optimization tools so as to maximize its performance at finite SNR.

\subsubsection{Coordinated Multi-Point (CoMP)} RSMA can be used for CoMP joint transmission where all base stations collaborate by sharing CSI and data. In such a setting, all antennas at the base station act together to form a giant BC or MAC (depending on whether the focus is on downlink or uplink) and precoders and power across streams can be optimized as a function of the CSI and path loss disparity subject to the per-cell power constraints. It was shown in \cite{mao2018networkmimo} that whenever there is little inter-user channel strength disparity but large inter-cell channel disparity, SDMA was a suitable option. On the other hand, whenever there is a large inter-user channel strength disparity but little inter-cell channel disparity, NOMA was a better option. In comparison, RSMA always bridges, generalizes, and outperforms existing SDMA and NOMA strategies. It was shown to be suited to any deployment with any inter-user and inter-cell channel disparities. 
Applications of RSMA to uplink CoMP would be worth investigating.

\subsubsection{Cloud and Fog-Radio Access Network (C/F-RAN)}  
C-RAN consists of multiple remote radio heads (RRH) connected through fronthaul links to a baseband unit that performs central processing. CoMP joint transmission can be implemented across all those RRH while accounting for the challenging feature that the fronthaul links have a limited capacity. Various RSMA approaches have been proposed for C-RAN: 1) RSMA transmit signal is compressed before being transmitted over the fronthaul \cite{cran2019wcl}; 2) the common and private streams of RSMA to be transmitted over each fronthaul are wisely selected so as to satisfy the fronthaul constraints \cite{Ahmad2018SPAWC,Ala2019IEEEAccess,Alaa2019uavCRAN,alaa2020EECRAN,alaa2020gRS,alaa2020powerMini,alaa2020cranimperfectCSIT,hassan2020fogRAN,Hassan2021FRAN,alaa2021cacheCRAN}. For a given fronthaul capacity constraint, both approaches have demosntrated that RSMA is more spectrally and energy efficient than SDMA and NOMA.

\subsubsection{Dynamic Resource Management and Cross-Layer Optimization} In RSMA-aided networks, network resource management has received more and more attention. Many of these works, however, only consider the short-term optimization of system resources without taking into account the long-term network operation constraints and objectives. Furthermore, they do not consider random traffic arrivals. To meet the explosive access demands of mobile devices, the multicast communication of a satellite and aerial-integrated network with RSMA is studied in \cite{Lin2021UAV}. Specifically, the unmanned aerial vehicle (UAV) sub-network uses RSMA to support massive access of internet of things (IoT) devices in a content delivery scenario. However, in practice, users may require different types of data traffic, such as high-quality video streaming, voice, video phone, online games, network broadcasting, and so on. This necessitates the cross-layer design of RSMA-aided system to cope with traffic exposure rate and long-term system constraints. It is recommended that joint adaptive source encoding and cross-layer resource management schemes be studied. For example, using a supervised learning-based approach for cross-layer resource allocation in a NOMA-aided system~\cite{tseng2022cross}, the execution time can be reduced by 98\% while ensuring that each user has at least one subcarrier. Given the outstanding characteristics of RSMA, more cross-layer (across physical - PHY, medium access control - MAC, and application - APPL - layers) dynamic resource management solutions and optimizations are expected to bring further improvement to the system performance. Though most of the RSMA optimization works focus on PHY and MAC layers (beamforming and power allocation, time and frequency resource allocation), some works apply RSMA for the wireless streaming of video and consider joint optimization across PHY, MAC, and APPL (hence, with the optimization of the encoding rate adaptation for video) \cite{Cui_2021}. This is an important research direction for RSMA, especially for mobile internet (where video takes more than 70\% of traffic) and emerging applications such as 360 video, autonomous driving, metaverse for 6G and beyond.

\subsubsection{Wireless Caching} The small storage capacity of edge devices combined with the high energy consumption of active caching shorten the standby period of user devices. By positioning caching devices adjacent to the user terminal, wireless caching networks (WCN) can reduce recurrent file transfers to achieve low power consumption caching. Furthermore, the spectrum efficiency can be increased by combining RSMA with WCN. Two caching policies can be examined, i.e., the most popular content (MPC) and the intelligent coded caching (CC) policy. In MPC policy, multiple popular files can be superimposed in the power domain to form a mixed file based on RSMA, which is broadcasted to cache devices by the base station \cite{Enrico2017cach,Piovano_caching_2019,Demarchou2022}. Another approach of cache placement is the coded caching (CC), where partitions of the files are stored instead~\cite{hachem2020coded}. Multiple cache-enabled receivers can be served using both caching policies within the help of RSMA~\cite{Enrico2017cach,Piovano_caching_2019,Demarchou2022}. It is shown that the caching gains can be improved significantly, the mutual benefit can be achieved through collaborative design of caching placements and RSMA~\cite{Enrico2017cach,Piovano_caching_2019,Demarchou2022}. In a way, RSMA for caching-aided multi-antenna BC expands upon, and inherits the main features of, RSMA used for the classical multi-antenna BC with imperfect CSIT. Additionally, using RSMA, the caching device can provide services to numerous users in the same time-frequency resource block. The quantity of caching files, their level of popularity, and their network resources all affect the key indicator of wireless caching networks, e.g., the system hit probability.

\subsubsection{Overloaded Cellular Internet of Things and Massive Access}
Cellular networks will have to cope with extensive IoT devices, and consequently serve simultaneously a large number of devices with heterogeneous demands and CSIT qualities. In \cite{enrico2016bruno,mao2021IoT}, RSMA is used to tackle such a scenario by considering an overloaded MISO downlink with two groups of CSIT qualities, namely, one group of users (representative of high-end devices) for which the
transmitter has partial knowledge of the CSI, the other group of users (representative of IoT devices) for which the transmitter only has knowledge of the statistical CSI. RSMA is shown to be DoF-optimal in such a setting with heterogeneous CSIT qualities, more efficient than various MA baselines, robust to CSIT inaccuracy, and flexible to cope with heterogeneous QoS rate constraints of all high-end and low-end users. 

\subsubsection{Joint Communication and Jamming} 
Thanks to its flexibility and robustness, RSMA can be used to efficiently communicate to information users (IUs) and simultaneously jam adversarial users (AUs) to disrupt their communications \cite{onur2021jamming,onur2021CR}. The precoders and power allocation to common and private streams can be optimized based on imperfect CSIT for IUs and statistical CSIT for AUs to maximize the sum-rate under jamming power constraints on the pilot subcarriers of AUs (as jamming pilot subcarriers is known to be very effective disruptive method). RSMA is shown to outperform significantly conventional MA schemes.

\subsubsection{Non-Orthogonal Unicast and Multicast (NOUM)}\label{NOUM_subsection} NOUM refers to mixed traffic services where a multicast (genuinely intended to multiple users, i.e., not user-specific) message is transmitted to multiple users and additionally one unicast (user-specific) message is transmitted per user. 
Conventionally OMA is used to transmit unicast and multicast services on different resources. This is suboptimal and a better approach is to superpose the multicast message on top of the unicast messages and use one SIC at each receiver to decode the multicast message first and then the intended unicast message. RSMA can do better by making a more efficient use of the SIC. Indeed, by splitting the unicast message into common and private parts and encoding jointly the common parts and the multicast message into a common stream to be decoded by all users, the SIC can be efficiently exploited for the dual purpose of separating multicast from unicast but also better manage interference between unicast streams \cite{mao2018rate,mao2019TCOM,mao2020DPCNOUM,Abanto_2022}. Taking the 2-user architecture of Fig. \ref{fig_2user_RSMA_1layerRS} as an example, RSMA in NOUM is obtained by encoding a multicast message $W_0$ (genuinely intended to both users) along with common parts $W_{\mathrm{c},1}$ and $W_{\mathrm{c},2}$ into a common stream. User-$k$ then decodes the common stream to retrieve $W_0$ and $W_{\mathrm{c},k}$, before decoding its private stream.
RSMA has been shown to outperform SDMA and NOMA counterparts in NOUM \cite{mao2018rate,mao2019TCOM,mao2020DPCNOUM}.

\subsubsection{Multigroup Multicast} This scenario considers $K$ users grouped into $G<K$ groups and a transmitter that delivers on the downlink one multicast message per group, i.e., all users in the same group are interested in the same multicast message. Conventional BC is a subset of that setting where there is only one user per group. Such a scenario can occur in broadcasting services, caching settings, satellite communications, multi-view video, virtual reality (360) video, online gaming, metaverse, etc. The challenge of multigroup multicast is that the number of users is often large compared to the number of transmit antennas, i.e., overloaded network, which creates severe multigroup interference issues. RSMA, thanks to its flexibility, is able to tackle that multigroup interference efficiently and outperform both SDMA and NOMA schemes significantly \cite{hamdi2017bruno,Tervo2018SPAWC,longfei2020satellite,yalcin2020RSmultigroup,hongzhi2020LLSmultigroup,hongzhi2021RSLDPC,hongzhi2020_TVT} and cope with various other multicast scenarios relevant to 6G \cite{Cui_2022}.

\subsubsection{Multibeam Satellite Communications} Multibeam satellite is often modeled by a multigroup multicast where each beam can be thought of one group. This comes from  the superframe/frame-based precoding assumption of multibeam satellite communication which is that one codeword encodes all users' messages in one beam. This is used in practice to increase the efficiency of the error correcting codes. Other challenges in satellite systems include the frequency reuse across multiple spot beams creating high levels of interference, the per-feed power constraints, the satellite channel subject to line-of-sight and large path loss, the imperfect CSIT and latency in CSI acquisition due to large round trip delay, the high doppler for low earth orbit satellite, and the overloaded settings (many terminals in each beam). RSMA has been found to be quite suited to tackle all those challenges thanks to its efficiency, flexibility, and robustness \cite{longfei2020satellite,RSSatellite2018ISWCS,longfei2020multibeam,si2021imperfectSatellite,miguel2018RSsatellite}.

\subsubsection{Unmanned Aerial Vehicles (UAV)-Assisted Networks} 

Due to the great mobility and flexibility, UAVs are able to provide services to users when they are outside of the base station's coverage area or when the user channel conditions are unfavorable. However, a UAV relies on its limited capacity battery to fly, hover and communicate, which leads to a limited endurance. Fortunately, RSMA can reduce the communication energy consumption of UAV. By acting as an aerial base station, the UAV can serve multiple ground users simultaneously using RSMA~\cite{singh2021outage,UAVRS2019ICC,jaafar2020UAV,jaafar2020UAV2}. RSMA for UAV leads to new joint cross-layer optimization problem of UAV location, transmit power allocation, bandwidth, and RSMA, accounting for CSI availability and traffic. RSMA and UAV locations can also be designed along with methods used for predicting the cellular traffic according to the analysis of previous data, therefore further saving transmit power consumption \cite{Lu_uav_ML_RSMA}.

\subsubsection{Space-Air-Ground/Satellite-Terrestrial Integrated Networks} 

In satellite-terrestrial (or in space-air-ground) integrated network, the satellite sub-network shares the same RF band with the terrestrial sub-network. A higher spectrum efficiency and throughput is achieved via dynamic spectrum access sharing to enhance spectrum utilization. However, severe interference in and between the sub-networks is induced by an aggressive frequency reuse, which calls for the use of efficient multi-antenna \cite{xiao2021bruno6G} and multiple access strategies \cite{longfei2021satelliteTerres,Lin2021Cognitive,Lin2021UAV}. In this context, RSMA has been shown to exhibit significant performance gains compared with various traditional transmission strategies such as SDMA and NOMA in various settings where only CSI is shared among sub-networks (to enable coordination of precoders) or where CSI and data are shared (to enable cooperation as in CoMP joint transmission across all antennas of both sub-networks) \cite{longfei2021satelliteTerres}.  

\subsubsection{Constructive Interference Exploitation/Symbol-Level Precoding} 
RSMA is conventionally studied with Gaussian inputs, but in practice finite constellations need to be used. 
Constructive interference (CI), also called symbol-level precoding, exploits the finite constellation such the information symbols are used, along with the CSI, in order to exploit the multi-user interference to increase the useful signal received power. In other words, CI designs the transmit precoders such that the resulting interference is constructive to the desired symbol, i.e., the interference signal pushes/moves the received symbols away from the decision thresholds of the constellation towards the direction of the desired symbol.
Interestingly, RSMA and CI techniques can be combined to further enhance the sum-rate achieved by RSMA with finite input alphabet \cite{abdel2019finite,Salem_2019}.

\subsubsection{Reliability and Low Latency } 
There are many reliability and latency-sensitive applications, such as industrial automation, smart grid and intelligent transportation. In order to reduce the transmission latency, shortpackets with finite blocklength codes are typically adopted. This brings a stringent latency requirement to the physical layer. Interestingly, the efficiency, robustness, flexibility benefits of RSMA over SDMA and NOMA in the infinite blocklength regime were also confirmed in the finite blocklength regime for both downlink and uplink, offering therefore the additional reliability and the low latency needed to enable those applications \cite{yunnuo2021FBL,yunnuo2022FBL,Jiawei2022}. In other words, RSMA can achieve a given performance requirement with a smaller blocklength than that needed by SDMA and NOMA. The significant performance gains of RSMA over SDMA and NOMA were also confirmed in link-level evaluations with practical codes and finite blocklengths \cite{Onur2020LLS,longfei2021LLS}.

\subsubsection{Integrated Radar Sensing and Communications} 

Integrated (radar) sensing and communications (ISAC) merges wireless communications and remote sensing into a single system, where both functionalities are combined via shared use of the spectrum, the hardware platform, and a joint signal processing framework. It also enables sensing capabilities of the network to help communications and inversely. The challenge is that a transmitter not only has to serve multiple users simultaneously but also has to satisfy radar performance requirements, lwhich leads to a tradeoff between communication and radar performance and calls for agile and versatile MA schemes. The flexibility and robustness of RSMA becomes particularly handy in this setting as it was shown that RSMA can provide a better communication-radar tradeoff than SDMA and NOMA schemes for a wide range of propagation conditions and radar metrics \cite{chengcheng2020radCom,xu2021rate,rafael2021radarsensing,JRCLoli2022,onur2021DAC,onur2022DFRC,CL2_2022,longfei2022wcnc}.

\subsubsection{Grant-based, Grant-Free and Semi-Grant Free Transmission and Massive Random Access}

Grant-based transmission has been a conventional approach to access the network when the network load is small. With the emergence of IoT devices and mMTC, hybrid grant-based (GB) and grant-free (GF) transmissions are needed to reduce latency. Grant-free access however leads to collisions that need to be managed. In such scenario, GF users meet opportunities to share wireless resources with GB users. In \cite{Hongwu2021}, RSMA was used in this setting where the GF users split their messages to realize distributed contentions and utilize transmit power most effectively for robust transmissions, meanwhile keeping themselves transparent to the GB user. RSMA was shown to significantly decrease outage probability and achieve full multi-user diversity gain without restricting the GB and GF users' target rates.

\subsubsection{Network Slicing} 
To guarantee the performance of heterogeneous services involving FeMBB, eURLLC, umMTC in 6G, network slicing is needed to allocate resources to different services. Network slicing can be done in an OMA fashion, which means that different services are isolated and allocated orthogonal (non-interfering) resources. However, as the number of users grows, OMA-based slicing may not be the optimal scheme for all scenarios, and a non-orthogonal scheme may achieve a better performance. Thanks to its message splitting and corresponding flexibility, RSMA has emerged as a superior MA scheme for network slicing, outperforming OMA and NOMA in many scenarios \cite{Santos2021,Yuanwen_2022}.

\subsubsection{Cognitive Radio} Cognitive radio may increase spectral efficiency through secondary spectrum sharing / dynamic spectrum access, transmitting under the interference temperature. Various setups can be considered where primary transmitter-receiver links 
communicate simultaneously with a group of cognitive secondary users that form one of classical multi-user channels such as MAC, IC, or BC \cite{petar2011RS}. In the BC, RSMA can for instance be employed at the secondary transmitter to communicate with secondary users while limiting the interference to primary users \cite{onur2021CR,Camana2020swiptRS}. Since the primary link remains oblivious to the secondary system operation in cognitive radio, another potential benefit of RSMA is whether the primary system does not need to time-share the channel with the secondary users, unlike the conventional spectral-gap filling approaches \cite{petar2011RS}.

\subsubsection{Optical and Visible Light Wireless Communications} RSMA is commonly studied for RF communications, but it can also be applied to other communication systems such as optical and visible light, though the constraints from those signals and systems need to be captured in RSMA design and optimization, e.g., visible light signals have peak and average optical power constraints (limited for eye safety and practical illumination requirement), are non-negative and real due to the
intensity modulation and direct detection technique \cite{naser2020vlc,siyu2020vlc,mashuai2021vlc}. Similarly to RF systems, RSMA outperforms SDMA and NOMA in optical and visible light communications \cite{naser2020vlc,siyu2020vlc,mashuai2021vlc}.

\subsubsection{Multi-carrier} Frequency domain using multi-carrier transmission (e.g., OFDM/OFDMA) can be combined with RSMA in the same way SDMA and OFDM/OFDMA work together in 4G and 5G, namely the spectrum is divided into subbands (made of contiguous or distributed subcarriers) and multiple users are paired together on one or multiple subbands using RSMA \cite{lihua2020multicarrier,onur2021CR,onur2021jamming,hongzhi2021RSLDPC}. Resource allocation, including user pairing per subband and power allocation to common and private streams, needs to be optimized. RSMA-OFDMA inherits the same benefits as SDMA-OFDMA since RSMA builds upon SDMA/MU-MIMO.

\subsubsection{Wireless Information and Power Transfer (WIPT)} WIPT has the similarity with ISAC that both systems need to use the spectrum to deliver two services: either sensing and communications, or power and communications. Interestingly, similarly to ISAC, the common stream in RSMA can be helpful to manage multi-user interference and at the same time boost the performance of the other service, i.e., sensing or power. Consequently, RSMA has been found more efficient than SDMA and NOMA in WIPT 
\cite{mao2019swipt,RSswiptIC2019CL}.

\subsubsection{Vortex Wave Communications/Orbital Angular Momentum} RSMA can be combined with orbital angular momentum (OAM) to benefit from the flexibility and robustness of RSMA and the additional degrees of freedom of OAM \cite{vortex_2022}.
\subsubsection{Mobile Edge Computing (MEC)} Equipped with powerful computing and storage capabilities, MEC can cope with the challenges of providing superior and latency-critical computing by enabling edge users to offload their tasks for nearby processing, therefore reducing the backhaul bottlenecks, network
delays, and transmission costs. Since RSMA can achieve the full rate boundary of the MAC, while NOMA can achieve only several separated points on the rate boundary, RSMA can be used more efficiently than NOMA to aid MEC where multiple users can offload their tasks while maintaining the QoS of each user \cite{Hongwu2022_MEC}.
\subsubsection{Mixed Criticality} Thanks to the efficiency of RSMA, QoS is enhanced \cite{mao2017rate,Maosurvey}. Thanks to its flexibility, RSMA can also deliver QoS enhancements in applications where, due to the diversity of users, services and applications in 6G, mixed criticality QoS levels are assigned to those users and services \cite{Reifert_2022_EE_MC,Reifert_2022_MC}.

\section{Myths}\label{section_myths}
\par \textbf{Myth 1: RSMA is a special (power-domain) NOMA}

It is actually the opposite with (power-domain) NOMA being a special RSMA technique. In the same way as decoding interference is a particular instance of RS (and has been known to be so since the 80s and the seminal works on RSMA \cite{TeHan1981,Rimo1996}), NOMA is a particular instance of RSMA, as illustrated in Fig. \ref{fig:bubble_interf_manag}, \ref{Fig_2userbubble} and \ref{Fig_ULRelation}. However, one needs to check more carefully at how the schemes have been built as NOMA is not a special case of all RSMA schemes. 
\par In uplink, NOMA simply relies on SIC. In other words, there is nothing special to NOMA as it is just an SIC receiver (similarly to the SIC used in spatial multiplexing for point to point MIMO). RSMA on the hand not only relies on SIC but also on splitting of the messages at the users. As shown in Section \ref{section_2user_ul}, by adjusting the split and the power allocation to the resulting streams, uplink RSMA boils down to uplink NOMA. 
\par In downlink SISO and MISO, all power-domain NOMA schemes are characterized by having at least one user being forced to fully decode the message(s) of other co-scheduled user(s) \cite{bruno2021MISONOMA}. In the two-user, NOMA (as well as SDMA, OMA, and physical-layer multicasting) is a subset of RSMA as shown in Section \ref{section_2user_dl}. In the general $K$-user MISO case as it would depend on the specific RSMA scheme used. 1-layer RS is a superset of SDMA since by turning off (i.e., allocating no power to) the common stream,  1-layer RS boils down to MU--LP. On the other hand, 1-layer RS is \textit{not} a superset of NOMA. 1-layer RS and NOMA are particular instances/schemes of the RSMA framework based on the generalized RS relying on multiple layers of SIC at each receiver, as discussed in Section \ref{section_Kuser} and \cite{bruno2021MISONOMA,Maosurvey}. 
\par In downlink MIMO, there is less research on RSMA. Nevertheless, RSMA is shown in \cite{bruno2021MISONOMA, anup2021MIMO} to outperform and be a superset of NOMA whenever at least one user is forced to fully decode the multiple streams of other co-scheduled users. More generally, as it appeared in Table \ref{fig_mapping_MIMO_BC} and related discussion, RSMA in a MIMO setting will always be a superset than NOMA because RSMA has the message splitting capability for each message (and therefore the related interference management capability), which does not feature in NOMA schemes. This implies that the optimization space of RSMA will be larger than that of NOMA.
\par
\begin{figure}[t!]
	\centering
		\includegraphics[width=0.4\textwidth]{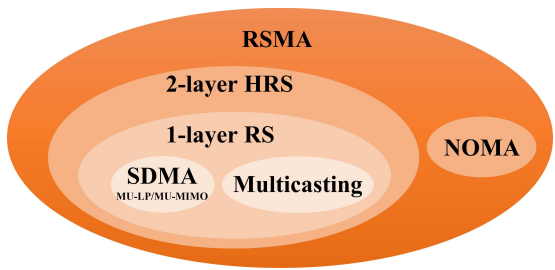}%
		\caption{Relationship between existing MA strategies and the $K$-user RSMA
framework.}
	\label{relations_kuser_fig}
\end{figure}

 The relationship between SDMA, NOMA, 1-layer RS, 2-layer (hierarchical) RS (as introduced first in \cite{Minbo2016MassiveMIMO} for FDD massive MIMO), and RSMA is further illustrated in Fig. \ref{relations_kuser_fig}.

\par \textbf{Myth 2: RSMA cannot outperform MU-MIMO}

MU-MIMO schemes can rely on linear or nonlinear precoding schemes \cite{clerckx2013mimo}. For both types of precoders, we can design linear precoded or non-linear precoded RSMA that are always a superset of MU-MIMO, and would therefore always achieve at least the same performance as MU-MIMO.

\par Taking $K$-user 1-layer RS for simplicity (but the discussion holds for other RSMA architectures), by decreasing the amount of power allocated to the common stream, 1-layer RS progressively converges to $K$-user SDMA/MU-MIMO and in the limit where no power is allocated to the common stream, $K$-user 1-layer RS swiftly boils down to $K$-user SDMA/MU-MIMO. Hence, 1-layer RS really builds upon SDMA/MU-MIMO and SDMA/MU-MIMO is a subscheme of 1-layer RS, which provides a guarantee to 1-layer RS that its rate and DoF are always the same or better than those of SDMA/MU-MIMO. The same observation holds for non-linear precoded RSMA as discussed in \cite{Andre2021THP,Flores2018ISWCS} for Tomlinson-{H}arashima precoded RSMA and in \cite{mao2019beyondDPC,mao2020DPCNOUM} for dirty paper coded RSMA. Note that since dirty paper coding achieves the capacity of MIMO Gaussian BC with perfect CSIT, applying dirty paper coded RSMA to a perfect CSIT setting would end up allocating zero power to the common streams, however non-zero power would be allocated to common streams to boost the performance in imperfect CSIT settings \cite{mao2019beyondDPC,mao2020DPCNOUM}. Other types of non-linear precoded RSMA schemes are left for further investigations. 

\par This is completely different from NOMA. NOMA does not build upon SDMA/MU-MIMO. With $G$ groups, $K$-user MISO NOMA can boil down to $G$-user SDMA by turning off the power to the weaker users in each group, but $K$-user MISO NOMA can mathematically never boil down to $K$-user SDMA \cite{bruno2021MISONOMA}. The rate/DoF of $K$-user NOMA can therefore be worse than that of $K$-user SDMA \cite{bruno2021MISONOMA}. 

\par \textbf{Myth 3:	RSMA is only beneficial for multi-antenna downlink}

RSMA is definitely beneficial in multi-antenna downlink, but RSMA is also beneficial in single-antenna downlink, single/multi-antenna uplink, in multi-cell, and in relaying.
\begin{figure}
	\centering
	\subfigure[{$R_k^{th}=0$}]{
		\centering
		\includegraphics[width=0.22\textwidth]{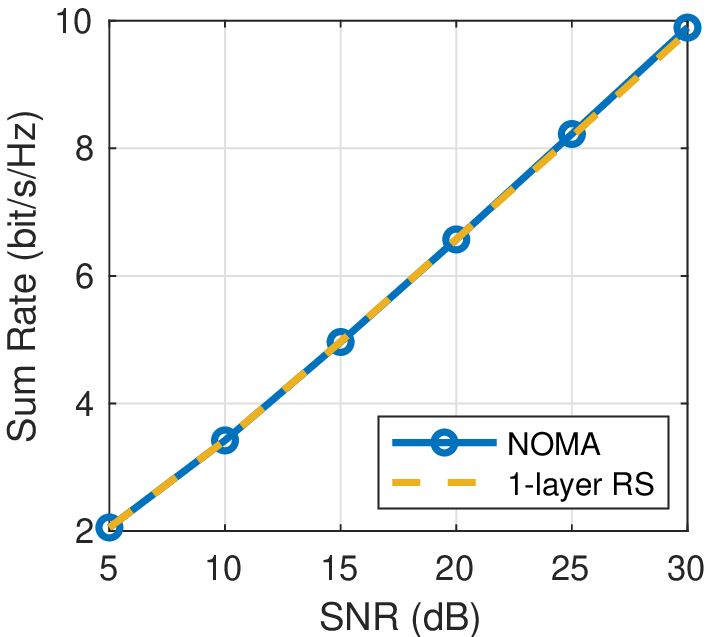}%
	}
	~
    \subfigure[{$R_k^{th}=0.2$ bit/s/Hz}]{
		\centering
		\includegraphics[width=0.22\textwidth]{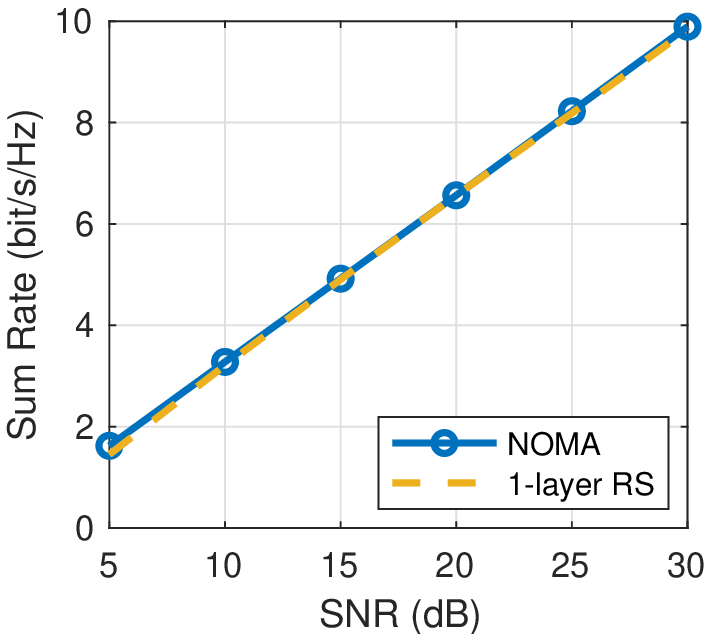}%
		}
	\caption{Sum-rate vs. SNR comparison in three-user SISO BC with perfect CSIT, $K=3, \sigma_1^2=1, \sigma_2^2=0.3, \sigma_3^2=0.1$.}
	\label{fig:SISOBC}
\end{figure}

\par In single-antenna downlink (SISO BC), the benefits of RSMA depends on whether the BC is degraded or non-degraded. In the degraded BC, RSMA boils down to NOMA (since NOMA is capacity achieving for degraded BC) but RSMA can be used to decrease the receiver complexity and still come close to the capacity region with a reduced number of SIC layers compared to NOMA \cite{mao2017rate}. 
We illustrate in Fig. \ref{fig:SISOBC} a three-user example to compare the sum rate of NOMA and 1-layer RS when users' channel variances are $\sigma_1^2=1, \sigma_2^2=0.3, \sigma_3^2=0.1$ and users are with and without QoS rate constraints. 
In the two subfigures, 1-layer RS respectively achieves 99.84\% and 97.65\% rate of NOMA while only a single layer of SIC is required at each user (instead of 2 for NOMA).
In the non-degraded BC, RSMA achieves a strictly larger rate region than NOMA (see Remark \ref{rem:non-degraded}). Examples for non-degraded SISO BC comprise (but are not limited to) multi-cell BC or BC with interference \cite{Rezvani2022}, BC with imperfect CSIT \cite{Lin2020}, RIS-assisted BC \cite{Bansal21}, IC with moderate or low interference \cite{TeHan1981}.
Intuitively the underlying feature of these examples is, that the channels to the receivers cannot be ordered. For all examples above, there exists reasons, why the channels, either because there are some carriers where one channel is better than the other and vice versa, or because on some channel realizations one user receives more or less interference than the other, or because the channel state is not perfectly known at the transmitter and therefore, the encoder cannot determine the optimal pre-coding order. In practice, we do not have perfect CSIT. Therefore, we will always have to operate on non-degraded SISO BCs.

\par In uplink (MAC), RSMA achieves every point at the boundary of the capacity region without time sharing \cite{Rimo1996}. As illustrated in Fig. \ref{fig:uplinkSE}, RSMA can also enhance the spectral efficiency when proportional user fairness is considered \cite{zhaohui2020ULRS}.
\par In multi-cell (IC), RSMA outperforms SC and SIC in the weak interference regime. It enables an enhanced  interference management of the intra-cell and inter-cell interference as the amount of intra/inter-cell interference to be canceled at the transmitters or decoded at the receivers can be flexibly adjusted. RSMA therefore achieves improved spectral  and energy efficiency over the conventional coordinated schemes without RSMA \cite{Jia2020SEEEtradeoff,wonjae2020multicell}. 
\par In relaying and cooperative systems, by enabling the users with strong channel strength to relay and forward the common stream to the users with weaker channel strength, not only can cooperative RSMA improves spectral and energy efficiency \cite{mao2019maxmin,CRSEE2022TVT}, but also offering substantial benefits in terms of coverage extension  \cite{Maosurvey}. 

\par \textbf{Myth 4: The common stream in RSMA is a multicast stream required for multiple users}

\par A common stream in RSMA is multicasted at the physical layer since it is to be decoded by multiple users. However the content of the common stream is not necessarily intended to those users. This is a difference from multicasting and broadcasting where a message is genuinely intended to multiple users, and therefore decoded by multiple users. In RSMA, the common stream is created for interference management purpose, not because the content of the common stream is intended to multiple users.
\par It is also possible to do multicasting on top of RSMA. This is the case where $K$ users want to receive unicast messages $W_1,\ldots,W_K$ (one for each user), but additionally a multicast message $W_0$ is transmitted and intended to all $K$ users. In that case, RSMA can be used to encode in a common stream the multicast message $W_0$ along with parts of the unicast messages $W_1,\ldots,W_K$, as discussed in NOUM subsection \ref{NOUM_subsection} \cite{mao2019TCOM,mao2020DPCNOUM}.
\par Recall that NOMA also has a common message/stream, though commonly not denoted using such terminology in the NOMA literature. Hence, the common message is not a message that is originally
intended for all users. It is required to be decoded by all users but is not necessarily intended for all users. 

\par \textbf{Myth 5:	As the common stream needs to be decoded by multiple users, it causes privacy/security issues}

Note that decoding the common stream at the physical layer does not imply the sharing of data as encryption is commonly implemented at higher layers and decryption is performed using user-specific codes. Same would go for other schemes relying on SIC and interference decoding such as NOMA. There is therefore no privacy/security issues as long as higher layer encryption is performed. However, from a physical layer security/secrecy perspective \cite{Wyner1975,Bloch2008}, the problem is different and RSMA can be designed to maximize the secrecy rate as discussed in Subsection \ref{PHY_layer_security}.

\par \textbf{Myth 6:	The message of each user is required to be split into one or multiple common parts and a private part}

\par In RSMA, the message of each user could be split into one or multiple common parts and a private part, but is not always required to be split. Whether the message of a single user or the messages of multiple users are split depends on the objective function. For instance in Example \ref{example_RS}, both messages $W_1$ and $W_2$ are split, but it could happen that only $W_1$ is split or only $W_2$ is split.

\par Splitting the message of a single user (as in \cite{RS2016hamdi}) or more users (as in \cite{mao2017rate}) at the transmitter has no impact on the performance if the objective is to maximize the WSR or EE (defined by sum rate dividing the sum transmit power) subject to transmit power constraint. In such case, the major question is whether RS is helpful or not, and how much of the total (sum) information should be carried by the common message regardless of how the common message is split.

\par However, if more user fairness is considered, i.e., when the objective is to maximize the minimum rate among users or/and subject to QoS rate constraints for each user, the choice of which users to split the messages at the transmitter will influence the final performance. The best method is to leave the possibility to split the messages of all users so as to provide rooms for allocating the rate of the common stream among users. 

\par \textbf{Myth 7:	RSMA has to sacrifice a higher receiver complexity in order to outperform (power-domain) NOMA}

Because RSMA does not enforce a given stream to be fully decoded or to be fully treated as noise, but rather split one or multiple messages such that a message is partially decoded by another users, RSMA can explore a wider space of communication schemes. This consequently leads to relatively simple schemes like 1-layer RS that relies on a single SIC to outperform multi-SIC NOMA schemes. It was for instance shown in \cite{bruno2021MISONOMA} how the DoF, and therefore rate, of 1-layer RS can be significantly larger than that of complicated NOMA schemes. Also RSMA builds upon SDMA/MU-MIMO and the addition of SIC layers comes with a performance enhancement. This contrasts with NOMA where the performance (DoF and rate) can degrade as we increase the number of SIC, as a consequence of the restrictive design philosophy of NOMA \cite{bruno2021MISONOMA}. The above discussion is further illustrated in Fig. \ref{fig:flexible}.

\par \textbf{Myth 8:	With more data streams to send from the transmitter, beamforming design and power control become very complicated in RSMA}

For any MA scheme, the larger the number of users and streams to serve, the higher the complexity for beamforming and power control. Hence this is not an issue specific to RSMA, but would hold for SDMA, NOMA, etc. Nevertheless, with RSMA, low complexity beamforming and power control can be perofrmed. For instance, in 1-layer RS, private streams could be precoded using zero-forcing beamforming (ZFBF) and have the power allocated uniformly (that would be much compliant with the way MU-MIMO is implemented in practice in 4G and 5G). The precoder of the common streams can be designed using low complexity techniques \cite{Maosurvey,Minbo2016MassiveMIMO,onur2021mobility,bruno2019wcl}. What remains to be designed is the power allocated to the common stream that could depend on the network load, propagation conditions, and metric \cite{bruno2019wcl,onur2021mobility,Minbo2016MassiveMIMO}.

\par \textbf{Myth 9: RSMA requires high SNR to achieve performance gain over NOMA or SDMA}

\begin{figure}[t!]
	\centering
		\includegraphics[width=0.4\textwidth]{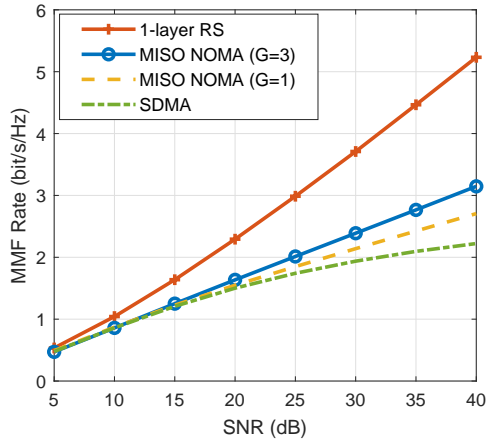}%
		\caption{Max-min fairness rate vs. SNR comparison in MISO BC with imperfect CSIT, $\alpha=0.5$, $M=5$, $K=6$, $\sigma_k^2=1, \forall k\in\mathcal{K}$. \cite{bruno2021MISONOMA}.}
	\label{fig:rateGainlowSNR}
\end{figure}

The gain of RSMA over SDMA and NOMA depends on many parameters, including network load, propagation conditions, objective function and QoS constraints, and CSIT quality. The more we favour fairness (e.g., MMF objective function, WSR, QoS constraints), the higher the gain of RSMA over a wide range of SNR. The lowest gain of RSMA would be experienced in massive MIMO regime with sum-rate maximization and accurate CSIT. In such scenario, RSMA would boil down to SDMA or allocate a very small amount of power to the common streams only at high SNR.  
We here illustrate a MMF rate comparison result when the network is overloaded and the CSIT is imperfect, as per Fig. \ref{fig:rateGainlowSNR}.  The relative rate gain of 1-layer RS over MISO NOMA ($G=3$) is 35\% when SNR is 15 dB, and 21\% when SNR is 10 dB, which is still non-negligible. Importantly, recall that this gain is achieved with a single SIC layer, while NOMA ($G=1$) and NOMA ($G=3$) require 5 and 1 SIC layers, respectively, and achieve much worse MMF rate.
\par \textbf{Myth 10:	RSMA only works when instantaneous channel state information is available at the transmitter}

RSMA can operate on any form of CSIT, such as instantaneous CSIT with high or low accuracy and various CSI acquisition mechanisms \cite{RS2016hamdi,anup2022_massive,anup2022_cfmassive}, delayed CSIT \cite{onur2021mobility}, or statistical CSIT based on for instance second order statistics of the channel, e.g., spatial covariance matrix \cite{minbo2017mmWave,Minbo2016MassiveMIMO,longfei2021wcnc}. 
A detailed summary of the imperfect CSIT models that have been adopted in the existing works of RSMA can be found in \cite{Maosurvey}.

\par \textbf{Myth 11: The gain of RSMA in massive MIMO is marginal}

The gain of RSMA in massive MIMO with perfect CSIT is negligible or inexistent since the transmitter can form pencil beams that would provide high beamforming gain and simultaneously eliminate multi-user interference. Hence private streams are sufficient and RSMA boils down to conventional SDMA-based massive MIMO.
\par As we depart from perfect CSIT and considers practical imperfection in CSIT acquisition, RSMA starts providing gains. In \cite{onur2021mobility}, RSMA was shown to provide significant gains in massive MIMO and to maintain multi-user connectivity in mobility conditions despite the delayed CSIT. In \cite{minbo2017mmWave,Minbo2016MassiveMIMO}, RSMA was applied to resolve the imperfect CSIT problem in FDD massive MIMO and was shown to provide performance gain over conventional massive MIMO based on SDMA processing. In \cite{anup2022_massive,anup2022_cfmassive}, RSMA was investigated in TDD massive MIMO and cell-free massive MIMO and was shown to be robust to pilot contamination and to provide significant gains whenever there is a likelihood of users being allocated the same uplink pilot sequence. This is motivated by mMTC where the probability of multiple users sharing the same pilot is very high. In \cite{AP2017bruno}, RSMA was used to enhance the robustness of massive MIMO in the presence of hardware impairments and in particular phase noise. RSMA was shown to outperform conventional massive MIMO in the presence of phase noise. In \cite{Asos2018MultiRelay}, RSMA was finally shown to be robust in multi-pair massive MIMO relay systems.

\section{Frequently Asked Questions}\label{section_asked_questions}

We here classify frequently asked questions about RSMA in three different categories, namely principles and benefits, standardization and implementation, and applications and interplay with other technologies.

\subsection{Principles and Benefits of RSMA}

\par \textbf{Question 1: What is the design principle of RSMA?}

In the downlink, the design principle of RSMA is much different to SDMA and NOMA. 

The NOMA design philosophy is based on having a stream to be fully decoded by another user. For instance, in the MISO case, NOMA forces one user to fully decode all streams in a group, i.e., its intended stream and the co-scheduled streams in the group. This leads to the strong constraint that the entire message of one of the users is mapped onto a common stream, e.g., $W_2$ mapped to $s_{\mathrm{c}}$ decoded by both user-1 and user-2 in Table \ref{fig_mapping_BC}.

This is radically different from SDMA design philosophy where messages are independently encoded into private streams and each receiver decodes its intended stream treating any residual multi-user interference as noise (even when the interference level is not weak enough to be treated as noise), as per Table \ref{fig_mapping_BC}. Interference is never decoded at the receivers due to the absence of common stream(s). 

In RSMA, a message of a given user is not forced to be treated as noise or be decoded by a user. Instead we have full flexibility on how to encode it and can evolve in the grey zone in-between. Hence, similarly to SDMA, in MISO downlink, $K$ private streams are enabled, but in contrast to SDMA, each user can partially decode the message of another user thanks to the presence of the common streams. In contrast to NOMA, no user is forced to fully decode the stream(s) of a co-scheduled user since all private streams are encoded independently and each receiver decodes its intended private stream treating any residual interference from the other private streams as noise. 

This comes with huge benefits as RSMA builds non-orthogonal transmission strategies upon SDMA (and therefore MU-MIMO) so that the performance benefits of SDMA are guaranteed but extra performance is observed by the use of SIC receivers. Indeed, a performance gain over SDMA is expected from a more complex receiver architecture. To achieve this, one should enable the versatility at the transmitter to encode messages such that parts of them can be decoded by all users using SIC while the remaining parts are decoded by their intended receivers and treated as noise by non-intended receivers. This is uniquely achieved with RSMA by providing flexibility in the message-to-streams mapping, as demonstrated in Tables \ref{fig_mapping_BC},\ref{fig_mapping_coop_BC}, and \ref{fig_mapping_MIMO_BC}. Indeed, instead of keeping a rigid mapping of a message into a predefined stream (as in SDMA and NOMA), RSMA allows each user to send part of a message in one or multiple common stream(s) and the rest in one of the $K$ private streams. By adjusting the power levels of the common and private streams, one can adjust the amount of interference that occurs on the private stream, so that its level can be weak enough to be treated as noise. This enables RSMA to manage multi-user interference by partially decoding the interference and treating the remaining interference as noise. 

In the uplink, the design philosophy of RSMA and NOMA is similar at the receivers since both aim at using SIC to decode all incoming streams. However, for a fixed number of users, the number of streams to decode at the receivers is different because of the specific features in the encoding operation at the transmitters between RSMA and NOMA. NOMA philosophy is to map a message into a stream such that in a two-user system, NOMA decodes user-1 before or after user-2. On the other hand, RSMA splits a message (say of user-1) into sub-messages, encodes them into independent streams, and superposes them. This split enables the receiver to decode part of user-1's message before user-2 and the other part of user-1's message after user-2.

\par \textbf{Question 2: What are the major benefits of RSMA?} 

\par \textit{Universal}: RSMA is a more general multiple access framework that unifies and generalizes (and consequently outperforms) OMA, SDMA (and multi-antenna) NOMA. OMA, SDMA, and NOMA, (and other schemes as physical-layer multicasting) are all particular instances of RSMA \cite{mao2017rate,bruno2019wcl}.
\par \textit{Flexible}: RSMA is flexible to cope with all user deployments (with a diversity of channel directions, channel strengths), network loads (underloaded and overloaded regimes), and interference levels (weak, medium, strong). RSMA automatically reduces to other MA techniques according to the channel conditions, i.e., it reduces to SDMA when user channels are orthogonal in the underloaded MISO BC with perfect CSIT. When the channels are aligned with certain channel strength disparities, it reduces to power-domain NOMA. For other channel conditions, RSMA takes advantage of the common streams and more efficiently manages multi-user interference by partially decoding the interference and partially treating the remaining interference as noise \cite{mao2017rate,bruno2019wcl}.
\par \textit{Robust}: RSMA is robust to any CSIT inaccuracy \cite{RSintro16bruno, RS2016hamdi}. This is very relevant in modern downlink multi-antenna deployments. While OMA, NOMA, SDMA, all incur a DoF loss in the presence of imperfect CSIT, RSMA is DoF-optimal and therefore, less sensitive to CSIT inaccuracy.
\par \textit{Spectrally efficient}: The spectral efficiency of RSMA is always larger than or equal to that of existing MA techniques. 
Considering a downlink without QoS constraints, the rate region of RSMA comes much closer to the capacity region (achieved by DPC) than SDMA and NOMA when CSIT is perfect \cite{mao2017rate}. When CSIT becomes imperfect CSIT, linearly precoded RSMA is able to achieve a larger rate region than complex DPC (and SDMA and NOMA) \cite{mao2019beyondDPC}. As RSMA achieves the optimal DoF in both perfect and imperfect CSIT \cite{enrico2017bruno}, it optimally exploits the spatial dimensions and the availability of CSIT. This contrasts with SDMA and NOMA that are suboptimal \cite{bruno2021MISONOMA}. Considering an uplink, RSMA outperforms NOMA without time sharing \cite{Rimo1996,Jiawei2022}. 
\par \textit{Energy efficient}: Thanks to its flexibility and universality, the EE of RSMA is also larger than or equal to that of existing MA techniques (OMA, SDMA, NOMA) in a wide range of user deployments \cite{mao2018EE,mao2019TCOM,Gui2020EESEtradeoff}.
\par \textit{Enhancing QoS and fairness}: RSMA exhibits an even larger performance gain over other MA techniques whenever each user is subject to a QoS rate constraint or whenever a higher weight is allocated to the user with a weaker channel condition \cite{mao2017rate,bruno2021MISONOMA,Maosurvey}. Therefore, the ability of a wireless network architecture to partially decode interference and partially treat interference as noise leads to enhanced QoS and user fairness. 
\par \textit{Reducing complexity}: RSMA has the double benefit of simultaneously boosting the performance and decreasing the complexity (at the transmitter and the receiver) compared with multi-antenna NOMA. Recall indeed that multi-antenna NOMA that requires user grouping, ordering, switching (between NOMA and SDMA) at the transmit scheduler and multiple layers of SIC at the receivers. On the other hand, 1-layer RS without any user ordering, grouping or dynamic switching at the transmit scheduler and with a single layer of SIC at each receiver is capable of achieving significant performance gain over NOMA \cite{bruno2021MISONOMA}. In contrast to SDMA that requires user pairing to pair users with semi-orthogonal channels, RSMA is suited to all channel conditions and it does not require complex user scheduling and pairing \cite{Minbo2016MassiveMIMO}. Moreover, RSMA is capable of further reducing CSI feedback overhead in the presence of quantized feedback \cite{chenxi2015finitefeedback}. RSMA nevertheless incurs a higher receiver complexity than SDMA due to the use of SIC.

\par \textit{Reducing latency and improving reliability}: 
Reducing the transmit packet size is widely known as one major approach of achieving low-latency communications, also known as, short-packet transmission \cite{shortPacket2016Petar}.
RSMA for short packet transmission has been investigated in \cite{yunnuo2021FBL,yunnuo2022FBL,Jiawei2022} for both downlink and uplink, and RSMA has been shown to use shorter blocklength and therefore lower latency to achieve equal sum rate and MMF rate with SDMA and NOMA.
It therefore has a great potential to enhance URLLC services in 6G \cite{onur2020sixG}.

\par \textbf{Question 3: What does the common message contain?}
From Example \ref{example_RS}, we note that the common message contains information bits from the original unicast (user-specific) messages of user-1 and user-2. In that example, both messages $W_1$ and $W_2$ are split to create the common message; hence the common message carries some bits from user-1 and from user-2. But we could have instances where only one of the users message is split, say $W_1$ only, and in such case, the common message carries some bits of user-1 only. The common message, whether it carries bits of one user or multiple users, is always decoded by multiple users.

\par \textbf{Question 4: Why do we combine the common parts of user messages into a single common message?}

Transmitter and receiver design is greatly simplified by combining the common parts of multiple users into a single common message, as shown with the two RSMA strategies of the 2-user downlink in Section \ref{MISO_RSMA_2user}. Moving to a $K$-user scenario, if each user's message is split into two parts without combining the messages, the sender would have to encode the $2K$ message and design $2K$ precoders. Each user needs more SIC layers and the decoding order needs to be optimized at the transmitter. In contrast, the 1-layer RS transmitter only encodes and precodes $K+1$ streams, and only one layer of SIC is needed for each user, hence no worry about decoding order. This greatly reduces the complexity of 1-layer RS.

\par \textbf{Question 5: How does the optimized rate allocation for the common stream guide the practical message split at the transmitter and the modulation and coding scheme?}

The reader is invited to consult \cite{Maosurvey} for detailed examples of practical message splits and modulation and coding schemes.

\par \textbf{Question 6: In what scenarios can RSMA achieve explicit performance gain over SDMA and NOMA?}

Let us start with the downlink. From a DoF perspective, RSMA achieves the same or higher DoF than SDMA and than NOMA with a lower number of SIC layers \cite{bruno2021MISONOMA}. This means that in the high SNR regime, RSMA will always outperform those two schemes and the gains would be larger as the objective functions accounts for fairness, the system gets more overloaded, or the quality of the CSIT degrades. A DoF gain also translates to a rate gain at finite SNR though the exact gain depends on the disparity of channel strengths among users and the angle between user channels.

In the low to medium SNR regime, the gains of RSMA over SDMA and NOMA will depend on the user channel orthogonality and the disparity of channel strength, as shown for a 2-user scenario in Fig. \ref{fig:universal}. SDMA favors orthogonal channels, accurate CSIT and similar channel strengths among users. NOMA favours aligned channels in each group and a large disparity of channel strengths. As we depart from those extremes, or as the CSIT quality degrades, RSMA provides explicit gains over SDMA and NOMA even more when constraints on fairness, QoS, or minimum rate are  imposed by the network \cite{bruno2019wcl,mao2017rate,Maosurvey}. 

RSMA was shown to reduce to SDMA in the presence of orthogonal channels (or close to orthogonal) and outperform SDMA otherwise. RSMA was shown to reduce to NOMA and achieve the same rate performance whenever: 1) the SNR is low, 2) the channels are closely aligned, 3) there is a sufficiently large disparity of channel gains, and 4) the CSIT is perfect. If all four conditions are met, NOMA, RSMA, and DPC schemes achieve very similar performance (if not the same performance). As we depart from those conditions, NOMA incurs a loss over RSMA. Same observations also hold in more general $K$-user settings \cite{mao2019beyondDPC}.

When it comes to the role played by the channel gain disparity among users, it is important to note that with RSMA, the larger the disparity of channel gains, the larger the benefits of using RSMA schemes with multiple SIC. In other words, given a cell size, the disparity of channel strengths among users could be statistically obtained using the conventional path loss model, and the designer can then decide how many SIC would be worth given the complexity that can be afforded. Nevertheless, 1-layer RSMA already brings significant gain even for realistic channel strength disparities. For instance, in Fig. \ref{fig:flexible}, with 10dB path loss difference and additional Rayleigh fading, 1-layer RS with a single layer of SIC outperforms NOMA with five SIC layers. Further gains could be obtained by using RSMA scheme with say two SIC, but this shows how powerful, RSMA is to efficiently make use of the SIC architecture, and therefore reduce receiver complexity \cite{bruno2021MISONOMA}.  

In the uplink, NOMA is heavily dependent on time sharing to achieve good performance and attain capacity. The gain of RSMA over NOMA is explicit whenever we cannot afford time sharing among users. RSMA achieves the capacity without the use of time sharing, which finds applications in scenarios where communication overhead
and stringent synchronization requirements due to the
coordination of the transmissions of all users is not possible \cite{Rimo1996}. This occurs for instance in services requiring grant-free access, which allow collisions to reduce the access latency stemming from the channel grant procedure \cite{Maosurvey,Hongwu2020ULRS,Hongwu2021}. Other uplink scenarios where RSMA outperforms NOMA is in uplink with finite blocklength \cite{Jiawei2022} or in network slicing \cite{Santos2021}.

\par The reader is also invited to consult the many applications and scenarios discussed in Section \ref{section_applications} where references have demonstrated that RSMA outperforms SDMA and NOMA. 

\par \textbf{Question 7: What can we learn from information theory about RSMA? How can information theory guide the modern study of RSMA?}

As re-visited in Section \ref{section_interference}, the roots of RS can be dated back to \cite{carleial1978RS,TeHan1981}, where the coding scheme was introduced for the SISO IC in the weak interference regime. At that time, the MAC capacity region was characterized completely by separate random coding and SIC at the receiver. The capacity region of the degraded BC was also known to be achieved by SC and SIC. Initial results on the IC showed that capacity region can be achieved by the same techniques in the case of very strong \cite{carleial1975} and strong \cite{Sato1981} interference. The basic idea to develop a coding scheme which allows the receives to decode part of the interference, to bridge the two extreme cases treat interference as noise (TIN) and decode interference, led to the development of RS. A very important ingredient in the development of the best achievable rate region for the IC is time sharing \cite{Sason2004}. Due to the remaining interference, the achievable rate region is non-convex. Only time sharing between different coding and decoding strategies including TIN, FDMA, and RS can guarantee the best achievable rate region. It must be stressed that there is a gap to the outer bound of the capacity of IC with weak interference. A characterization of the capacity region within a finite number of bits is derived in \cite{Tse2008}. The achievable scheme in this work is based on a simplified RS approach. 

In the BC and IC, the RS approach was introduced in recent works in the context for multi-antenna settings (MISO and MIMO) with imperfect CSIT . Though the Gaussian MIMO BC with perfect CSIT is known and achieved by DPC \cite{capacityRegion2006HW}, the capacity and capacity-achieving schemes of those channels with imperfect CSIT are still unknown, but RS is known to play a crucial role in achieving optimality in DoF \cite{RS2016hamdi,enrico2017bruno,chenxi2017bruno,DoF2013SYang,hamdi2019spawc,Davoodi2021DoF}, generalized DoF \cite{AG2017Gdof,Davoodi_gdof_2018,Davoodi_gdof_2020}, and constant gap \cite{SYang2018SPAWC,yang2018itw}.

In the MAC, the RS approach was first introduced in \cite{Rimo1996}. The motivation behind this development was to apply single-user coding without requiring synchronization among users. There the term RSMA was coined. Already in \cite{Rimo1996}, the important cases with fading and interference were considered to bring the proposed coding and decoding scheme to practical applications. 

In multihop-multiflow communication, e.g., in a $2\times2\times2$ setting comprised of 2 sources, 2 relays, and 2 destinations, the combination of RS
with decode-and-forward and amplify-and-forward schemes plays a crucial role in achieving the fundamental limits \cite{Wang_2022}.

RS is also a crucial ingredient in the theoretical and practical principles of the broadcast approach to communication over state-dependent channels and networks \cite{Tajer_2021}. This is relevant in scenarios where the transmitters have access to only the probabilistic description of the time-varying states while remaining unaware of their instantaneous realizations \cite{Tajer_2021}.

Later the term multiple access was used also for scenarios in which several users share a link including BC. The term (power-domain) NOMA was introduced mainly for the downlink transmission corresponding to BC setup. It corresponds to SC SIC \cite{Vaezi2019} and achieves the capacity region in degraded BC \cite{Tcover1972}. The term RSMA was also applied to the BC in \cite{mao2017rate} where it can provide significant achievable rate gains. 

Network information theory \cite{EL2011networkIT} provides the solid basis for system modeling, a taxonomy of known results with achievable encoding and decoding schemes with unique taxonomy, and a toolbox of methods and schemes for the modern study of multi-user communication systems. All currently standardized and developed transceiver schemes have their roots in the fundamental information theoretic results - even if the name and the terminology might have changed. 

\subsection{Standardization and Implementation of RSMA}

\par \textbf{Question 8: What is the status of RSMA standardization? Why would RSMA succeed in 6G when NOMA was not well received in 5G?}

RSMA is still very new in 3GPP and has not been discussed by standard bodies yet. The machinery required for RSMA is nevertheless already partially being studied, discussed and developed in 3GPP. Indeed, past 3GPP study/work items such as MU-MIMO, full-dimensional MIMO, coordinated multi-point (CoMP), multi-user superposition transmission (MUST), NOMA, network-assisted interference cancellation and suppression (NAICS), multicast and broadcast services, can be leveraged to design RSMA. However the key novelty of RSMA, namely relying on message split, has not been discussed in standardization bodies. Inversely, RSMA, once introduced in the standard, would address numerous issues and therefore boost the performance of all those work items.  

NOMA was heavily investigated in 5G but not so well received at the end. From a theoretical point of view, it is predictible that (power-domain) NOMA would not fly given its deficiencies and the lack of clear gains (and even loss) over MU-MIMO \cite{bruno2021MISONOMA,Behrooz2020}. In contrast, RSMA does not suffer from those issues as it really builds upon SDMA/MU-MIMO. Hence 6G could envision a single transmission mode based on RSMA as a replacement or an enhancement of the MU-MIMO transmission mode used in 4G and 5G, but also play numerous other roles in the entire air interface, such as enabling efficient simultaneous unicast-multicast-broadcast transmissions, with numerous new applications in automotive driving, VR, 360 video, metaverse, etc.  

\par \textbf{Question 9: Does RSMA create more complications for implementation?}
\par The 1-layer RS strategy and its benefits in terms of implementation and complexity over SDMA and NOMA have been already discussed above. There are nevertheless other challenges to overcome to make RSMA practical. In most of the RSMA works, its transmitter-side design assumes Gaussian inputs, and it can be tricky to fit a Gaussian-optimized RSMA-based precoder into a real physical layer, where it often deals with the finite blocklength and finite constellations, or with pre-standardized MCS. Interestingly, recent efforts have been made to make the 1-layer RS strategy work with the state-of-the-art channel codes and modulation and significant throughput performance have been observed using realistic link-level simulations \cite{Onur2020LLS,hongzhi2020LLS,longfei2021LLS,anup2021MIMO,JRCLoli2022,DLloli2022}, though more work is needed in this area. 

\par Moving to other RSMA schemes, like the generalized RS strategy \cite{mao2017rate}, can be complex to implement especially for a large number of users. Nevertheless, as a generalized framework of RSMA, it embraces SDMA, NOMA, physical-layer multicasting, OMA as special cases, and suggests a novel method to softly bridge existing MA techniques without using naive hard switching.

Moreover, the generalized RS is applicable to the scenarios with relatively small $K$ and it achieves non-negligible performance gain over existing MA techniques. Hence, the transmitter could schedule a small number of users in each resource block.

Another use of the generalized RS is to act as a benchmark to demonstrate the performance of other low-complexity RS strategies such as 1-layer RS and 2-layer HRS. It enables to identify which common streams are effective or ineffective, and consequently trim the generalized RS into a low complexity RS scheme that would rely on a subset of the common streams. From our experience, in many applications, low-complexity RS strategies achieve performance reasonably close to that of generalized RS while their complexities are much lower than the generalized RS and NOMA strategies. This demonstrates that by departing from the extremes of treating interference as noise and fully decoding interference, one can find alternative MA strategies that are spectrally and energy efficient and simultaneously computationally efficient (relatively low complexity and small number of SIC layers). It helps us draw the conclusion that 1-layer RS is a good alternative to the generalized RS in many practical scenarios.

Another benefit of generalized RS is the ability to come up with a set of schemes whose performance improve as the number of SIC layers increases. This is helpful to figure out the performance gap vs complexity tradeoff between low-complexity RS strategies and generalized RS and understand whether the addition of one or multiple common streams is worth the complexity increase. This contrasts with NOMA where a larger number of SIC layers can lead to a lower performance \cite{bruno2021MISONOMA}. 

Therefore, the generalized RS is a significant strategy in the framework of RSMA.

\subsection{Applications and Interplay between RSMA and other Wireless Technologies}

\par \textbf{Question 10: Can RSMA be integrated with other MA techniques such as OFDMA, SDMA, power-domain NOMA, and code-domain NOMA?}

RSMA can definitely be combined with OFDMA in the same way as it is done with SDMA/MU-MIMO in 4G and 5G, namely a group of users are paired and served using RSMA on a given resource block or subband. Though much remains to be investigated in OFDMA-RSMA design, some research on the user grouping and power allocation optimization has been initiated in \cite{hongzhi2020_TVT,hongzhi2021RSLDPC}.

Interplay between SDMA or power-domain NOMA and RSMA would not bring benefits, since SDMA is always part of any RSMA scheme, namely when it comes to the transmission and reception of the private streams, and power-domain NOMA is part of the generalized RS architecture of RSMA.

On the other hand, there is no effort so far on the interplay between code-domain NOMA and RSMA and this is an area of interest to see whether we can further enhance RSMA performance by bringing the code-domain dimension. Code-domain NOMA (CD-NOMA) employs carefully designed interleavers and/or code sequences to multiplex users. The idea was inspired by the traditional CDMA \cite{Proakis08} or interleaver-division multiple access (IDMA) \cite{idmaliping}. Some well-known examples of CD-NOMA include sparse code multiple access (SCMA) \cite{Nikopour13} and non-orthogonal coded access (NOCA) \cite{R1165019}. In SCMA, each user is assigned a sparse codeword according to its message. To exploit the sparsity introduced by the codeword, the receiver adopts message passing algorithm for multi-user detection. In contrast, NOCA assigns each user a dense spreading signature to fully utilize the available time-frequency resources. The receiver exploits the low cross-correlation properties of spreading sequences for interference mitigation/suppression. Nevertheless, it is possible to incorporate the design principle of CD-NOMA into RSMA. Notice that the split common and private messages in RSMA can be seen as virtual users. Each of the virtual user can be assigned a dedicated sequence for enabling code-domain multiplexing. Some interesting research directions on code-domain RSMA can include but not limited to: design of sparse/dense codewords, detection and decoding architectures, resource allocation for optimizing achievable rate and energy efficiency.

It should also be noted that the RSMA framework can be expanded in the time or frequency domains to get a space-time or space-frequency RSMA framework as in \cite{chenxi2015finitefeedback,chenxi2017bruno}, as discussed in Section \ref{ST_RSMA_section}. This is particularly relevant when the CSIT quality changes across users and time or frequency or when we deal with asymmetric downlink or multi-cell framework where the receivers have a different number of receive antennas.

\par \textbf{Question 11: Can RSMA be integrated with emerging waveform, e.g., orthogonal time frequency space (OTFS), orthogonal delay-Doppler division multiplexing (ODDM)?}

OTFS was recently proposed as a new two-dimensional modulation scheme \cite{Hadani17} \cite{hadani2018orthogonal}, which multiplexes information symbols in the delay-Doppler (DD) plane (or domain) rather than the time-frequency (TF) domain as for conventional multi-carrier modulation or OFDM schemes in the current 4G/5G cellular and WiFi networks. The DD domain symbol multiplexing enables a direct coupling of the transmitted symbols with the channel's delay-Doppler spread function, which has nice properties, such as quasi-static, compact, and sparse, to be exploited to achieve a low channel estimation overhead and full channel diversity with low complexity receivers \cite{pilot} \cite{2010.03344} \cite{8424569} \cite{thaj2020low} \cite{otfsmmseicc22}. 
OTFS has also stimulated additional research on delay-Doppler plane modulation with orthogonal pulses, such as orthogonal delay-Doppler division multiplexing (ODDM) \cite{oddm} \cite{oddmicc22}, which can achieve orthogonality with respect to the channels' delay and Doppler resolutions that are generally much smaller than the symbol duration and subcarrier spacing in conventional OFDM.

ODDM, or general delay-Doppler plane multi-carrier modulation, is a promising waveform for future wireless systems, particularly on doubly-selective channels. This type of schemes not only provides robust performance in high-mobility channels, they can also be a viable choice for the future ISAC, due to that their transmit/receive pulses and the corresponding ambiguity function satisfy the orthogonality property with respect to the delay-Doppler resolutions. Similar to OFDMA, ODDM itself can be employed as an orthogonal multiple access scheme. 

With this property, it is natural to have RSMA combined with OTFS/ODDM waveform in MISO or MIMO systems. The combination of RSMA and OTFS/ODDM can provide high system design flexibility in terms of its resource allocation and optimization, in the mean time having a potential to offering high spectral efficiency, signal localisation, integrated sensing and communication capabilities.  

\par \textbf{Question 12: Can RSMA use discrete signaling without SIC?}

To analyze the performance of RSMA, such as achievable rate and energy efficiency, Gaussian signaling is often assumed. As we know, Gaussian signaling is the optimal signaling in many channels, e.g., point-to-point Gaussian channels, Gaussian multiple access channels, and Gaussian broadcast channels. Hence, assuming Gaussian signaling becomes natural in RSMA. Under this assumption, RSMA can handle interference with different strengths effectively.

In practical communication systems, discrete signaling, i.e., coded modulation, is used. In addition, for some applications with stringent requirements on decoding complexity and latency, e.g., downlink URLLC services, TIN is more favorable than interference decoding. If Gaussian signaling is assumed, low-complexity TIN is only optimal when the interference is very weak, e.g., see Table \ref{tab:lessonFromIT}. Indeed, despite being the optimal input distribution for many channels, Gaussian signaling is also the worst noise. On the other hand, discrete signaling can behave differently from Gaussian signaling when being treated as noise. When the interference strength is not very weak, it is possible for discrete signaling with TIN to achieve a strictly larger achievable rate region than that for Gaussian signaling with TIN.

Discrete signaling with TIN, i.e., without SIC, has been investigated in recent works, e.g., \cite{8291591,8517129,8731926,9535131}. In \cite{8291591}, a lattice partition based NOMA scheme was proposed for the single antenna $K$-user Gaussian broadcast channel. It was rigorously proved that the scheme based on discrete signaling and TIN is capable of achieving the whole capacity region to within a constant gap independent of the number of users and channel parameters. The same results hold true for the aforementioned channel model with only statistical CSI available at the transmitter side \cite{8517129}. By further exploiting the algebraic properties of lattices, the full diversity of the broadcast channel with block fading and close-to-perfect SIC error performance can be attained for each user with TIN decoding \cite{8731926}. Finally, in \cite{9535131}, it was shown in the first time that the capacity region of the Gaussian (asymmetric) interference channel can be achieved to within a constant gap by discrete signaling and TIN.

To unleash the full potential of RSMA in practical communication systems, it is important to exploit the properties of discrete inputs with low-complexity TIN in RSMA. In light of the above works \cite{8291591,8517129,8731926,9535131}, discrete signaling with TIN should also benefit RSMA to achieve a better tradeoff between performance and complexity.

\par \textbf{Question 13: How can machine/federated learning help RSMA? What is the role of machine/federated learning in RSMA design? }

Machine learning (ML) and federated learning (FL) can be used on multiple fronts. 

\par ML can be used at the receiver of RSMA. In \cite{DLloli2022}, a model-based deep learning (MBDL) method was used to propose new and practical RSMA receiver designs exploiting the conventional SIC receiver and the robustness and model agnosticism of deep learning techniques. Thanks to its ability to generate on demand non-linear symbol detection boundaries
in a pure data-driven manner, the MBDL receiver was shown to significantly outperform conventional SIC receiver with imperfect CSIR. 
\par ML can be used at the transmitter of RSMA to optimize the resource allocation, power
allocation, task offloading (as in MEC), and beamforming design as an alternative to conventional convex optimization methods. In \cite{hieu2021}, the power allocation for each transmit stream was designed using a deep reinforcement learning algorithm. It was shown that RSMA achieves a significant performance gain over SDMA. 
\par Different from the centralized ML method, FL uploads trained model parameters rather than raw data~\cite{li2020federated}. However, when training involves wireless edge devices at the edge network, communications could become a significant problem. In conventional FL, TDMA is typically used for uplink transmission. However, the central cloud has to wait until it receives information from all the user. RSMA enables multiple users to upload information by sharing the uplink channel at the same time. Thus, by incorporating RSMA in the FL framework, it is expected that the aggregation latency can be reduced while maintaining model training quality. For example, multiple IoT devices in fog radio access networks can cooperate to perform a FL task by repeatedly uploading locally updated models to a cloud server. To overcome the performance limitations due to finite capacity front-haul links, a rate-splitting transmission scheme at IoT devices can be used~\cite{park2022completion}. With flexible hybrid edge and cloud decoding strategy achieved by RSMA, we can reduce the completion time of FL while maintaining a specific target global accuracy. 
\par Not only should the PHY layer benefit from the
integration of RSMA and ML, but the focus should also be on cross-layer design such as aiding network orchestration, to truly assist wireless communication for intelligent 6G \cite{CL1_2022}. For instance, ML has been used for UAV deployment in combination with RSMA transmission \cite{Lu_uav_ML_RSMA}. RSMA was shown to require less power compared to other MA schemes.


\par \textbf{Question 14: How would RSMA work in Millimeter-wave and Terahertz networks?}

Early research on millimeter-wave  communication focused on the strong directionality benefits that could be attained using beamforming.  The general idea was to communicate with each user using a narrow spatial beam.  It was thought that the small beamwidths would in turn generate minimal interference between users, which would allow for extremely efficient SDMA designs.

The results of millimeter-wave 5G deployments have begun to change this thinking.  The millimeter-wave benefits to 5G NR has been relatively disappointing, with few devices being scheduled on the large, under-utilized frequency bands.  It is thought that this problem is a result of the beam blockage issues, which appear to be a much bigger problem in a large-scale deployment than was initially thought. 

To make millimeter-wave and higher frequencies practical for future broadband wireless access, it is likely that many of the tight spatial directivity design philosophies will need to change.  Beamwidths may need to increase, and multi-beam communication may become a necessity.  This will likely result in interference becoming the dominant limiting factor as it is in sub-6 GHz frequencies.  

Even in scenarios where blockage is unlikely, the ever-increasing number of antennas will likely not lead to beamwidths predicted by theory.  Ideally, the number of feedback bits should scale linearly with the number of users \cite{au2007performance,ding2007rotation,NJindalMIMO2006,auyeungtrellis}.  From a standardization perspective, this is almost impossible to maintain in the long-term.  This will result in a broadening of beams and a mismatch between the  beamformers/precoders used and the CSI.  This effect will cause millimeter-wave and terahertz systems to have interference challenges.

There has already been some initial work on the application of RSMA at millimeter-wave frequency
\cite{minbo2017mmWave}.  This work focused on the CSI feedback mismatch issue and showed a number of benefits that could simplify implementations in practice. Recently, RSMA has been shown to expand the coverage of terahertz systems \cite{Junil_2022_THz}.

\section{Conclusions}\label{conclusions}
This paper has provided a tutorial on RSMA. To demonstrate how powerful RSMA is, the tutorial has departed from the oversimplisitc OMA vs NOMA discussion held in 5G and has re-centered the discussion and design of multiple access techniques around the key role of interference management. Building upon first principles, the tutorial has demonstrated how existing multiple access techniques are fundamentally limited by their inherent interference management strategy, namely orthogonalization in OMA, treat interference as noise in SDMA, and decode interference in NOMA. In contrast RSMA schemes build upon the RS principle which enables to partially decode interference and partially treat interference as noise. Consequently RSMA has been shown to provide unique benefits, including enhanced spectral, energy and computation \textit{efficiency}; \textit{universality} by unifying and generalizing OMA, SDMA, NOMA, physical-layer multicasting; \textit{flexibility} by coping with any interference levels, network loads, services, traffic, user deployments; \textit{robustness} to inaccurate channel state information (CSI) and \textit{resilience} to mixed-critical quality of service; \textit{reliability} under short channel codes and \textit{low latency}.

\par Future systems will see a growing demand for spectrum intensive applications and for integrated wireless systems. Moreover they will have to face growing concerns for congested and contested electromagnetic environments (in both civilian and defense networks). These will push network designers to adopt sophisticated interference management, multi-user, and multiple access techniques. Thanks to its deep root in information theory, numerous benefits and applications, and its superiority over previous generation multiple access techniques (OMA, SDMA, NOMA), RSMA will play a growing and underpinning role in next generation communication systems. It is hoped that the RSMA techniques presented in this article will help inspiring future research in this exciting new area and pave the way for designing and implementing RSMA in 6G and beyond.

 \bibliographystyle{IEEEtran}
 \bibliography{IEEEabrv,reference}

\begin{IEEEbiography}[{\includegraphics[width=1in,height=1.25in,clip,keepaspectratio]{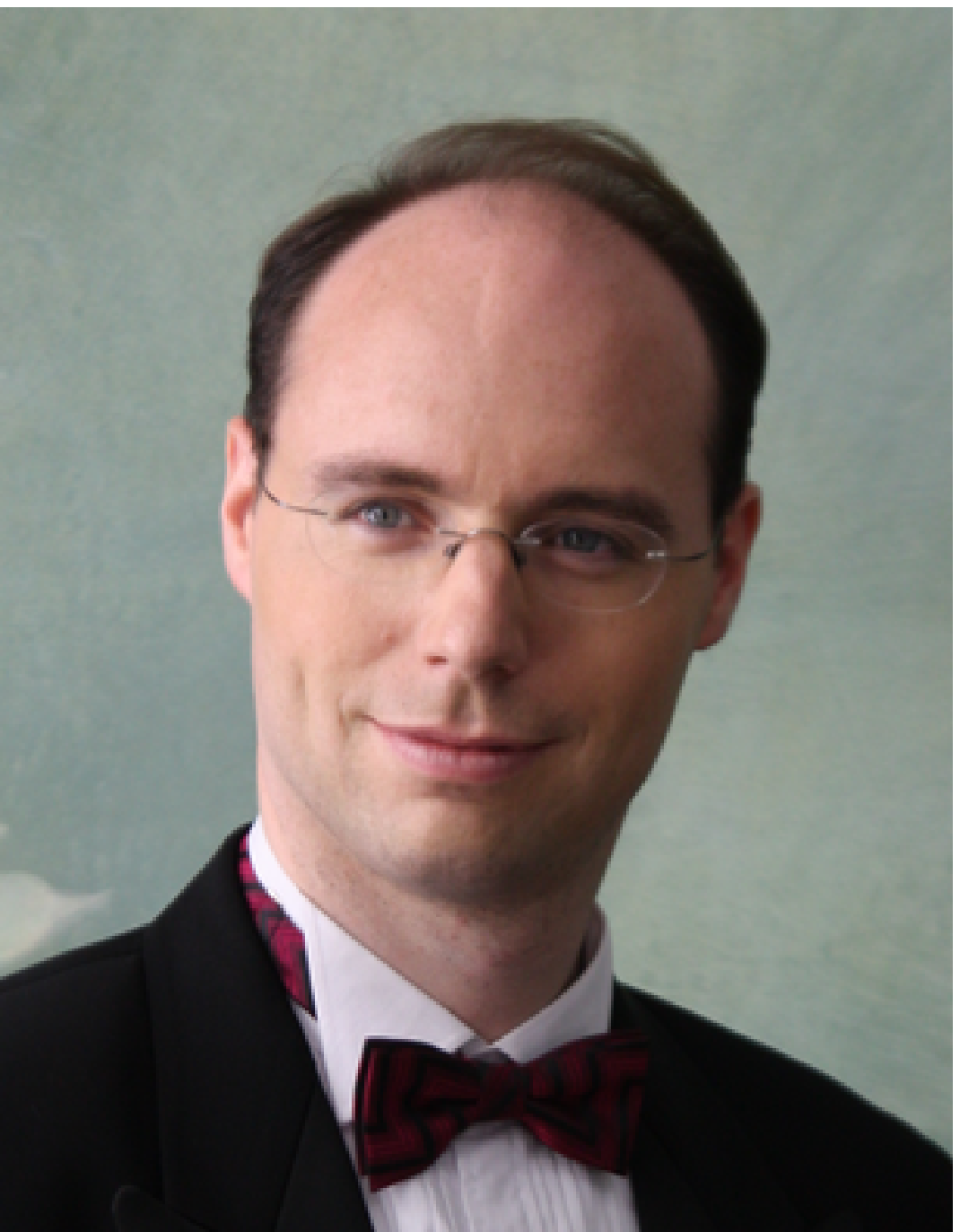}}]
{Bruno Clerckx} is a Professor, the Head of the Wireless Communications and Signal Processing Lab, and the Deputy Head of the Communications and Signal Processing Group, within the Electrical and Electronic Engineering Department, Imperial College London, London, U.K. He is also the Chief Technology Officer (CTO) of Silicon Austria Labs (SAL) where he is responsible for all research areas of Austria's top research center for electronic based systems. He received the MSc and Ph.D. degrees in Electrical Engineering from Université Catholique de Louvain, Belgium, and the Doctor of Science (DSc) degree from Imperial College London, UK. 
He has authored two books on “MIMO Wireless Communications” and “MIMO Wireless Networks”, 250 peer-reviewed international research papers, and 150 standards contributions, and is the inventor of 80 issued or pending patents among which 15 have been adopted in the specifications of 4G standards and are used by billions of devices worldwide. His research spans the general area of wireless communications and signal processing for wireless networks. He served as an editor or guest editor for the IEEE TRANSACTIONS ON COMMUNICATIONS, the IEEE TRANSACTIONS ON WIRELESS COMMUNICATIONS, and the IEEE TRANSACTIONS ON SIGNAL PROCESSING, EURASIP Journal on Wireless Communications and Networking, IEEE ACCESS, the IEEE JOURNAL ON SELECTED AREAS IN COMMUNICATIONS, the IEEE JOURNAL OF SELECTED TOPICS IN SIGNAL PROCESSING, the PROCEEDINGS OF THE IEEE, and the IEEE Open Journal of the Communications Society. He was an Editor for the 3GPP LTE-Advanced Standard Technical Report on CoMP. 
He received the prestigious Blondel Medal 2021 from France for exceptional work contributing to the progress of Science and Electrical and Electronic Industries, the 2021 Adolphe Wetrems Prize in mathematical and physical sciences from Royal Academy of Belgium, multiple awards from Samsung, IEEE best student paper award, and the EURASIP (European Association for Signal Processing) best paper award 2022. He is a Fellow of the IEEE and the IET, and an IEEE Communications Society Distinguished Lecturer 2021-2023.
 
\end{IEEEbiography}

\begin{IEEEbiography}[{\includegraphics[width=1in,height=1.25in,clip,keepaspectratio]{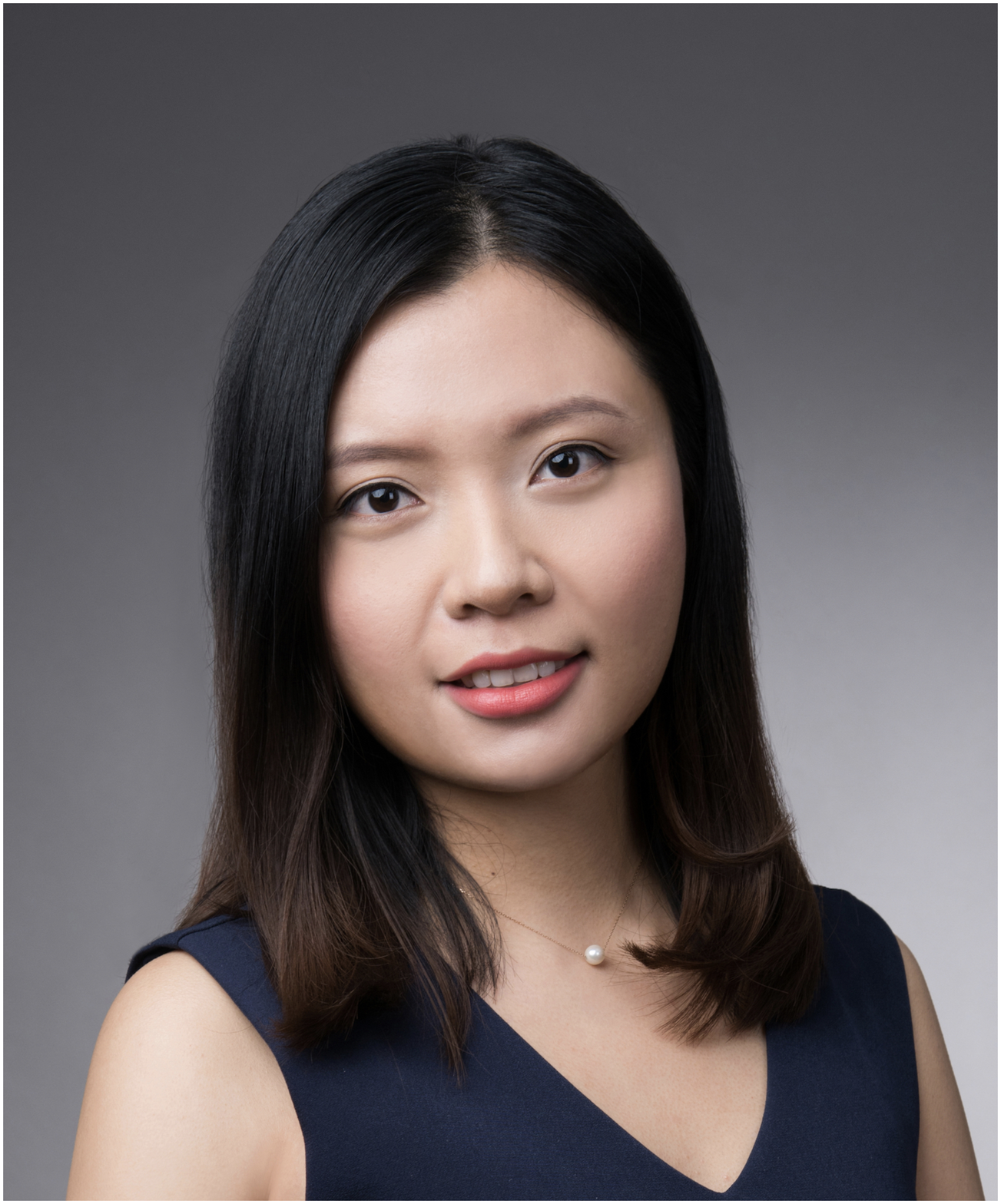}}]
{Yijie Mao} is an Assistant Professor at the School of Information Science and Technology, ShanghaiTech University (Shanghai, China). She received the B.Eng. degree from the Beijing University of Posts and Telecommunications (Beijing, China), the B.Eng. degree (Hons.) from the Queen Mary University of London (London, United Kingdom) in 2014, and the Ph.D. degree from the Electrical and Electronic Engineering Department, the University of Hong Kong (Hong Kong, China) in 2018. She was a Postdoctoral Research Fellow at the University of Hong Kong from 2018 to 2019 and a postdoctoral research associate with the Communications and Signal Processing Group, Department of the Electrical and Electronic Engineering at the Imperial College London (London, United Kingdom) from 2019 to 2021. Her research interests include the design of future wireless communications and artificial intelligence-empowered wireless networks. Dr. Mao receives the Best Paper Award of EURASIP Journal on Wireless Communications and Networking 2022 and the Exemplary Reviewer for IEEE Transactions on Communications 2021. She is currently serving as a guest editor for special issues of IEEE Journal on Selected Areas in Communications and IEEE Open Journal of the Communications Society. She has been a vice-chair of IEEE ComSoc WTC SIG on rate-splitting multiple access (RSMA) and a workshop co-chair for 2020-2022 IEEE ICC, 2021-2022 IEEE WCNC, 2020-2022 IEEE PIMRC, and 2022 IEEE SECON, and she has been a Technical Program Committee member of many symposia on wireless communication for several leading international IEEE conferences.
\end{IEEEbiography}

\begin{IEEEbiography}[{\includegraphics[width=1in,height=1.25in,clip,keepaspectratio]{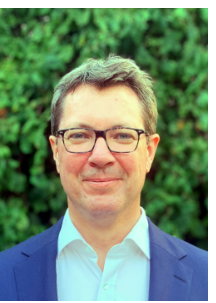}}]
{Eduard A. Jorswieck} is managing director of the Institute of Communications Technology and the head of the Chair for Information Theory and Communications Systems and Full Professor at Technische Universität Braunschweig, Brunswick, Germany. From 2008 until 2019, he was the head of the Chair of Communications Theory and Full Professor at TU Dresden, Germany. Eduard’s main research interests are in the broad area of communications. He has co-authored some 160 journal papers, 15 book chapters, 4 monographs, and more than 300 conference papers. Since 2017, he serves as Editor-in-Chief of the Springer EURASIP Journal on Wireless Communications and Networking. He currently serves as editor for IEEE Transactions on Communications. He has served on the editorial boards for IEEE Transactions on Signal Processing, IEEE Transactions on Wireless Communications, IEEE Signal Processing Letters, and IEEE Transactions on Information Forensics and Security. In 2006, he received the IEEE Signal Processing Society Best Paper Award.
\end{IEEEbiography}

\begin{IEEEbiography}[{\includegraphics[width=1in,height=1.25in,clip,keepaspectratio]{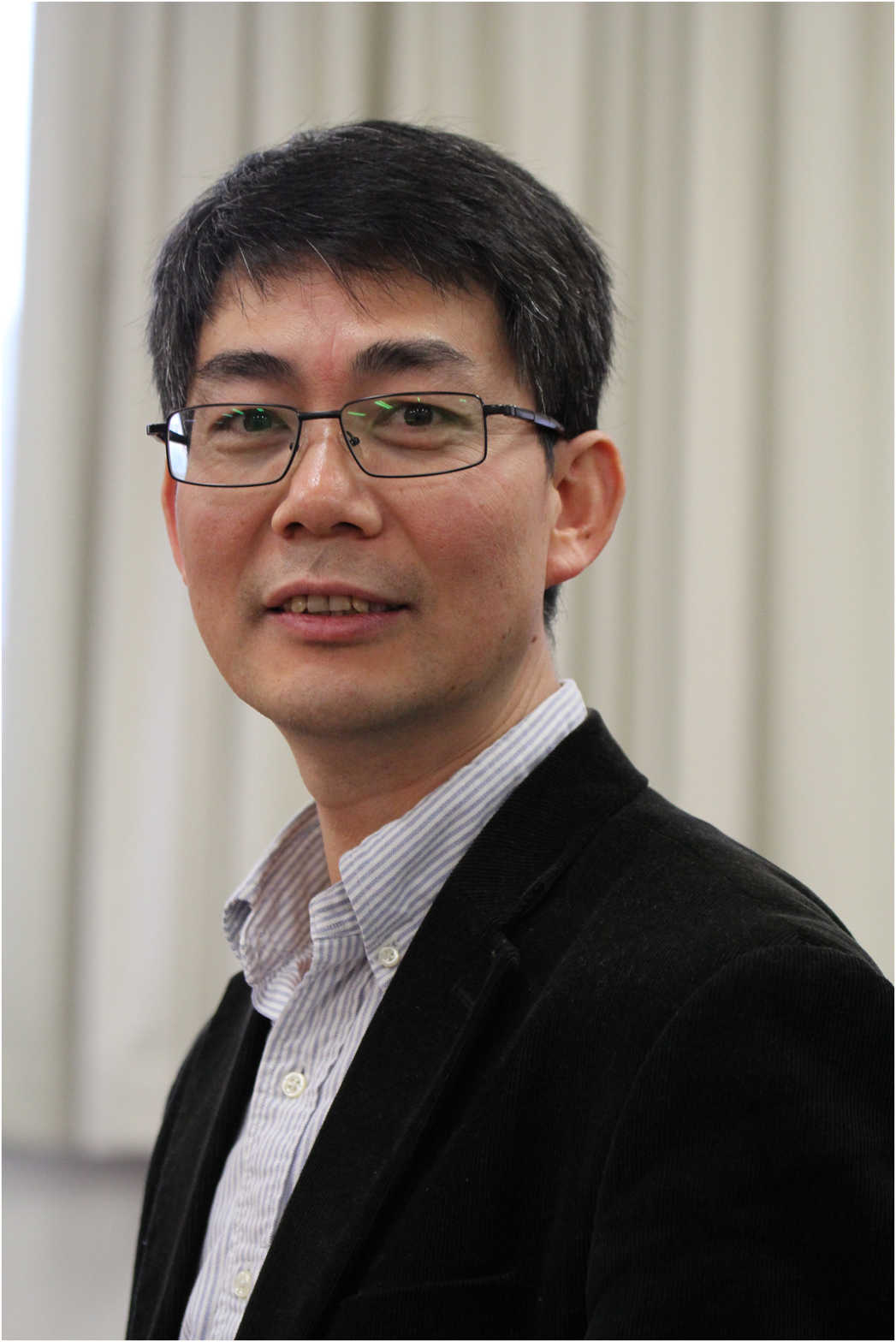}}]
{Jinhong Yuan} (M'02--SM'11--F'16) is a Professor and Head of Telecommunication Group with the School of Electrical Engineering and Telecommunications, The university of New South Wales, Sydney, Australia. He has published two books, five book chapters, over 300 papers in telecommunications journals and conference proceedings, and 50 industrial reports. He is a co-inventor of one patent on MIMO systems and four patents on low-density-parity-check codes. He has co-authored four Best Paper Awards and one Best Poster Award, including the Best Paper Award from the IEEE International Conference on Communications, Kansas City, USA, in 2018, the Best Paper Award from IEEE Wireless Communications and Networking Conference, Cancun, Mexico, in 2011, and the Best Paper Award from the IEEE International Symposium on Wireless Communications Systems, Trondheim, Norway, in 2007. He is an IEEE Fellow and currently serving as an Associate Editor for the IEEE Transactions on Wireless Communications and IEEE Transactions on Communications. He served as the IEEE NSW Chapter Chair of Joint Communications/Signal Processions/Ocean Engineering Chapter during 2011-2014 and served as an Associate Editor for the IEEE Transactions on Communications during 2012-2017. His current research interests include error control coding and information theory, communication theory, and wireless communications.
\end{IEEEbiography}

\begin{IEEEbiography}[{\includegraphics[width=1in,height=1.25in,clip,keepaspectratio]{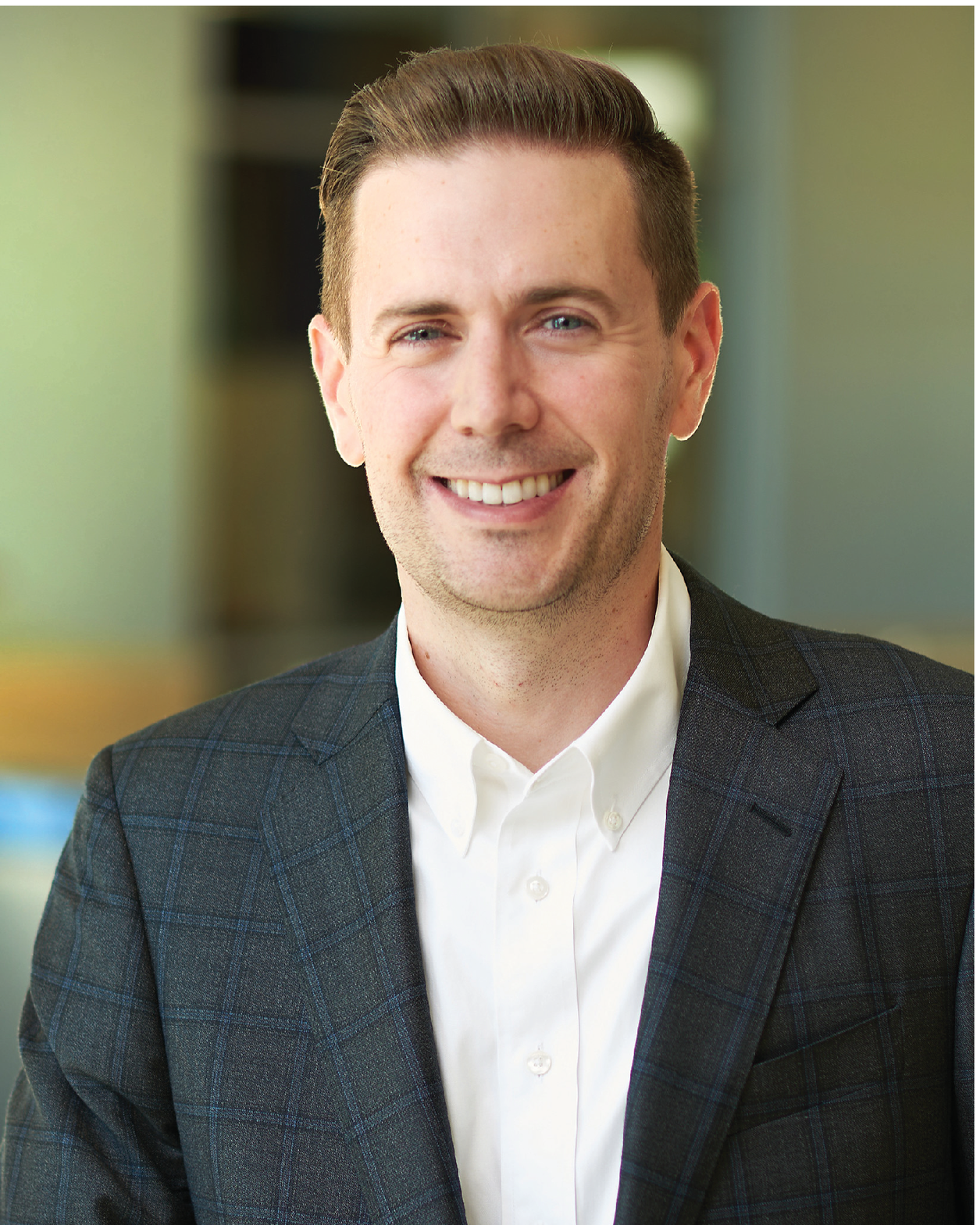}}]
{David J. Love} (S’98 - M’05 - SM'09 - F'15) is the Nick Trbovich Professor of Electrical and Computer Engineering  at Purdue University.  His research interests are in the design and analysis of broadband wireless communication systems, beyond 5G wireless systems, multiple-input multiple-output (MIMO) communications, millimeter wave wireless, software defined radios and wireless networks, coding theory, and MIMO array processing.  He holds 32 issued US patents.   He served as a Senior Editor for IEEE Signal Processing Magazine,  Editor for the IEEE Transactions on Communications,  Associate Editor for the IEEE Transactions on Signal Processing, and  Guest Editor for special issues of the IEEE Journal on Selected Areas in Communications and the EURASIP Journal on Wireless Communications and Networking. Since 2022, he is a Fellow of the  American Association for the Advancement of Science (AAAS).
Along with his co-authors, he won best paper awards from the IEEE Communications Society (2016 Stephen O. Rice Prize and 2020 Fred Ellersick Prize), the IEEE Signal Processing Society (2015 IEEE Signal Processing Society Best Paper Award), and the IEEE Vehicular Technology Society (2010 Jack Neubauer Memorial Award).

\end{IEEEbiography}

\begin{IEEEbiography}[{\includegraphics[width=1in,height=1.25in,clip,keepaspectratio]{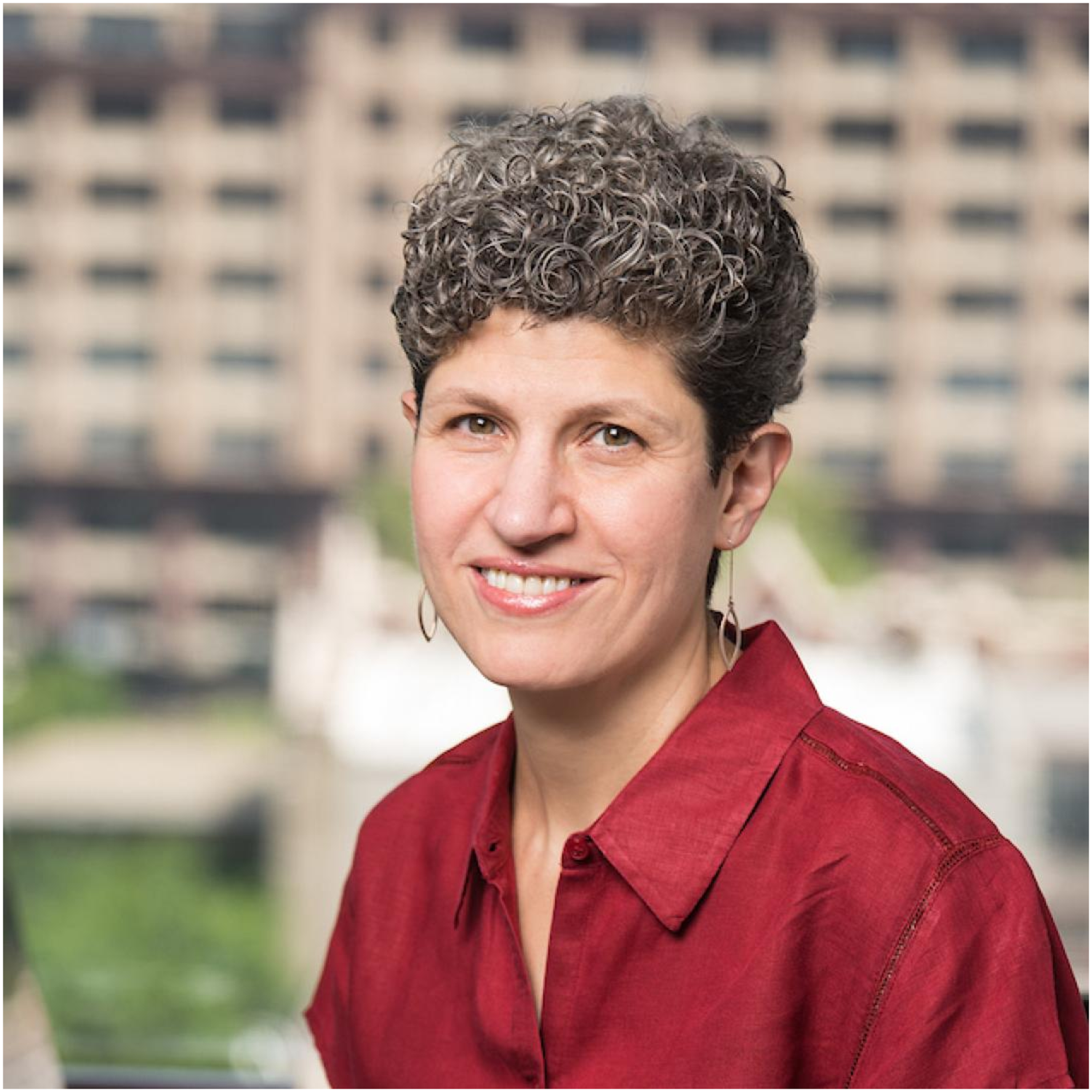}}]
{Elza Erkip}  is an Institute Professor in the Electrical and Computer Engineering Department at New York University Tandon School of Engineering. She received the B.S. degree in Electrical and Electronics Engineering from Middle East Technical University, Ankara, Turkey, and the M.S. and Ph.D. degrees in Electrical Engineering from Stanford University, Stanford, CA, USA.  Her research interests are in information theory, communication theory, and wireless communications.

Dr. Erkip is a member of the Science Academy of Turkey and is a Fellow of the IEEE. She received the NSF CAREER award in 2001, the IEEE Communications Society WICE Outstanding Achievement Award in 2016, the IEEE Communications Society Communication Theory Technical Committee (CTTC) Technical Achievement Award in 2018, and the IEEE Communications Society Edwin Howard Armstrong Achievement Award in 2021. She was the Padovani Lecturer of the IEEE Information Theory Society in 2022. Her paper awards include the IEEE Communications Society Stephen O. Rice Paper Prize in 2004,  the IEEE Communications Society Award for Advances in Communication in 2013 and  the IEEE Communications Society Best Tutorial Paper Award in 2019.  She was a member of the Board of Governors of the IEEE Information Theory Society 2012-2020, where she was the President in 2018.   She was a Distinguished Lecturer of the IEEE Information Theory Society from 2013 to 2014.

\end{IEEEbiography}

\begin{IEEEbiography}[{\includegraphics[width=1in,height=1.25in,clip,keepaspectratio]{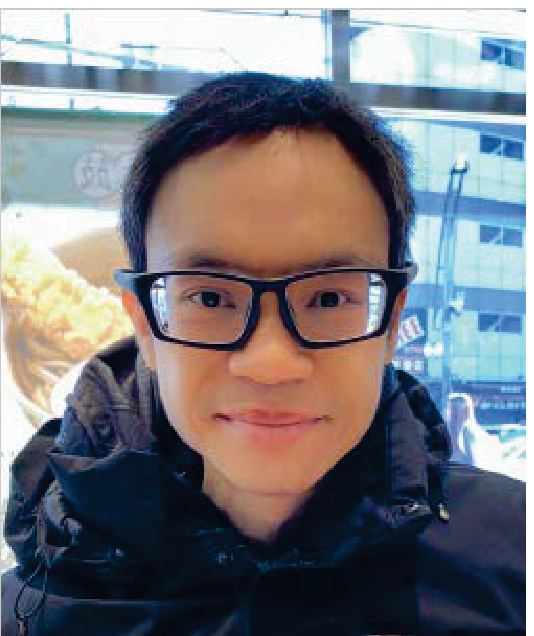}}]
{Dusit Niyato} is the President's Chair Professor in the School of Computer Science and Engineering, at Nanyang Technological University, Singapore. He received B.Eng. from King Mongkuts Institute of Technology Ladkrabang (KMITL), Thailand in 1999 and Ph.D. in Electrical and Computer Engineering from the University of Manitoba, Canada in 2008. His research interests are in the areas of sustainability, edge intelligence, decentralized machine learning, and incentive mechanism design.
\end{IEEEbiography}

\end{document}